\documentclass[%
 reprint,
superscriptaddress,
 amsmath,amssymb,
 aps,
 nofootinbib,
prx,
]{revtex4-1}
\usepackage{layouts}
\usepackage{mathtools}
\usepackage{graphicx}
\usepackage{dcolumn}
\usepackage{bm}
\usepackage{gensymb}
\usepackage[breaklinks=true]{hyperref}
\usepackage{multirow}
\hypersetup{
colorlinks   = true, 
urlcolor     = blue, 
linkcolor    = blue, 
citecolor    = red 
}
\usepackage{cleveref}
\usepackage{comment}
\usepackage[normalem]{ulem}
\usepackage{xcolor}

\usepackage{xy}
\xyoption{matrix}
\xyoption{frame}
\xyoption{arrow}
\xyoption{arc}

\usepackage{ifpdf}
\ifpdf
\else
\PackageWarningNoLine{Qcircuit}{Qcircuit is loading in Postscript mode.  The Xy-pic options ps and dvips will be loaded.  If you wish to use other Postscript drivers for Xy-pic, you must modify the code in Qcircuit.tex}
\xyoption{ps}
\xyoption{dvips}
\fi

\entrymodifiers={!C\entrybox}

\newcommand{\bra}[1]{{\left\langle{#1}\right\vert}}
\newcommand{\ket}[1]{{\left\vert{#1}\right\rangle}}
\newcommand{\qw}[1][-1]{\ar @{-} [0,#1]}
\newcommand{\qwx}[1][-1]{\ar @{-} [#1,0]}

\newcommand{\gate}[1]{*+<.6em>{#1} \POS ="i","i"+UR;"i"+UL **\dir{-};"i"+DL **\dir{-};"i"+DR **\dir{-};"i"+UR **\dir{-},"i" \qw}

\newcommand{\measureD}[1]{*{\xy*+=<0em,.1em>{#1}="e";"e"+UR+<0em,.25em>;"e"+UL+<-.5em,.25em> **\dir{-};"e"+DL+<-.5em,-.25em> **\dir{-};"e"+DR+<0em,-.25em> **\dir{-};{"e"+UR+<0em,.25em>\ellipse^{}};"e"+C:,+(0,1)*{} \endxy} \qw}

\newcommand{\control}{*!<0em,.025em>-=-<.2em>{\bullet}}

\newcommand{\ctrl}[1]{\control \qwx[#1] \qw}

\newcommand{\multigate}[2]{*+<1em,.9em>{\hphantom{#2}} \POS [0,0]="i",[0,0].[#1,0]="e",!C *{#2},"e"+UR;"e"+UL **\dir{-};"e"+DL **\dir{-};"e"+DR **\dir{-};"e"+UR **\dir{-},"i" \qw}
    
\newcommand{\ghost}[1]{*+<1em,.9em>{\hphantom{#1}} \qw}

\newcommand{\rstick}[1]{*!L!<-.5em,0em>=<0em>{#1}}
\newcommand{\lstick}[1]{*!R!<.5em,0em>=<0em>{#1}}

\newcommand{\Qcircuit}{\xymatrix @*=<0em>}

\crefname{equation}{Eq.}{Eqs.}
\Crefname{equation}{Eq.}{Eqs.}
\crefname{figure}{Fig.}{Figs.}
\Crefname{figure}{Fig.}{Figs.}
\crefname{section}{Sec.}{Secs.}
\Crefname{section}{Sec.}{Secs.}
\crefname{appendix}{Appendix}{Appendices}
\Crefname{appendix}{Appendix}{Appendices}
\crefname{table}{Table}{Tables}
\Crefname{table}{Table}{Tables}

\newcommand{\braket}[1]{\langle #1 \rangle}
\newcommand{\ip}[2]{\langle{#1}|{#2}\rangle}
\newcommand{\op}[2]{\ket{#1}\bra{#2}}
\newcommand{\expt}[1]{\langle{#1}\rangle}
\newcommand{\dg}{^\dagger}

\newcommand{\tr}{\text{tr}}
\newcommand{\e}{e}

\newcommand{\dd}{d}

\newcommand{\half}{\frac{1}{2}}

\newcommand{\hn}{\hat{n}}

\newcommand{\CROT}{\textsc{crot}}
\newcommand{\CNOT}{\textsc{cnot}}

\newcommand{\SUM}{\textsc{sum}}

\newcommand{\optimal}{\textsc{optimal}}
\newcommand{\pg}{\textsc{pretty good}}
\newcommand{\phase}{\textsc{phase}}

\newcommand{\smallfrac}[2]{{\mbox{$\frac{#1}{#2}$}}}
\newcommand{\CN}{\hat{R}_N}
\newcommand{\ZN}{\hat{Z}_N}
\newcommand{\XN}{\hat{X}_N}
\newcommand{\SN}{\hat{S}_N}
\newcommand{\TN}{\hat{T}_N}

\newcommand{\TransN}{\hat{\Sigma}_N} 

\newcommand{\lbar}{\bar}
\newcommand{\I}{\lbar{I} }
\newcommand{\X}{\lbar X}

\newcommand{\Z}{\lbar Z}
\newcommand{\Hd}{\lbar H}
\newcommand{\CZ}{\lbar{C}_Z}
\newcommand{\Had}{\lbar H}
\newcommand{\Sgate}{\lbar S}
\newcommand{\Tgate}{\lbar T}

\newcommand{\Pauli}{\lbar{P}}
\newcommand{\Clifford}{\lbar{C}}
\newcommand{\MX}{\mathcal{M}_X}
\newcommand{\MZZL}{\mathcal{M}_{Z_LZ_L}}

\newcommand{\diag}{\text{diag}}

\newcommand{\fc}{f} 

\newcommand{\PB}{\textnormal{\ensuremath{\mathtt{pb}}}}
\newcommand{\cat}{\textnormal{\ensuremath{\mathtt{cat}}}}
\newcommand{\scat}{\textnormal{\ensuremath{\mathtt{scat}}}}
\newcommand{\bin}{\textnormal{\ensuremath{\mathtt{bin}}}}
\newcommand{\triv}{\textnormal{\ensuremath{\mathtt{triv}}}}
\newcommand{\can}{\textnormal{\ensuremath{\mathtt{phase}}}}
\newcommand{\code}{\textnormal{\ensuremath{\mathtt{code}}}}
\newcommand{\ncode}{\ensuremath{\bar n_\code}}

\newcommand{\ON}{\textnormal{\ensuremath{\mathtt{0N}}}}
\newcommand{\gkp}{\textnormal{\ensuremath{\mathtt{gkp}}}}
\newcommand{\gkps}{\textnormal{\ensuremath{\mathtt{gkps}}}}
\newcommand{\gkph}{\textnormal{\ensuremath{\mathtt{gkph}}}}

\newcommand{\dtheta}{\Delta_N(\theta)}
\newcommand{\dthetahet}{\Delta^\text{het}_N(\theta)}

\makeatletter
\def\@bibdataout@aps{%
 \immediate\write\@bibdataout{%
  @CONTROL{%
   apsrev41Control,author="08",editor="1",pages="0",title="0",year="1",eprint="1"%
  }%
 }%
 \if@filesw
  \immediate\write\@auxout{\string\citation{apsrev41Control}}%
 \fi
}%
\makeatother 

\begin{document}

\title{
Quantum computing with rotation-symmetric bosonic codes
}

\author{Arne L. Grimsmo}\email{arne.grimsmo@sydney.edu.au}
\affiliation{Centre for Engineered Quantum Systems, School of Physics, The University of Sydney, Sydney, Australia}
\author{Joshua Combes}
\affiliation{Centre  for  Engineered  Quantum  Systems,  School  of  Mathematics  and  Physics, The  University  of  Queensland,  St.  Lucia  QLD  4072,  Australia}
\author{Ben Q. Baragiola}
\affiliation{Centre for Quantum Computation and Communication Technology, School of Science, RMIT University, Melbourne, Victoria 3001, Australia}

\date{\today}

\begin{abstract}
Bosonic rotation codes, introduced here, are a broad class of bosonic error-correcting codes based on phase-space rotation symmetry.
We present a universal quantum computing scheme applicable to a subset of this class---number-phase codes---which includes the well-known cat and binomial codes, among many others.
The entangling gate in our scheme is code-agnostic and can be used to interface different rotation-symmetric encodings.
In addition to a universal set of operations, we propose a teleportation-based error correction scheme that allows recoveries to be tracked entirely in software. Focusing on cat and binomial codes as examples, we compute average gate fidelities for error correction under simultaneous loss and dephasing noise and show numerically that the error-correction scheme is close to optimal for error-free ancillae and ideal measurements.
Finally, we present a scheme for fault-tolerant, universal quantum computing based on concatenation of number-phase codes and Bacon-Shor subsystem codes.
\end{abstract}

\maketitle

\section{\label{sec:intro} Introduction}

Encoding quantum information into bosonic systems \cite{ChuaLeunYama97, CochMilbMunr99, Gottesman01}
is an alternative to conventional error-correcting codes based on discrete two- (or few-) state systems.  
The infinite Hilbert space of a single bosonic mode provides a playground for redundant digital encodings that can be tailored to a specific task \cite{LundRalpHase08, Mirrahimi14, Michael16, Puri:2017aa, Puri:2018aa, Noh:2018aa, Linshu-Li:2019aa}.
A natural application is to use a bosonic code at the ground level in a concatenated error-correction scheme to suppress errors below the fault-tolerance threshold of a conventional qubit-based code~\cite{Gottesman01,Menicucci14}, potentially reducing the total overhead.
Decoders that exploit the continuous-variable nature of bosonic codes can improve the fault-tolerance threshold~\cite{Fukui:2017aa, Vuillot2018, Noh:2019aa} and reduce the number of physical qubits required \cite{Fukui:2018aa}. 
From a hardware perspective, well controlled, low loss bosonic modes occur in many quantum technology platforms, such as electromagnetic modes in optical cavities~\cite{Kimble:1998aa, Mabuchi2002} and free space \cite{Yokoyama:2013aa}, superconducting circuits and microwave cavities~\cite{Wallraff2004,Devoret2013,Barends2014,Haroche2006, Reagor2013}, and motional modes in ion traps~\cite{Leibfried:2003aa, Monroe1164, Fluhmann:2019aa}.

Gottesman, Kitaev, and Preskill (GKP) introduced a seminal scheme for quantum computing with bosonic codes, based on states with discrete translation symmetry in phase space~\cite{Gottesman01}.
While GKP codewords were recently prepared and manipulated in the laboratory~\cite{Fluhmann:2019aa}, experimental progress with other bosonic codes---especially the cat~\cite{CochMilbMunr99, LeghKircVlas13, Mirrahimi14, BergLooc16} and binomial code families~\cite{Michael16}---is more advanced.
In a breakthrough experiment,
Ofek \emph{et al.}~\cite{Ofek16} demonstrated enhancement in the lifetime of a cat-code qubit compared to an unencoded qubit using the same hardware---the so-called break-even point for error correction. This was the first error-correction scheme to achieve this milestone in an experiment. A similar experiment using binomial codes also came very close to break even~\cite{Hu:2019aa}. Initial steps towards fault-tolerant error correction with these codes have been made recently~\cite{Rosenblum2018}.

In this work we show how cat and binomial codes belong to a larger family of bosonic codes characterized by discrete rotation symmetry in phase space, in analogy to the discrete translation symmetry of GKP codes.
Specifically,
we consider codes where a single qubit is encoded into a subspace in which the discrete rotation operator
$\ZN = \exp \left(i \frac{\pi}{N}\hat n  \right)$
acts as logical $\Z$, where $\hat n$ is the Fock-space number operator.
We refer to these codes as \emph{bosonic rotation codes}. 
The parameter $N$ here quantifies the degree of discrete rotation symmetry for the code:
It immediately follows that the operator $\CN = \ZN^2$ acts as the identity on the codespace, \emph{i.e.}, any state in the codespace is $N$-fold rotation symmetric in the sense that it is invariant under a rotation by $2\pi/N$.

A consequence of $N$-fold rotation symmetry is that an encoded state $\ket{\psi_N}$ has support only on every $N$th Fock state, $\ket{\psi_N} = \sum_{k=0}^\infty c_{kN}\ket{kN}$.
The degree of rotation symmetry $N$ thus quantifies the magnitude of a detectable shift in number and sets a code distance in Fock space.
A complementary notion of distance for phase-space rotations, $\pi/N$, quantifies detectable rotation errors and  reveals a trade-off between detectable number-shift and rotation errors.

A special subset of the bosonic rotation codes we introduce are
\emph{number-phase codes}, which are rotated superpositions of states with small phase uncertainty~\cite{Holevo11}. We show that vanishing phase uncertainty is related to (approximate) number-translation symmetry.
In the limit of vanishing phase uncertainty, the number-phase codes are analogous to ideal GKP codes,
 with number and phase playing dual roles in place of position and momentum. 
 Interestingly, we show that cat codes, binomial codes, and the shift-resistant qudit codes introduced in Ref.~\cite{Gottesman01} each approach this ideal limit of vanishing phase uncertainty as their average excitation number increases.

We present a scheme for universal quantum computation for number-phase codes, where the workhorse is an entangling controlled-rotation ($\CROT$) gate based on a cross-Kerr interaction ($\sim \hat n \otimes \hat n$).
A properly tuned $\CROT$ serves as a logical controlled-$Z$, $\hat C_Z = \text{diag}(1,1,1,-1)$, between two number-phase codes.
Notably, codes with different degree of rotation symmetry and of different type (\emph{e.g.} cat and binomial) can be fully entangled by the $\CROT$ gate.
Similarly, a self-Kerr interaction ($\sim \hat n^2$) can be used to enact the single-qubit phase gate $\hat S=\diag(1, i)$. Augmented with preparation of dual-basis codewords $\ket{+_N}$ and destructive measurements in the dual basis, all Clifford gates can be executed. Universality is achieved using injected magic states.
The gates we introduce have favorable properties in terms of fault tolerance: Errors are amplified and spread in a very limited way, such that the ``size'' of an error after a gate is proportional to the ``size'' prior to the gate.

We also introduce a new error-correction scheme for the number-phase codes discussed above. 
The error correction is based on teleporting the encoded information into a new, fresh ancilla~\cite{Knill05}. A remarkable feature of this approach is that recovery operations can be tracked entirely in software. In essence, the need for potentially difficult and highly nonlinear operations to restore the codespace has been replaced by preparation of ancillae in logical basis states. 

We perform a numerical study of error-correction performance in the presence of simultaneous loss and dephasing noise on encoded qubits, while assuming noiseless ancillae.
A critical question for a bosonic error-correcting code is whether it performs better than the ``trivial encoding''---a qubit encoded in Fock states $\ket 0$ and $\ket 1$---under the same noise model. The point where the error-corrected qubit performs as well as the trivial encoding is referred to as the break-even point for error correction.
For the studied noise model, we find that both cat and binomial codes go beyond break even by several orders of magnitude for degrees of rotation symmetry in the range $N=2$--$4$, and dimensionless noise strength for simultaneous loss and dephasing in the range $\kappa t = 10^{-3}$--$10^{-2}$.
Remarkably, we also find that the teleportation-based error correction scheme we introduce performs nearly as well as the optimal recovery map found numerically, for noise-free ancillae and idealized measurements.

Finally, we outline a fault-tolerant and universal scheme based on concatenating number-phase codes with Bacon-Shor subsystem codes~\cite{Bacon2006,Poulin2005,Aliferis08,brooks2013fault}. The concatenation serves three purposes: To deal with errors in state preparation which are hard to check for, to suppress errors in noisy measurements by exploiting correlations across multiple measurement outcomes, and to further suppress the logical error rate by dealing with errors that are too large for the bosonic code to handle.
Our specific example using a Bacon-Shor code illustrates the broader point that a fault-tolerant scheme should be tailored to strengths and weaknesses of the bosonic code at the ground level.

This paper is laid out as follows. 
In \cref{sec:gcb} we detail the structure of bosonic rotation codes and describe how to distinguish the codewords in the computational and dual bases using number and phase measurements, respectively. We introduce a measure of modular phase uncertainty that quantifies embedded phase-measurement error in a given bosonic rotation code. 
In \cref{sec:CyclicStabilizerCodes} we define number-phase codes, a family of bosonic rotation codes with small modular phase uncertainty, and we give examples of number-phase codes, including cat and binomial codes. 
In \cref{sec:gates} we present a universal scheme for quantum computing with number-phase codes. 
In \cref{sec:numbermodN} we give a method to perform a modular excitation number measurement, $\hat n \bmod N$, which can be used to detect number errors and to prepare bosonic rotation codewords.
In \cref{sec:errorsanderrorprop} we study in detail error propagation and
show that the gates in our set map small errors to small errors and are in that sense fault tolerant. 
In \cref{sec:EC} we lay out a scheme for teleportation-based error correction that can be used for any number-phase code, and we 
numerically test the performance of our error-correction procedure for cat and binomial codes under simultaneous loss and dephasing.
In~\cref{sec:concat} we outline a fault-tolerant scheme for quantum computing based on concatenation of number-phase codes and Bacon-Shor subsystem codes.
Finally, in \cref{sec:conclusion}, we summarize our results
and highlight open problems.

\section{\label{sec:gcb} Bosonic rotation codes}

\begin{figure*}
  \centering 
  \includegraphics[width=0.94\textwidth]{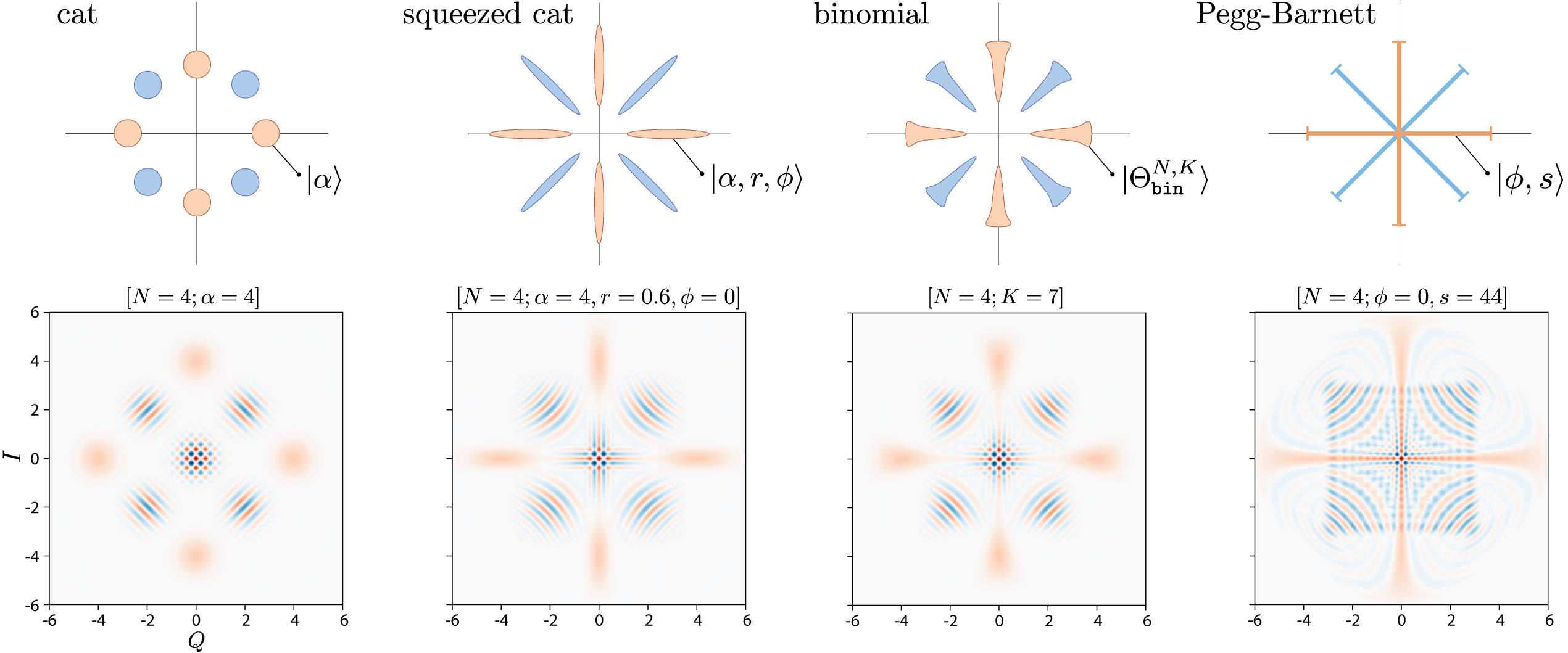}
  \caption{\label{fig:summary}
  Graphical summary of several $N=4$ rotation codes: cat and squeezed cat (\cref{Appendix:squeezedcat}), binomial (\cref{appendix:binomial}), and Pegg-Barnett (\cref{appendix:PB}). Logical codewords are $+1$ eigenstates of the discrete rotation operator, \cref{eq:C_N}, and exhibit $N$-fold rotation symmetry. Top row: ball-and-stick diagrams illustrating $\ket{+_N}$ (orange) and $\ket{-_N}$ (blue). Indicated on each is the primitive $\ket{\Theta}$ for the code.
  Bottom row: Wigner functions for the $\ket{+_N}$ state, $W_{\ket{+_N} }(
  \alpha)$. Red/blue are positive/negative, the color scale on each plot is different, and $Q = \frac{1}{2} (\alpha + \alpha^*)$ and $I = \frac{1}{2i} (\alpha - \alpha^*)$ are the real and imaginary parts of $\alpha$.
  }
\end{figure*}

Single-mode bosonic codes utilize the large Hilbert space of a bosonic mode to protect logical quantum information. There are countless ways to embed a qubit, with each encoding comprising a two-dimensional subspace spanned by states serving as the logical codewords, and the remaining Hilbert space providing the freedom to detect and correct errors. 
In many physical implementations, the dominant sources of noise are loss, gain, and dephasing. Various bosonic qubit encodings have been designed specifically to counteract these noise processes, notably cat~\cite{CochMilbMunr99, LeghKircVlas13, Mirrahimi14, BergLooc16} and binomial codes~\cite{Michael16}, which share the key property that the codewords exhibit discrete rotational symmetry. 

We use discrete rotation symmetry to define a class of single-mode codes called bosonic rotation codes. First, we say that a code has discrete $N$-fold rotation symmetry if any state $\ket\psi$ in the code subspace (codespace) is a $+1$ eigenstate of the discrete rotation operator\footnote{Whenever we talk about a code with $N$-fold rotation symmetry we will always implicitly assume that the code does not have any degree of rotation symmetry \emph{higher} than $N$.}
\begin{equation}\label{eq:C_N}
        \CN = e^{i \frac{2\pi}{N}\hat n },
\end{equation}
where $\hat n = \hat a\dg \hat a$ is the Fock-space number operator ($[\hat a,\hat a\dg]=1$). 
It follows that the square root of $\CN$ preserves the rotation symmetry and acts as a Hermitian operator that squares to the identity on the codespace. We define an order-$N$ \emph{bosonic rotation code}, or rotation code for short, to be a code where the operator
\begin{equation}\label{eq:Z_N}
    \ZN \coloneqq \hat R_{2N} = \e^{i \frac{\pi}{N}\hn},
\end{equation}
acts as logical $\Z$.
The choice of $\Z$ over $\X$ is just a convention, but we emphasize that a code where $\ZN$ acts as a non-Pauli operator is \emph{not} a rotation code according to our definition. An example of the latter is a square-lattice GKP code where $\Z_2$ acts as a logical Hadamard. This distinction is important when discussing ``easy'' and ``hard'' gates for different codes. We briefly comment on the rotation symmetry of GKP codes in~\cref{sec:GKP}.

The logical codewords for any order-$N$ rotation code can be constructed from discrete rotated superpositions of a normalized \emph{primitive} state $\ket{\Theta}$. In fact, we have that two states $\ket{0_{N,\Theta}}$ and $\ket{1_{N,\Theta}}$ satisfy $\ZN \ket{j_{N,\Theta}} = (-1)^j\ket{j_{N,\Theta}}$ if and only if the two states can be written in the following form:
\begin{subequations} \label{eq:01codewords}
    \begin{align} \label{eq:comp0_rot}
        \ket{0_{N, \Theta}}
            =& \frac{1}{\sqrt{ \mathcal{N}_0} } \sum_{m=0}^{2N-1} e^{i \frac{m \pi}{N} \hat{n}} \ket{\Theta}, \\
        \ket{1_{N, \Theta}}
            =& \frac{1}{\sqrt{ \mathcal{N}_1} } \sum_{m=0}^{2N-1} (-1)^m e^{i \frac{m \pi}{N} \hat{n}} \ket{\Theta}, \label{eq:comp1_rot}
    \end{align}
\end{subequations}
where $\mathcal N_i$ are normalization constants. 
There is a technical constraint on $\ket\Theta$ for the codewords to be non-zero, which we specify precisely in~\cref{sec:FockStructure}.
When the rotated primitives appearing in~\cref{eq:01codewords} are orthogonal, $\bra{\Theta}  (\ZN)^m \ket{\Theta} = 0$ for $0 < m < 2N$, then $\mathcal{N}_0 = \mathcal{N}_1 = 2N$. Generally, however, they are not orthogonal, and the normalization constants are different, $\mathcal{N}_0 \neq \mathcal{N}_1$. Regardless, the codewords themselves are exactly orthogonal for any valid $\ket \Theta$, and any state in the codespace, $\ket{\psi_N} = a \ket{0_{N}} + b \ket{1_{N}}$, has $N$-fold rotation symmetry, $\CN \ket{\psi_N} = \ket{\psi_N}$. Here and henceforth we suppress the $\Theta$-subscript when referring to a generic rotation code of order $N$ unless it is required for clarity. 

From the above, we recognize that there are many different rotation codes with the same degree of rotation symmetry $N$. While a given primitive $\ket \Theta$ uniquely defines an order-$N$ rotation code, there are many possible primitives that give rise to the same code.
Several examples of rotation codes are shown in~\cref{fig:summary}. The best known example are cat codes, where the primitive is a coherent state $\ket{\Theta_{\rm \cat}} = \ket{\alpha}$~\cite{Mirrahimi14}.\footnote{It is also common to define nonorthogonal codewords for cat codes, as in, \emph{e.g.}, Ref.~\cite{LundRalpHase08}, but we here follow the more recent convention from, \emph{e.g.}, Ref.~\cite{Mirrahimi14}.}
Another simple example are the ``\ON{} codes'' 
$\ket{0_{\ON}} = \ket 0$, $\ket{1_{\ON}} = \ket N$~\cite{Sabapathy:2018aa, Elder:2019aa}, which can be constructed, \emph{e.g.}, from $\ket{\Theta_{\ON}} = (\ket 0 + \ket N)/\sqrt 2$.

The form of~\cref{eq:01codewords} is reminiscent of GKP codes whose codewords are superpositions of translated squeezed states~\cite{Gottesman01}. However, due the natural $2\pi$ periodicity of rotations, only a finite number of superpositions are required, and the resulting codeword always remains physical. This is because the rotation operator in \cref{eq:Z_N} conserves excitation number, unlike the displacement operator used to define the GKP codewords.  In~\cref{sec:CyclicStabilizerCodes}, We make a tighter analogy between GKP codes and a subclass of rotation codes called number-phase codes.

\subsection{Fock-space structure} \label{sec:FockStructure}

Discrete rotational symmetry enforces a spacing of the codewords in Fock space, which underpins the codes' robustness to loss and gain errors.
This can be seen by acting the operator $\ZN$ on an arbitrary state $\ket{\psi} = \sum_n a_n \ket n$: $\ZN \ket{\psi} = \sum_n \e^{i\pi n/N} a_n \ket{n}$. Clearly $\ket\psi$ is a $+1$ eigenstate of $\ZN$ if and only if $a_n = 0$ for all $n\neq 2kN$ for integer $k$. Similarly, $\ket \psi$ is a $-1$ eigenstate if and only if $a_n=0$ for all $n \neq (2k+1)N$. This leads to the following  general form for any $\pm 1$ eigenstates of $\ZN$:
\begin{subequations}\label{eq:01logical}
\begin{align}
    \ket{0_N} ={}& \sum_{k=0}^\infty \fc_{2kN} \ket{2kN}, \label{eq:01logical_0} \\
    \ket{1_N} ={}& \sum_{k=0}^\infty \fc_{(2k+1)N} \ket{(2k+1)N} \label{eq:01logical_1}.
\end{align}
\end{subequations}
The coefficients $\fc_{kN}$ in \cref{eq:01logical} are related to the Fock-state coefficients of the primitive $\ket\Theta = \sum_n c_n \ket n$ in~\cref{eq:01codewords} as follows: $\fc_{2kN} = c_{2kN}/\sqrt{\mathcal{N}_0^*}$ and $\fc_{(2k+1)N} = c_{(2k+1)N}/\sqrt{\mathcal{N}_1^*}$, where $\mathcal{N}_0^* = \sum_k |c_{2kN}|^2$ and $\mathcal{N}_1^* = \sum_k |c_{(2k+1)N}|^2$.
The normalization factors introduced here are related to the normalization constants in~\cref{eq:01codewords} as $\mathcal{N}^*_i =  \mathcal{N}_i/(2N)^2$.
There are no further restrictions on the coefficients $\fc_{kN}$ apart from normalization.

To explicitly connect the two representations of rotation-symmetric codewords, \cref{eq:01codewords} and \cref{eq:01logical}, we use the relation for a Kroenecker comb,
\begin{equation} \label{eq:delta}
    \frac{1}{M}\sum_{m=0}^{M-1} \e^{i \frac{2\pi mn}{M}} = \sum_{k=0}^\infty \delta_{n,kM} \quad \text{for $n =0,1,2,\dots$},
\end{equation}
to construct a projector onto the set of Fock states $\ket{2kN+\ell}$ for $k=0,1,2\dots$:\footnote{Note that the parity operator is $\hat{\Pi}_2^0-\hat{\Pi}_2^1$.}
\begin{equation} \label{eq:projector_ell}
    \begin{aligned}
        \hat{\Pi}_{2N}^\ell \coloneqq {}& \sum_{k=0}^\infty \op{2kN+\ell}{2kN+\ell} \\
        ={}& \frac{1}{2N}\sum_{m=0}^{2N-1} \left(\e^{-i\frac{\pi \ell}{N}} \ZN\right)^m.
    \end{aligned}
\end{equation}
The factors $\e^{i\pi \ell/N}$ with $\ell \in \{0,1,\dots,2N-1\}$ are the complex $2N$th roots of unity.
Acting the projector on a primitive $\ket\Theta = \sum_{n} c_n \ket n$ with at least one non-zero $c_{2kN+\ell}$,
\begin{equation}\label{eq:generaleigenstate}
    \hat\Pi_{2N}^\ell \ket{\Theta} = \sum_{k=0}^\infty c_{2kN+\ell}\ket{2kN+\ell},
\end{equation}
produces an (unnormalized) eigenstate of the operator $\ZN$ with eigenvalue 
$\e^{i\pi \ell/N}$.\footnote{This can be checked using~\cref{eq:projector_ell} and the fact that $(\ZN)^{2N} = (\ZN)^0 = \hat I$, $(\e^{i\pi \ell/N})^{2N} = (\e^{i\pi \ell/N})^0 = 1$.}

It is now straightforward to see that \cref{eq:01codewords} and \cref{eq:01logical} are equivalent, since we can write
\begin{subequations}\label{eq:01logical_projector}
\begin{align}
    \ket{0_N} ={}& \frac{\hat{\Pi}_{2N}^0 \ket{\Theta}}{ \sqrt{\mathcal{N}_0^*}},\\
    \ket{1_N} ={}& \frac{ \hat{\Pi}_{2N}^N \ket{\Theta}}{ \sqrt{\mathcal{N}_1^*}},
\end{align}
\end{subequations}
with $\mathcal{N}_0^* = \bra{\Theta} \hat{\Pi}^0_{2N} \ket{\Theta}$ and $\mathcal{N}_1^* = \bra{\Theta} \hat{\Pi}^N_{2N} \ket{\Theta}$. With the help of \cref{eq:projector_ell} we recognize these as \cref{eq:comp0_rot,eq:comp1_rot}. For the codewords to be well-defined it is now clear that the primitive must have support on at least one of the $\ket{2kN}$ Fock states and one of the $\ket{(2k+1)N}$ Fock states. This is the only constraint on $\ket\Theta$, and the orthogonality of the codewords is then easily seen from the fact that they have support on strictly different Fock states.

The dual-basis codewords, $\ket{\pm_N}$, are constructed as usual via superpositions of the computational basis codewords, $\ket{\pm_N} = (\ket{0_N} \pm \ket{1_N})/\sqrt{2}$ leading to Fock space representations
	\begin{subequations} \label{eq:pmlogical}
	\begin{align} 
    \ket{+_N} = & \frac{1}{\sqrt{2}}\sum_{k=0}^\infty  \fc_{kN} \ket{kN}\label{eq:plus},\\
    \ket{-_N} = & \frac{1}{\sqrt{2}} \sum_{k=0}^\infty  (-1)^k \fc_{kN} \ket{kN}\label{eq:minus}.
	\end{align}
	\end{subequations}
Both $\ket{\pm_N}$ have support on the full set of $\ket{kN}$ Fock states.
From~\cref{eq:01logical_0,eq:plus} it is clear that any logical $\ket{0_N}$ state is also a logical $\ket{+_{2N}}$ for a code with twice the rotation symmetry, which is not necessarily the \emph{same} code. In other words, for a given codeword $\ket{0_{N,\Theta}}$ defined by a primitive $\ket{\Theta}$, we have that
\begin{equation}\label{eq:zerplusrelation}
    \ket{0_{N,\Theta}} = \ket{+_{2N,\Theta'}},
\end{equation}
for some primitive $\ket{\Theta'}$. 
Whenever the codeword normalization constant in~\cref{eq:01codewords} satisfy $\mathcal N_0=\mathcal N_1$, the above holds with $\ket{\Theta}=\ket{\Theta'}$. See~\cref{sec:conjugateprimitive} for further details.

While it is evident that $\hat{Z}_N \ket{j_N} =(-1)^j \ket{j_N}$ for $j=0,1$, it is not generally possible to find a similarly simple operator that acts as a logical $\X$ within the codespace.
In~\cref{sec:CyclicStabilizerCodes} we present a set of rotation codes whose logical $\X$ is approximately given by a number-translation operator. The absence of a simple logical $\X$ does fundamentally pose a problem for quantum computing, since universality can be achieved without this operator as shown in~\cref{sec:gates}.

A useful quantity for comparing codes is the average excitation number of the code,
    \begin{align} \label{eq:code_energy}
        \ncode \coloneqq \tfrac{1}{2} \tr \big[ \hat{\Pi}_\code\hat n \big],
     \end{align}
where $ \hat{\Pi}_\code \coloneqq \ket{0_N}\bra{0_N} + \ket{1_N}\bra{1_N}$ is a two-dimensional projector onto the codespace~\cite{Albert17}.\footnote{Encoded logical gates are denoted by an overbar, \emph{e.g.} $\Z$, while operators projected onto the codespace are denoted, following Ref. \cite{Albert17}, by the subscript ``code".} By construction the average excitation number is the same for each of the dual-basis codewords, and $\ncode = \bra{\pm_N} \hat{n} \ket{\pm_N}$. Since the rotation operator commutes with the number operator, $[\CN, \hat{n}]=0$, rotations doe not increase the excitation number. The computational-basis codewords have different average excitation number due to different phases in the superpositions in~\cref{eq:01codewords}, but it remains bounded.
This is in contrast to translation-based GKP codes that suffer from the following: either the code states are unphysical as they have infinite excitations, or the discrete translational symmetry is spoiled by energy constraints.

The \ON{} codes introduced previously are the order-$N$ rotation codes with the lowest possible mean excitation number, $\bar{n}_\ON=\frac{N}{2}$.
It is worth noting that the \emph{trivial encoding} where a qubit is encoded in the two lowest Fock states, 
	\begin{equation} \label{eq:trivialencoding}
    		\ket{0_\triv} = \ket{0}, \quad
       		\ket{1_\triv} = \ket{1} \, ,
	\end{equation}
is a rotation code with $N=1$.
The trivial encoding serves as a useful benchmark against which other codes can be compared~\cite{Ofek16, Albert17, Hu:2019aa}.

\subsubsection{The phase and Fock grids}

\begin{figure}
  \centering
  \includegraphics{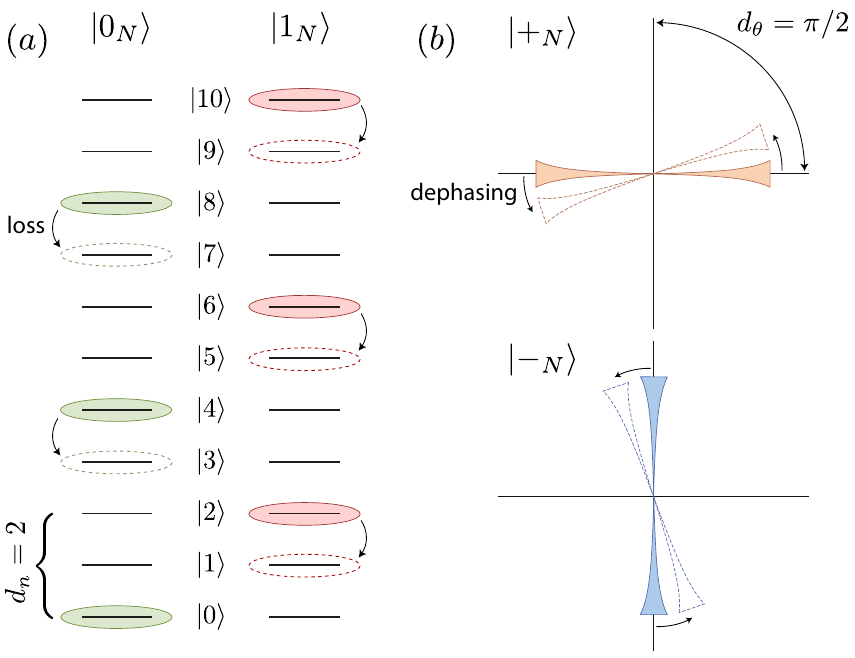}
  \caption{\label{fig:errorstructure} Graphical summary of the Fock-space and phase-space structure of codewords for an $N=2$ rotation code.  (a) The computational-basis codewords $\ket{0_N}$ and $\ket{1_N}$ have support on every $2kN$ and $(2k+1)N$ Fock states for $k=0,1,2\dots$, respectively. Up to $N-1$ loss or gain errors can in principle be detected. (b) The dual codewords $\ket{\pm_N}$ are related by a rotation in phase space by $\pi/2$. Rotation errors small compared to $d_\theta = \pi/2$ are detectable with a code-dependent uncertainty. }
\end{figure}

With a given primitive $\ket\Theta$,
the order of rotation symmetry $N$ parameterizes a code both in phase space and in Fock space, via~\cref{eq:01codewords,eq:01logical}. From these two forms of the codewords, it is natural to define a \emph{number distance} and \emph{rotational distance} of a code as\footnote{In Ref.~\cite{Albert17} and related works, cat and binomial codes are described by a spacing parameter $S$, which is related to code distance by $S =d_n-1$. $S$ is the largest detectable number error.}
\begin{equation} \label{eq:codedistances}
    d_n \coloneqq N , \quad \quad 
    d_\theta \coloneqq \pi/N.
\end{equation}
respectively. 
Number and phase-space rotation play dual roles: codes of increasing $N$ are further separated in Fock space but more closely separated in terms of rotations.
For an order-$N$ rotation code, we refer to the set of angles $\{ m d_\theta \}$ for $m=0,1,\dots2N-1$ as the \emph{phase grid} and the set of Fock states $\{ \ket{kN} \}$ for $k = 0,1, \dots$ as the \emph{Fock grid}. These are in analogy to the position- and momentum-space grids (or more general quadrature-grids) on which GKP codes are defined. They broadly characterize the types of error to which the code is naturally resilient. 
A number-shift error smaller than $d_n$ (\emph{e.g.} $\sim \hat a^k$ or $\sim (\hat a\dg ){}^k$ with $k<N$) is detectable.
For rotation errors the boundary between detectable and undetectable errors is not as sharp in general, but a good code should be able to detect
 a rotation which is small compared to $d_\theta$ (\emph{e.g.} $\e^{i \theta \hat{n}} $ with $\theta<d_\theta$). This basic idea is illustrated in \cref{fig:errorstructure}. We return to a more detailed discussion of ``small'' vs. ``large'' errors in~\cref{sec:errorsanderrorprop}.

\subsection{Distinguishing the codewords} \label{sec:distinguishing}

The structure of rotation codes in Fock space, \cref{eq:01logical}, and in terms of rotations, \cref{eq:01codewords}, suggests a natural way to distinguish codewords in the computational and dual bases using number and phase measurements, respectively. Phase measurements play a crucial role in the quantum computing scheme introduced in~\cref{sec:gates}.
		
\subsubsection{Number measurement}
The computational-basis codewords $\ket{0_N}$ and $\ket{1_N}$ can be distinguished (destructively) by measuring excitation number as seen clearly in \cref{eq:01logical}. For an order $N$ code, a number measurement returns an outcome $kN$ for integer $k$, where even $k$ corresponds to $\ket{0_N}$ and odd $k$ to $\ket{1_N}$. In the presence of errors such a measurement can still correctly identify the damaged codewords. For example, if we assume that upwards and downwards shifts in number (\emph{e.g.} loss and gain) are equally likely, a simple decoder to identify the codeword from a number measurement outcome $n$ is to round to the nearest $kN$ and declare the outcome ``$0$'' if $k$ is even and ``$1$'' if $k$ is odd. In practice the noise model motivates the decoding scheme. Under pure loss, for example, the decoder should always round \emph{up} to the nearest $kN$.

\subsubsection{\label{sec:phaseest}Phase measurement}
An approach to distinguishing $\ket{\pm_{N,\Theta}}$ with a natural robustness to rotation errors relies on \emph{phase estimation}. In the absence of errors, the task can be formulated as follows: Given the codeword $\ket{+_{N,\Theta}}$ for a rotation code, we wish to determine $\theta$ in $\e^{i\theta \hat n}\ket{+_{N,\Theta}}$. If $\theta \pmod{2\pi/N} = 0$ the state is $\ket{+_{N,\Theta}}$ while if $\theta \pmod{2\pi/N} = \pi/N$, the state is $\ket{-_{N,\Theta}}$.
In the presence of rotation errors, the phase estimation problem is generalized to estimating a continuous variable $\theta \in [0,2\pi)$.
A decoding procedure rounds $\theta$ to the nearest $m\pi/N$. Then, if $m$ is even declare the outcome ``$+$,'' and if $m$ is odd declare the outcome ``$-$.'' This is exactly analogous to the excitation number decoding scheme described above. Again, the decoding scheme should in general be adapted to the system's noise model.

Holevo~\cite{Holevo11} and Helstrom~\cite{Helstrom69} introduced an optimal measurement to distinguish states in the one-parameter family $\e^{i\theta\hat n}\ket{\psi_0}$ for the situation where no prior information about $\theta$ is assumed (and for a specific class of ``deviation functions'' quantifying the deviation of the estimated value from the true value).
The POVM elements of this measurement depend on a fiducial state $\ket{\psi_0}=\sum_n c_n\ket n$ that defines the phase estimation problem. 
They can be written in the Fock basis (Theorem 4.4.1 in Ref.~\cite{Holevo11}),
\begin{equation}\label{eq:POVM_Holevo}
  \hat M^\text{can}(\theta) = \frac{1}{2\pi} \sum_{m,n=0}^\infty \gamma_m\gamma_n^* \e^{i(m-n)\theta} \ket m \bra n,
\end{equation}
with completeness relation $\int_0^{2\pi} \dd\theta \, \hat M^\text{can}(\theta) = \hat I$,
and $\gamma_n = c_n/|c_n|$ for $c_n \neq 0$ and otherwise an arbitrary complex number with $|\gamma_n|=1$. In the present context $\ket{\psi_0} = \ket{+_{N,\Theta}} = \frac{1}{\sqrt 2}\sum_{k} \fc_{kN} \ket{kN}$.
We refer to the set of POVMs in~\cref{eq:POVM_Holevo} as the canonical phase measurement.

For an order-$N$ rotation code, the relation $\CN \ket{+_{N,\Theta}} = \ket{+_{N,\Theta}}$ implies that phase measurements only acquire information about the phase $\theta$ modulo $2\pi/N$. To quantify the uncertainty in a phase estimation measurement where we only care about the modular phase, we slightly modify a phase-uncertainty measure introduced by Holevo~\cite{Holevo11}. We adopt a measure of modular phase uncertainty by considering the $2\pi/N$ periodic random variable $\e^{iN\theta}$
under the probability distribution $\mu(\theta) = \tr \big[\hat M^\text{can}(\theta)\ket{+_{N,\Theta}}\bra{+_{N,\Theta}} \big]$ for $\theta$.
We can then define an uncertainty measure
\begin{equation} \label{eq:phaseuncertainty_relation}
  \dtheta \coloneqq \frac{\braket{\Delta \e^{iN\theta}}}{|\braket{\e^{iN\theta}}|^2} = \frac{1}{|\braket{\e^{iN\theta}}|^2} - 1,
\end{equation}
where $\braket{\e^{iN\theta}} := \int_0^{2\pi} \e^{iN\theta} \mu(\theta) \dd\theta$ is the mean modular phase,
and $\braket{\Delta \e^{iN\theta}} \coloneqq \int_0^{2\pi} |\e^{iN\theta} - \braket{\e^{iN\theta}}|^2 \mu(\theta) \dd\theta$.

Even in the absence of loss of phase information due to external noise, the underlying primitive endows a code with some level of modular phase uncertainty.  This can be found from \cref{eq:phaseuncertainty_relation} using the following form for the mean modular phase,
\begin{equation}\label{eq:modularphase}
  \braket{\e^{iN\theta}} = \int_0^{2\pi} \e^{iN\theta} \mu(\theta) \dd\theta = \half \sum_{k=0}^\infty | \fc_{kN} \fc_{(k+1)N}|,
\end{equation}
where $f_{kN}$ are the Fock-grid coefficents of the code, \cref{eq:01logical}, as before.
Henceforth, we use $\dtheta$ to quantify the \emph{embedded} phase uncertainty in the codewords. This embedded uncertainty is in analogy to the embedded position/momentum errors in finitely squeezed GKP codewords \cite{Gottesman01,MoteBaraGilc17}. 
Note that, in general, the embedded phase uncertainty depends on the order of rotation symmetry.
In \cref{sec:CyclicStabilizerCodes} we show that certain families of rotation codes have meaningful limits where $\dtheta \to 0$, which is akin to the limit of infinite squeezing for ideal GKP codes.

Note that because the canonical measurement~\cref{eq:POVM_Holevo} assumes no prior knowledge about $\theta$, it might be sub-optimal if such knowledge exists. If rotation errors are small, we expect that $\theta$ is close to one of the values $m\pi/N$, and it might be possible to exploit this knowledge in a phase-estimation measurement. Moreover, since the measurement is defined with respect to the ideal, undamaged codeword~$\ket{+_{N,\Theta}}$, it might perform less than ideal in the presence of general noise that includes errors beyond pure rotation errors. For the numerical simulations of error correction circuits in \cref{sec:ECnumerics}, we compare the canonical phase measurement to Pretty Good Measurements, as introduced in Ref.~\cite{Belavkin1975aa,Belavkin1975bb, Hausladen94} (see also~\cref{sec:prettygood}), which explicitly takes information about the noise model into account.

When all $\gamma_n$ are real in~\cref{eq:POVM_Holevo}, the measurement can be approximated using adaptive homodyne measurements~\cite{Wiseman98} and has been the subject of much theoretical and experimental study~\cite{Higgins07,Berni15,Daryanoosh17,Berry00,Holevo84,Wiseman09, LeighMartin:2019aa}. One can similarly define phase-uncertainty measures analogous to~\cref{eq:phaseuncertainty_relation} for other phase-estimation schemes.
As an example, heterodyne measurement also measures the phase, but at the cost of larger uncertainty. An ideal heterodyne measurement is described by measurement operators $\ket{\alpha}\bra{\alpha}$, $\ket\alpha$ a coherent state, with completeness relation $\int \frac{\dd^2\alpha}{\pi} \ket\alpha\bra\alpha = \hat I$. The estimated phase for an outcome $\alpha$ is simply $\theta = \arg(\alpha)$.
We can define a phase uncertainty $\dthetahet$ for heterodyne measurement analogously to the canonical phase uncertainty defined above.
A heterodyne measurement gives larger uncertainty in the estimated phase than the canonical phase measurement; however, heterodyne measurement is nevertheless a likely practical candidate for distinguishing the logical codewords of the number-phase codes introduced in the next section.

\begin{figure}
  \centering
  \includegraphics{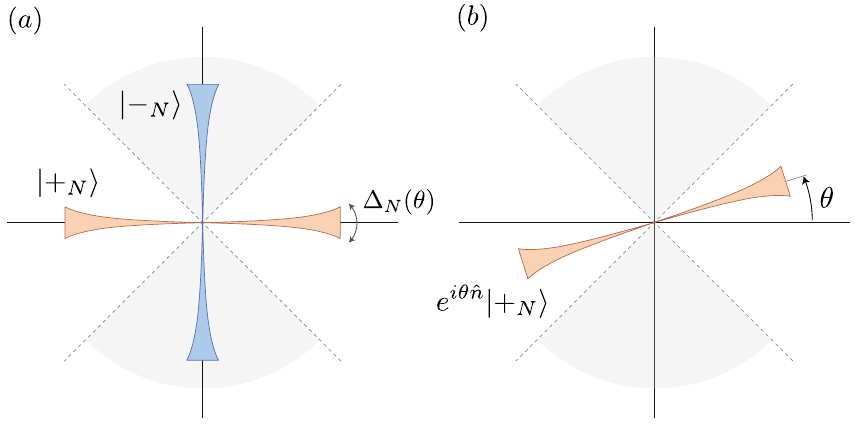}
  \caption{\label{fig:phasestructure}
Distinguishing the $\ket{\pm_N}$ codewords. By design $\ket{-_N} = \e^{i\frac{\pi}{N} \hat n}\ket{+_N}$, such that distinguishing the codewords can be viewed as a phase estimation problem.
The phase precision, $\dtheta$, depends on the measurement scheme, with a canonical scheme given by the POVM in~\cref{eq:POVM_Holevo}. 
(a)~Measuring $\theta$
to distinguish the codewords for an $N=2$ code ($d_\theta = \pi/2$). If the measurement result falls in a white wedge, we declare ``+" and in a gray wedge, we declare ``$-$". (b)~The same procedure for a codeword that has undergone a rotation error, $e^{i \theta \hat{n}} \ket{+_N}$. For large rotations, $\theta \sim d_\theta/2$, the logical state will be misidentified.
If the noise model is biased in such a way that clockwise rotations are more likely than counter clockwise (or vice versa), the decoder should be biased accordingly by rotating the white and gray wedges.
}
\end{figure}

In general, the phase uncertainty in a measurement is a function of both the codewords and the choice of measurement scheme. As long as the phase fluctuations in the measurement results are small compared to the rotational distance $d_\theta = \pi/N$, the two codewords $\ket{\pm_{N,\Theta}}$ can be distinguished faithfully. This idea is illustrated in~\cref{fig:phasestructure}.

\section{\label{sec:CyclicStabilizerCodes}Number-Phase Codes}

The embedded phase uncertainty in \cref{eq:phaseuncertainty_relation}, $\Delta_N(\theta)$, is minimized when the distribution of Fock-grid coefficients is completely flat, \emph{i.e.}, $|\fc_{kN}| = |\fc_{(k+1)N}|$ for all $k$, such that $|\braket{\e^{iN\phi}}| = 1$ and $\Delta_N(\theta) = 0$, c.f.~\cref{eq:phaseuncertainty_relation,eq:modularphase}. However, such a state is not normalizable, rendering this limit unphysical. Formally, we can consider families of physical codes that in some appropriate limit satisfy $\Delta_N(\theta)\to 0$. As we will show, this is satisfied for a class of rotation codes very much analogous to GKP codes, with number and phase playing roles akin to position and momentum. Perhaps somewhat surprisingly, the previously studied cat and binomial codes both belong to this class.\footnote{We note that GKP codes defined for a quantum rotor, \emph{i.e.}, a particle moving on a circle described by angular position $\hat \varphi$ and angular momentum $\hat N$, have been studied previously~\cite{Gottesman01,Raynal:2010aa}.  However, the strictly non-negative spectrum of the number operator, $\hat n$, implies that no unitary operator $\e^{i\hat \phi}$ exists such that $[\hat n, \e^{i\hat \phi}] = \e^{i\hat \phi}$~\cite{PeggBarn89}. In this sense number and phase do not constitute a proper conjugate variable pair. See, \emph{e.g.}, Ref.~\cite{Levy1976} for a discussion on angle and phase operators in quantum mechanics, and Refs.~\cite{PeggBarn88,PeggBarn89} for a proposal to define unitary and hermitian phase operators as a limit of a sequence of regularized operators.}

\begin{table*}
    \begin{tabular}{  c | c | c || c || c | c | c}
         code  &  $\ket{\Theta}$ &   $\text{$\fc_{kN}$}$ &  \text{\ON{} code} & limit & $\text{limiting $\fc_{kN}$}$ & limiting mean modular phase $\expt{\e^{i\theta N}}$ \\  \hline
        cat &  $\ket{\alpha}$  &
         $ \sqrt{ \frac{2}{\mathcal{N}^*_i} } \frac{\e^{-|\alpha|^2/2} \alpha^{kN}}{\sqrt{(kN)!}}$  & 
         $\alpha \rightarrow 0$  &
         	$\alpha \rightarrow \infty$ &
		$ \big( \frac{2 N^2}{\pi\alpha^2} \big)^\frac{1}{4} \exp \left[\frac{-(kN-\alpha^2)^2}{4\alpha^2} \right]$  &
		$ e^{-\frac{N^2}{8\alpha^2}} \vartheta_3 \Big[ \frac{N\pi}{2}(1-\frac{2\alpha^2}{N}), \e^{- 2\pi^2 \frac{\alpha^2}{N^2} } \Big] $  \\ 
	[4pt] 
        binomial & $\big| \Theta_{\bin}^{N,K} \big\rangle$  &
         $\sqrt{ \frac{1}{2^{K-1}} \binom{K}{k} }$  & 
          	$K = 1$  &
                $K \rightarrow \infty$ &
                $(\frac{8}{\pi K} )^{\frac{1}{4}} \exp \left[ \frac{-(k-K/2)^2}{K}\right] $ & 
        		$\e^{-\frac{1}{2K}} \vartheta_3 \Big[ \frac{\pi}{2}(1-K),\e^{-\frac{K\pi^2}{2}} \Big] $   \\ [4pt] 
	    Pegg-Barnett &
		 $\ket{\phi=0,s}$  &
          	$ \sqrt{2/\lceil s/N \rceil}  $  &
		 $s = N+1$  &
                 $s \rightarrow \infty$ &
                 $ \sqrt{2/\lceil s/N \rceil}$ & 
                 $1-1/\lceil s/N \rceil $  \\ [4pt] 
    \end{tabular}
\caption{  \label{code_table}
Three codes---cat, binomial and Pegg-Barnett---whose phase uncertainty vanishes in an appropriate limit (see \cref{sec:CodeZoo} for definitions and details of each code). The second and third columns give a primitive and the Fock-grid coefficients [\cref{eq:01logical}]
(for cat codes $\mathcal{N}^*_i$ is the normalization factor $\mathcal{N}^*_0$ for even $k$ and $\mathcal{N}^*_1$ for odd $k$). The fourth column shows the limit where each code becomes the \ON{} encoding (which is the trivial encoding, \cref{eq:trivialencoding}, for $N=1$), and the fifth gives the limit where $\Delta_N(\theta) \rightarrow 0$. The sixth column gives an asymptotic form of the Fock-grid coefficients, and the last column gives an asymptotic form for the mean modular phase. For each code the mean modular phase uncertainty approaches zero, $\Delta_N(\theta) \rightarrow 0$. The function $\vartheta_3(r, \tau)$ is the Jacobi elliptic theta function of the third kind.
}
\end{table*}

\subsection{Ideal number-phase codes}
In analogy with ideal GKP codes defined in terms of superpositions of position (or momentum) eigenstates, we define a family of order-$N$ rotation codes as:
\begin{subequations}\label{eq:canonicalcode}
\begin{align}
  \ket{0_{N,\,\can}} ={}& \sum_{m=0}^{2N-1} \ket{\phi = \frac{m\pi}{N}},\\
  \ket{1_{N,\,\can}} ={}& \sum_{m=0}^{2N-1} (-1)^m \ket{\phi = \frac{m\pi}{N}},
\end{align}
\end{subequations}
where $\ket{\phi} \coloneqq \frac{1}{\sqrt{2\pi}}\sum_{n=0}^\infty \e^{in\phi}\ket n$ are known as phase states and are not normalizable~\cite{Wiseman1995,Jacobs2008}.
In the Fock basis, the codewords are simply $\ket{0_N} \propto \sum_k \ket{2kN}$ and $\ket{1_N} \propto \sum_k \ket{(2k+1)N}$.
Note that we can write the POVM for the canonical phase measurement in \cref{eq:POVM_Holevo} (for real coefficients $\gamma_n=1$) as $\hat M^\text{can}(\phi) = \ket \phi \bra \phi$.
\cref{eq:canonicalcode} follows from~\cref{eq:01codewords} by replacing $\ket\Theta \propto \ket{\phi=0}$.

Any state in the codespace spanned by~\cref{eq:canonicalcode} is, of course, a $+1$ eigenstate of $\CN$. The codewords are also $+1$ eigenstates of a number-translation operator
\begin{equation}\label{eq:transstabilizer}
    \TransN \coloneqq \sum_{n=0}^\infty \ket{n}\bra{n+2N}.
\end{equation}
We have that $\CN\TransN=\TransN\CN$, and the operator
\begin{equation}\label{eq:XN}
  \XN \coloneqq \hat{\Sigma}_{N/2} = \sum_{n=0}^\infty \ket{n}\bra{n+N}
\end{equation}
satisfies $\hat X_N \hat Z_N = - \hat Z_N \hat X_N$, while both $\ZN$ and $\XN$ commute with $\CN$ and $\TransN$. Finally, for the code~\cref{eq:canonicalcode}, we have that $\XN\ket{\pm_N} = \pm\ket{\pm_N}$, and this operator thus acts as logical $\X$.

\begin{figure}[b]
  \centering
  \includegraphics[width=1\columnwidth]{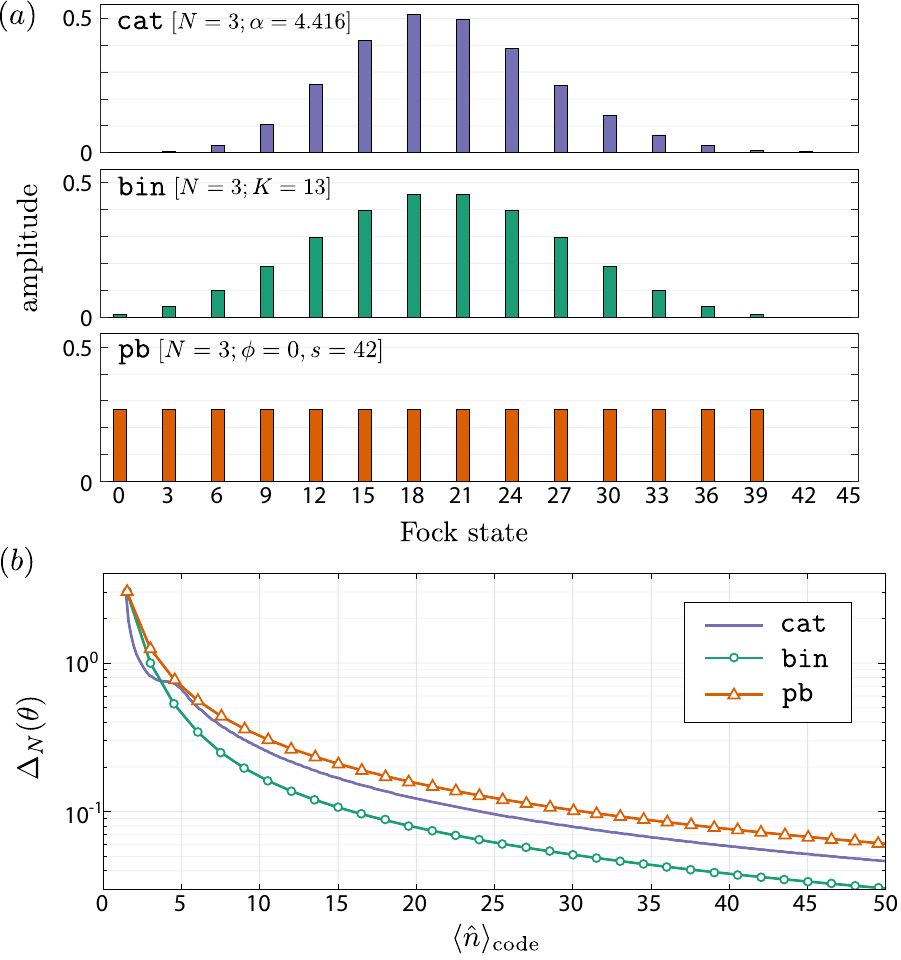}
  \caption{\label{fig:FockDist} (a) Magnitude of the Fock-state amplitudes for a $\ket{+_N}$ state for $N=3$ cat, binomial, and Pegg-Barnett codes with parameters that give $\ncode = 19.5$ for each. The amplitudes are nonzero only on Fock states $\ket{3k}$ for $k=0,1,2,\dots$. (b) The embedded phase uncertainty, \cref{eq:phaseuncertainty_relation}, for each family of $N=3$ codes as $\ncode$ increases. For each code a proper asymptotic limit, given in \cref{code_table}, yields no embedded phase uncertainty, $\Delta_N(\theta) \rightarrow 0$. In the opposite regime, $\ncode \rightarrow \smallfrac{3}{2}$, all three codes limit to the 03 encoding.
Note that $N=3$ codes with $\ncode < \smallfrac{3}{2}$ do not exist and that binomial and Pegg-Barnett codes are defined only at discrete excitation numbers; connecting lines are guides for the eye. 
}
\end{figure}

\subsection{Approximate number-phase codes}

Ideal number-phase codes as defined in~\cref{eq:canonicalcode} are unphysical as they require infinite excitation number and are not normalizable. 
Nevertheless, certain families of rotation code approach ideal number-phase codes, in a sense to be defined below, in limits of large excitation number.

We first recognize that for any rotation code with positive, real Fock-grid coefficients $\{ \fc_{kN} \} $ [~\cref{eq:01logical}], we have $\braket{\pm_{N,\Theta}|\XN|\pm_{N,\theta}} = \pm\braket{\e^{iN\theta}}$. As a consequence,
\begin{equation}\label{eq:phasecodelimit}
    \braket{\pm_{N,\Theta}|\XN|\pm_{N,\theta}} \to \pm 1
\end{equation}
if and only if $\dtheta \to 0$, where $\dtheta$ and $\braket{\e^{iN\theta}}$ are defined in~\cref{eq:phaseuncertainty_relation,eq:modularphase}, respectively. It follows that the condition of vanishing phase uncertainty defines a family of codes that, in this limit, are stabilized by $\hat{\Sigma}_N$ [\cref{eq:transstabilizer}] and for which $\hat{X}_N$ [\cref{eq:XN}] acts as logical $\X$.

An interesting property of number-phase codes arises from the fact that in the limit of~\cref{eq:phasecodelimit}, the two codewords normalization constants are equal, $\mathcal N_0=\mathcal N_1$, in~\cref{eq:01codewords}. It follows that the dual-basis codewords also become equal-weight superpositions of the primitive [see~\cref{eq:pmcodewordslimit}],
\begin{subequations}
    \begin{align}
        \ket{+_{N,\Theta}} &\rightarrow \frac{1}{\sqrt{N}} \sum_{m=0}^{N-1} e^{i \frac{2 m\pi}{N} \hat{n}} \ket{\Theta}, \\
        \ket{-_{N,\Theta}} &\rightarrow \frac{1}{\sqrt{N}} \sum_{m=0}^{N-1} e^{i \frac{(2m+1)\pi}{N} \hat{n}} \ket{\Theta}. 
    \end{align}
\end{subequations}
A stronger number-phase duality is apparent for number-phase codes than for generic rotation codes: the dual-basis codewords, $\ket{\pm_N}$, are separated from each other by exactly $d_\theta=\pi/N$ in terms of phase-space rotations but have support on the same Fock states, c.f.~\cref{eq:pmlogical}. Conversely, the computational-basis codewords, $\ket{0_N/1_N}$ are separated from each other by $d_n=N$ in Fock space but supported on the same set of phases $\{m\pi/N\}$.

There are many examples of rotation codes that limit to ideal number-phase codes in the sense of~\cref{eq:phasecodelimit}. 
A straightforward example are codes constructed from normalized Pegg-Barnett phase states, $\ket{\phi,s} \coloneqq \frac{1}{\sqrt{s}}\sum_{n=0}^{s-1}\e^{in \phi}\ket{n}$~\cite{PeggBarn88,BarnPegg89,PeggBarn89}, as the primitive $\ket{\Theta}$. As the parameter that sets their truncation in Fock space becomes large, $s \to \infty$, Pegg-Barnett codes approach ideal number-phase codes (up to normalization).
With the constraint $s=p\times 2N$ for integer $p\ge 1$, the Pegg-Barnett codes are identical to the shift-resistant qudit codes introduced in Ref.~\cite{Gottesman01}, embedded in the infinite Hilbert space of a bosonic mode.\footnote{Using the notation of Ref.~\cite{Gottesman01}, Pegg-Barnett codes as defined here with $s=p\times 2N$ map onto shift-resistant qudit codes with $d=s$, $n=2$, $r_1=N$, and $r_2=p$.}
Further details about Pegg-Barnett codes can be found in \cref{appendix:PB}.

Perhaps more surprisingly, cat codes (\cref{Appendix:squeezedcat}) and binomial codes (\cref{appendix:binomial}) also satisfy~\cref{eq:phasecodelimit} in the limit of large $\alpha$ and large truncation $K$, respectively.
\Cref{code_table} summarizes the three examples of number-phase codes we discuss here. For these codes, the embedded phase uncertainty approaches zero when the relevant code parameter is large enough, as shown in \cref{fig:FockDist} for $N=3$ codes.

Colloquially, we refer to any family of rotation codes that satisfies $\dtheta \to 0$ in an appropriate limit as a number-phase code. 
In practice, the specific form of the Fock-grid coefficients $\{ f_{kN} \}$, can make a significant difference in the embedded phase uncertainty for small to moderate excitation numbers. Phase measurements play a critical role in the quantum computing scheme of \cref{sec:gates} and the error-correction procedures of \cref{sec:EC}. As a consequence, although different number-phase codes behave identically for large excitation numbers, they can have different performance at experimentally relevant excitation numbers~\cite{Albert17}. This is apparent in~\cref{sec:ECnumerics}, where we compare the error-correction performance of cat and binomial codes.

\section{\label{sec:gates}Symmetry as a resource: Quantum computing with number-phase codes}

We introduce a quantum computing scheme based on three basic ingredients: Preparation of codewords in the state $\ket{+_N}$, a small number of unitary gates, and (destructive) measurement in the logical $\X$ basis.
Our focus is on operations that are, in a certain sense, fault-tolerant.
For the unitary gates this means that the gates should not amplify errors too badly, a notion that is made precise later in~\cref{sec:errorsanderrorprop}, while for the measurement it means that the codewords $\ket{\pm_N}$ should be distinguishable in the presence of small errors. 
As already discussed in~\cref{sec:phaseest}, the latter can be achieved with phase measurements, given that the codes have small inherent phase uncertainty.
Vanishing phase uncertainty is equivalent to (approximate) translation symmetry in Fock space, as we showed in~\cref{sec:CyclicStabilizerCodes}, such that the existence of a robust $\X$-basis measurement can be said to be a consequence of this number-translation symmetry.
On the other hand, it turns out that the unitary gates we introduce only rely on rotation symmetry. 
We show that, given a supply of encoded basis states $\ket{+_N}$, these two symmetries imply the ability to enact any gate in the logical Clifford group, using physically reasonable operations that are fault-tolerant in the above sense.
The situation can be compared to GKP codes, where (approximate)
discrete quadrature translation symmetry leads to any logical Clifford being (approximately) implementable by a Gaussian operation~\cite{Gottesman01}.

As for the state preparation step, which is in general code dependent, flexible techniques have already been developed in various physical platforms, in particular in circuit QED~\cite{Ofek16,Hu:2019aa} and ion-traps~\cite{Fluhmann:2018aa}, that can be applied to prepare codewords for a wide range of codes. It is beyond the scope of the present work to discuss state preparation in detail, but we briefly comment on a few approaches in~\cref{sec:stateprep}. All of the known approaches are limited by noise in the preparation procedure, and one can not rule out codewords with large errors.
The issue of faulty state preparation is postponed until~\cref{sec:concat}, and for the time being we simply assume that $\ket{+_N}$ states can be prepared.

Due to the existence of a robust $\X$-basis measurement, our main focus will be on approximate number-phase codes from here on. That said, the unitary gates introduced below apply to \emph{any} rotation code, such that the scheme can be extended to any such code where a practical $\X$-basis measurement can be found.

\subsection{Universal operations for number-phase codes}

The quantum computing scheme introduced in the following is based on the universal set
\begin{equation} \label{eq:QCscheme}
    \left\{ \Sgate, \CZ \right\} \cup \left\{\mathcal P_{\ket{+_N}}, \mathcal P_{\ket{T_N}}, \MX \right\} ,
\end{equation}
where
$\Sgate$ is an encoded version of the phase gate
$\hat S = \text{diag}(1,i)$, 
$\CZ$ an encoded $\hat C_Z=\text{diag}(1,1,1,-1)$,
$\mathcal P_{\ket{\psi_N}}$ stands for preparation of an encoded state $\ket{\psi_N}$, and
$\MX$ stands for measurement in the logical $\ket{\pm_N}$ basis.
The state $\ket{T_N} \propto \ket{0_N} + \e^{i\pi/4}\ket{1_N}$ is used for gate-teleportation of the non-Clifford $\hat T = \text{diag}(1,\e^{i\pi/4})$ gate. 

A remarkable feature of the following scheme is that the required unitary gates are agnostic to the specific details of the rotation code in question. The encoded $\{\Sgate, \CZ\}$-gates depend only on the degree of rotation symmetry and consequently apply equally to all rotation codes.
In particular, this allows entangling operations between \emph{different} rotation codes, \emph{e.g.}, cat and binomial, and teleportation from one code to another.

\subsubsection{Unitary gates}

We have already shown that logical $\Z $ for an order-$N$ rotation code can be implemented through a simple rotation $\hat Z_N = \e^{i\pi \hat n/N}$.
The next gate in our set is the following two-mode gate that implements a controlled rotation between an order-$N$ and an order-$M$ rotation code \cite{CochMilbMunr99,Zhang:2017aa}:
\begin{equation} \label{eq:CROT}
  \CROT_{NM} \coloneqq \e^{i\frac{\pi}{NM} \hat n \otimes \hat n},
\end{equation}
where $\hat n$ are number operators for the two respective modes.
To understand its action on the codespace, we recognize that 
$\CROT_{NM} \ket{kN}\otimes\ket{lM} = \e^{i\pi k l} \ket{kN}\otimes\ket{lM}$
for Fock states $\ket{kN}$ and $\ket{lM}$. 
Since $kl$ is even unless both $k$ and $l$ are odd, it follows from~\cref{eq:01logical} that the action on an encoded state $\ket{i_N}\otimes\ket{j_M}$ is
\begin{equation}
  \CROT_{NM} \ket{i_N}\otimes\ket{j_M} = (-1)^{ij} \ket{i_N}\otimes\ket{j_M}
\end{equation}
for $i,j=0,1$.
Thus, the $\CROT$ gate acts as a $\CZ$ gate on the codespace of \emph{any} two rotation codes.

A logical $\hat S = \text{diag}(1, i)$ can similarly be enacted by a quartic (in $\hat{a}, \hat{a}^\dagger$) single-mode Hamiltonian,
\begin{equation}
    \Sgate = \SN \coloneqq \e^{i\frac{\pi}{2N^2}\hat n^2}.\label{eq:S_N}
\end{equation}
The action of $\SN$ on the codespace can be seen by acting on the Fock states in \cref{eq:01logical}: $\SN \ket{2kN} = \e^{\frac{i\pi}{2} 4k^2}\ket{2kN} = \ket{2kN}$, since $2k^2$ is even.  On the other hand, $\SN \ket{(2k+1)N} = \e^{\frac{i\pi}{2} (4k^2 + 4k + 1)}\ket{(2k+1)N} =i\ket{(2k+1)N}$. 

In principle, Hamiltonians of even higher order in $\hat n$ can be used to generate diagonal non-Clifford gates. In particular, a logical $\bar T$ gate could be enacted by
\begin{equation}
    \TN \coloneqq \e^{i \frac{\pi}{4 N^4} \hat{n}^4},
\end{equation}
as is easily checked by acting on the codewords $\ket{0_N}$, $\ket{1_N}$ in~\cref{eq:01logical}. However, in addition to the experimental challenge of engineering an octic Hamiltonian, the $\TN$ gate can amplify errors in damaging ways as we show in~\cref{sec:errorsanderrorprop}. It is therefore unlikely to be of practical use and we propose a scheme based on gate teleportation instead.

\subsubsection{Teleported gates}

The $\CROT$ gate and logical $\MX$ measurements, together with appropriately prepared ancillae, allow us to complete a universal set of operation with a gate-teleported logical Hadamard $\bar{H}$ and $\bar{T}$-gate.

An ancilla prepared in $\ket{+_M}$ allows execution of the Hadamard gate, using the following teleportation circuit~\cite{Kitaev06protected,Aliferis08,Webster15}:
\begin{equation}\label{eq:logicalteleport}
  \begin{aligned}
  \Qcircuit @C=1em @R=1.3em {
    \lstick{\ket{\psi_N}} & \multigate{1}{\rotatebox{270}{\CROT}} & \measureD{\MX} & \pm \\
    \lstick{\ket{+_M}}    & \ghost{\rotatebox{270}{\CROT}}        & \rstick{\X^i\Hd\ket{\psi_M},} \qw
 }
  \end{aligned}
\end{equation}
where $\ket{\psi_N} = a\ket{0_N} + b\ket{1_N}$ and $\ket{\psi_M} = a\ket{0_M} + b \ket{1_M}$ represents the same encoded qubit state (but not necessarily in the same code). The measurement is in the $\ket{\pm_N}$ basis, and we use $i=0$ for the outcome ``$+$'' and $i=1$ for the outcome ``$-$.'' The notation $\X^i\Hd$ thus means that depending on the outcome, an $\Hd$ (``$+$'') or $\X\Hd$ (``$-$'') is applied to the data.
Since our scheme for quantum computation is based only on Clifford gates and magic-state injection, rather than correcting the conditional Pauli $\X$, the simplest solution is to keep track of it in a Pauli frame~\cite{Knill05,Chamberland17}.
Alternatively, the process can be repeated until the desired outcome is achieved~\cite{Kitaev06protected}. 

The set $\{\CZ,\Hd,\Sgate\}$ generates the Clifford group. To achieve universality, we teleport
the gate $\hat T=\diag(1,\e^{i\pi/4})$ by consuming an ancilla prepared in $\ket{T_M} = (\ket{0_M}+\e^{i\pi/4}\ket{1_M})/\sqrt{2}$:\footnote{
Note that in this case, because $\Tgate$ is not a Clifford gate, the correction for the ``$-$'' outcome ($i=1$) is non-Pauli. The correction can be done by applying $\X\Sgate$ for this outcome (using that $\X\Sgate \Tgate\X = -\e^{-i\pi/4}\Tgate$, where the overall phase factor is unimportant). In fact, since $\X\Sgate$ is equivalent to $\Sgate$ up to a change of Pauli frame, we can simply apply $\Sgate$~\cite{terhal2015quantum}. Alternatively we can repeat the protocol until the desired outcome is achieved.    
}
\begin{equation}\label{eq:Tteleport}
  \begin{aligned}
  \Qcircuit @C=1em @R=1.3em {
    \lstick{\ket{\psi_N}} & \gate{\Hd} & \multigate{1}{\rotatebox{270}{\CROT}} & \measureD{\MX} & \pm \\
    \lstick{\ket{T_M}}   & \qw        & \ghost{\rotatebox{270}{\CROT}}        & \rstick{\Tgate(\X)^i\ket{\psi_M}.} \qw
 }
  \end{aligned}
\end{equation}
In order to teleport the $\bar{T}$ gate, we need an encoded $\ket{T_M}$ state. Arbitrary encoded states can be injected assuming we have universal control over two-level ancillae in the trivial encoding $\ket{0_\triv}=\ket 0$, $\ket{1_\triv}=\ket 1$:
The circuit in~\cref{eq:logicalteleport} can be used to teleport an arbitrary ancilla state into a rotation code, where the top rail represents the two-level ancilla with $N=1$. The procedure is not fault-tolerant, because an error on the ancilla propagates into the rotation code. 
Preparation of $\ket{T}$ states can be made fault-tolerant using magic-state distillation, assuming high-quality Clifford operations are available~\cite{Bravyi2005}.\footnote{An unfortunate clash in nomenclature has led to the state $\ket{T}$ that teleports the gate $\hat T$ being called an $H$-type (not $T$-type) magic state, because it is equivalent to a Hadamard eigenstate up to Clifford gates. Further, in its original setting in Ref.~\cite{Bravyi2005}, $\hat T$ was used to denote a different gate: that whose eigenstates are $T$-type magic states.}
As we discuss in~\cref{sec:concat}, this can be achieved by concatenation with a second code and performing state distillation at the level of the top code.

If needed, we can also execute an $\hat S = \diag(1,i)$ gate in a similar fashion by consuming an ancilla prepared in $\ket{+i_M} = (\ket{0_M} + i\ket{1_M})/\sqrt{2}$:
\begin{equation}\label{eq:Steleport}
  \begin{aligned}
  \Qcircuit @C=1em @R=1.3em {
    \lstick{\ket{\psi_N}} & \gate{\Hd} & \multigate{1}{\rotatebox{270}{\CROT}} & \measureD{\MX} & \pm \\
    \lstick{\ket{+i_M}}   & \qw        & \ghost{\rotatebox{270}{\CROT}}        & \rstick{\Sgate\X^i\ket{\psi_M}.} \qw
 }.
  \end{aligned}
\end{equation}
Again, the outcome dependent factor $\X^i$ can be kept track of using a Pauli frame. This can be an alternative to using the gate~\cref{eq:S_N} directly, which might be more practical experimentally than controlling a self-Kerr Hamiltonian $\sim \hat n^2$.

\subsection{\label{sec:stateprep} State preparation}

State preparation is a crucial ingredient for quantum computing with bosonic codes and can also be the most difficult part. Here, the logical $\ket{+_N}$ state plays a key role in both the gate teleportation and state injection needed for teleported $\bar{T}$ gates. 
A large literature exists on state preparation of GKP codewords~\cite{TravMilb02, TerhWeig16, MoteBaraGilc17, Eaton:2019aa, shi2019fault}, while relatively less theoretical work has been done for rotation codes, and the optimal way to prepare codewords will in general depend on the code in question. This is therefore a broad topic, and we do not address it in detail in this work.

One experimental approach to preparing logical basis states involves a strong dispersive interaction between the mode and an ancilla qubit (in circuit QED these could be a cavity mode and a transmon qubit). Combined with displacement operations, such an interaction is in principle universal~\cite{Heeres2015,Krastanov2015}, 
in the sense that any state of the mode can be prepared in the absence of errors. 
Optimal control can be used to generate a sequence of control operations that 
maximizes the fidelity of a target state in the presence of errors, which has been performed in the circuit QED setting~\cite{Ofek16,Hu:2019aa}.
An alternate approach could be to prepare a primitive $\ket \Theta$, and then measure excitation number mod $N$, using the measurement introduced below in~\cref{sec:numbermodN}.
For either of these approaches the fidelity of the state preparation is limited by noise in the preparation procedure, and one can not rule out corrupted codewords with large errors. We postpone a discussion of how to deal with faulty state preparation until~\cref{sec:concat}.

\subsubsection{\label{sec:breeding}Breeding rotation symmetry}
We here present a method to ``breed" logical states for an order-$N$ rotation code from code states with lower order rotation symmetry. 
In analogy to a scheme for producing GKP states from Ref.~\cite{TravMilb02}, each stage of the breeding consists of coupling the bosonic mode to an ancilla mode followed by measurement of the ancilla. When successful, a $\ket{0_{2N,\Theta}}$ state is produced from a $\ket{0_{N,\Theta}}$ state, and the whole process can be repeated as required. 

We begin with a codeword $\ket{0_{N,\Theta}}$ and an ancilla qubit prepared in a state $\ket{+_M}$ coupled via $\CROT$ interaction
\begin{align} \label{eq:statecouplingHam}
	\CROT_{2N,M} = \e^{i\frac{\pi}{2NM} \hat n \otimes \hat n}.
\end{align}
Critically, we note that the ancilla could be a two-level system in the state $\ket{+_{\triv}} = (\ket 0 + \ket 1)/\sqrt 2$ $(M=1)$ or encoded in a rotation code of any order $M$.
The ancilla is then measured in the $\ket{\pm_M}$-basis.
When the ``+" outcome is obtained, the state becomes superposed with a rotated version to create a codeword of higher rotation symmetry,
	\begin{equation} \label{eq:KrausBreeding}
		\smallfrac{1}{2} \big(\hat{I} + \hat Z_{2N}\big) \ket{0_{N,\Theta}} = \sqrt{\mathcal{P}_{+}} \ket{ 0_{2N,\Theta} },
	\end{equation}
with outcome probability $\mathcal{P}_{+} = \frac{1}{2} \big( 1 + \bra{0_{N,\Theta} } \hat{Z}_{2N}\ket{ 0_{N,\Theta} } \big)$.
Beginning from a primitive $\ket{\Theta}$, $n$ successful rounds of breeding produce a logical state with $2^n$-fold rotation symmetry, $\ket{0_{2^n,\Theta}}$.
Just as in Ref.~\cite{TravMilb02}, the success probability falls off exponentially in $n$, but this could be improved by making use of higher-dimensional ancillae such as in Ref.~\cite{MoteBaraGilc17}.
The relation, \cref{eq:KrausBreeding}, can likewise be used on a low-order $\ket{+_{N,\Theta}}$ (or a dual primitive, see \cref{sec:conjugateprimitive}) to breed a higher order $\ket{+_{2N,\Theta}}$ state. 

It is also worth noting here that for number-phase codes discussed in~\cref{sec:CyclicStabilizerCodes}, which are our main focus,
we have the approximate relation (see also~\cref{sec:conjugateprimitive})
\begin{equation}
     \ket{0_{N,\Theta}} \simeq \ket{+_{2N,\Theta}}.
\end{equation}
Thus, preparation of $\ket{0_{N,\Theta}}$ states is equivalent to preparation of approximate $\ket{+_{2N,\Theta}}$ states for these codes.

\section{\label{sec:numbermodN}Modular number measurement}
 
The $\CROT$ gate is a fundamental building block for the universal gate set, state breeding, and the error-correction scheme we introduce in~\cref{sec:EC}. 
Here, we show that the $\CROT$ gate can also be used to perform a non-destructive modular measurement of excitation number $\hat{n}$. 
In addition to being a novel measurement for a bosonic mode, the modular number measurement can be used both to detect number-shift errors on an encoded state and to conditionally prepare $N$-fold rotation-symmetric codewords from a primitive $\ket{\Theta}$. The measurement we present here is a natural generalization of previous schemes where a two-level ancilla (transmon qubit) was used to detect number mod $2$ for cat and binomial codes~\cite{Ofek16,Hu:2019aa}. 

\begin{figure}
\centering
\includegraphics[width=0.95\columnwidth]{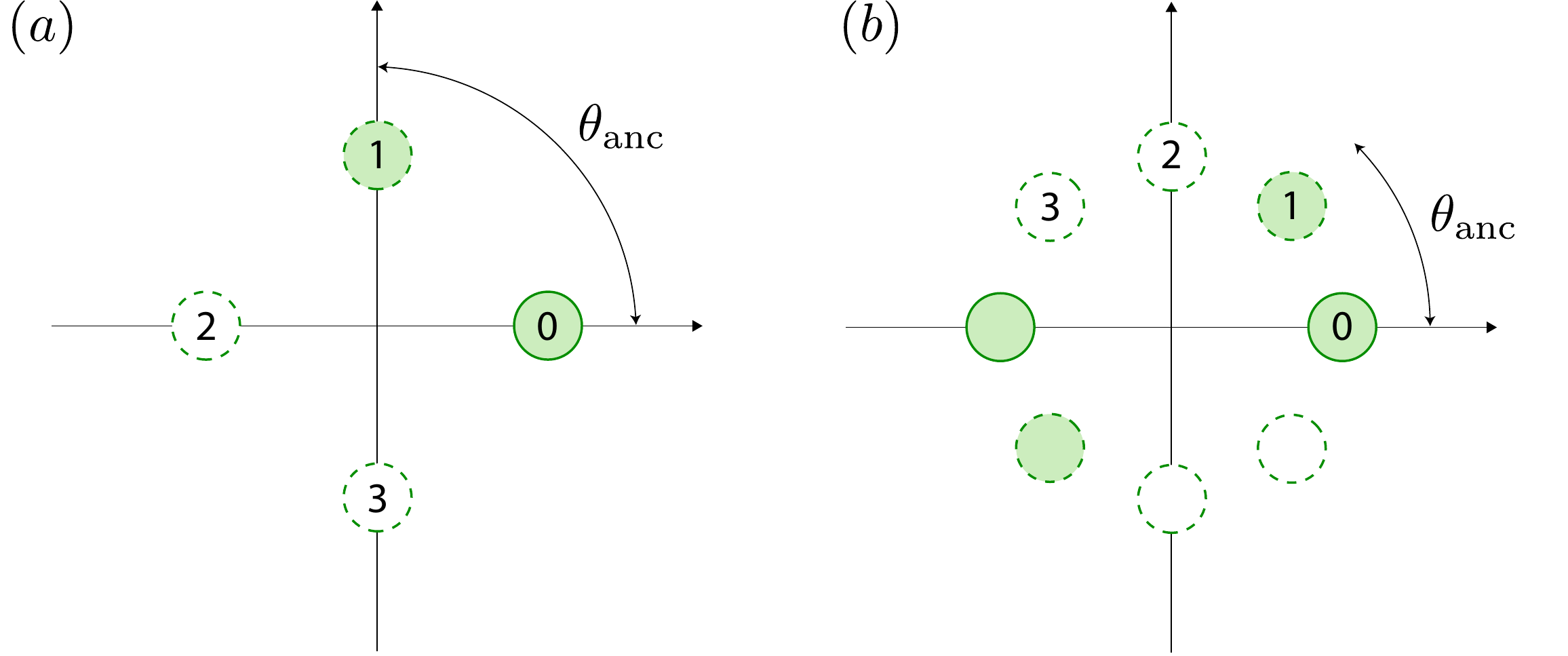}
\caption{\label{fig:numbermodN}
Measurement of $\hat n \bmod 4$ using an ancilla prepared in (a) a coherent state and (b) a two-lobe cat-state. 
Shown are the initial ancilla states (solid), and each circle is a potential rotation of the ancilla that depends on the data-rail state being measured. They are labeled by outcomes $\ell$ for a measurement of $\hat{n} \text{ mod }4$. In both (a) and (b) the specific ancilla rotation shown, $\theta_{\rm anc}$, corresponds to an $\ell=1$ outcome indicating that the data rail is off the order-$N$ Fock grid.
}
\end{figure}

A non-destructive measurement of $\hat n \bmod N$ can be performed with the following circuit:
\begin{equation}\label{eq:numbersyndrome}
  \begin{aligned}
  \Qcircuit @C=1em @R=1.3em {
                        & \gate{\CN} & \qw \\
     \lstick{\ket{+_M}} & \ctrl{-1}     & \measureD{\MX} & \lstick{,}
  }
  \end{aligned}
\end{equation}
where the controlled-$\CN$ gate is defined as
\begin{equation} \label{eq:CROTnumbermodN}
    \hat C_{R_N} := \CROT_{NM/2} = \e^{i \frac{2\pi}{NM} \hat n \otimes \hat n},
\end{equation}
and the measurement is to be understood as a phase measurement as described in~\cref{sec:phaseest}.
To see how this circuit works, consider the action of the $\hat C_{R_N}$ gate on a Fock state, $\ket{n}$, for the data rail and $\ket{+_M}$ for the ancilla:
\begin{equation}
  \begin{aligned}
    \hat C_{R_N}\ket{n}\otimes\ket{+_M} 
     ={}& \ket{n}\otimes\e^{i \frac{\ell}{N}\frac{2\pi}{M}\hat n} \ket{+_M},   
  \end{aligned}
\end{equation}
where $\ell \coloneqq n \bmod{N}$, and
we used that $\e^{i\frac{2\pi p}{M}\hat n}\ket{+_M} = (\hat R_M)^p \ket{+_M} = \ket{+_M}$ for any integer $p$.
The net result is a rotation of the ancilla state $\ket{+_M}$ by an angle $\theta_{\rm anc} = 2\pi \ell/(N M)$, which takes the ancilla out of its codespace. 
This rotation can be detected using a destructive phase measurement of the ancilla with a resolution set by the ancilla's embedded phase uncertainty $\Delta_M(\theta)$ relative to $2\pi/(NM)$. 
For illustration, \cref{fig:numbermodN} shows a $\hat n\bmod 4$ measurement using two different ancillae: (a) a coherent state and (b) a two-lobe cat state. These can be interpreted (for large enough $\alpha)$ as a $\ket{+_M}$ codeword for $M=1$ and $M=2$, respectively. The motivation for using ancillae with higher $M$ is the fact that errors may occur \emph{during} the $\hat C_{R_N}$ gate, as discussed below.

After the measurement, the value of $\ell$ is known, and the data rail is (in the ideal case) projected into an $N$-fold rotation-symmetric subspace: $\hat{\rho} \rightarrow \hat{\Pi}^\ell_N \hat{\rho} \hat{\Pi}^\ell_N$, where the projector $\hat{\Pi}^\ell_N$ is given in \cref{eq:projector_ell}. Thus, a $\hat n \bmod N$ measurement can be used to prepare codewords for an $N$-fold rotation code in the following way: 
First, prepare the primitive associated with the code, $\ket \Theta$, and then measure excitation number mod $2N$.
Conditional on ancilla phase being undisturbed ($\ell = 0$), the primitive undergoes the transformation $\hat\Pi_{2N}^0 \ket{\Theta} = \ket{0_{N,\Theta}}$. Similarly, an ancilla outcome corresponding to $\ell = N$ conditionally prepares $\hat\Pi_{2N}^N \ket{\Theta} = \ket{1_{N,\Theta}}$.

\subsubsection{Error propagation in modular number measurement}

Errors on the ancilla can propagate to the data rail during the $\hat n \bmod N$ measurement.
Rotation errors on the ancilla do not propagate to the data rail (because they commute with the $\CROT$ gate) and are not problematic beyond reducing the measurement fidelity. However, any ancilla error that does not commute with $\hat n$, 
such as loss or gain, can induce a rotation error on the data rail (see~\cref{sec:errorsanderrorprop} for a detailed discussion of error propagation). 
For example, a loss error at a random time during the measurement leads to a rotation error of the data rail by an angle $\theta_\text{data} \in \left[0,\frac{2\pi}{NM}\right)$.
Thus, using an ancilla with small $M$ brings along the danger of propagating a large rotation error back to the data qubit.
In particular, for $M=1$ ancilla, such as the coherent state in \cref{fig:numbermodN}(a), a loss error at an unknown time during the $C_{R_N}$-gate completely randomizes the data rail's phase.
The maximum magnitude of the rotation error induced by a single ancilla loss/gain event scales as $1/M$, and thus can be reduced by using large-$M$ ancillae. 
The drawback of using higher-$M$ ancilla is larger measurement uncertainty in the phase measurement in \cref{eq:numbersyndrome}. This can be seen in \cref{fig:numbermodN}(b) for the two-lobe cat state ancilla.

We note that an improved number-parity syndrome for $N=2$ rotation codes was recently demonstrated experimentally, where the ancilla was encoded in a simple rotation code using three levels of a transmon qubit~\cite{Rosenblum2018}. 
In place of a $\CROT$ gate the entangling interaction, $\e^{i\pi\hat n \otimes (\ket 1 \bra 1 + \ket 2\bra 2)}$ was used, by engineering the level-dependent cross-Kerr interactions using sideband drives. 
The interaction exactly commutes with the dominant ancilla error $\ket 1\bra 2$, such that there is no back-action from the ancilla onto the data at all for this error. This syndrome detection scheme can be said to be fault-tolerant to a single loss event on the ancilla~\cite{Rosenblum2018}.
Further suppression of state preparation errors can be achieved using a concatenated error-correction scheme, as discussed in~\cref{sec:concat}.

\section{\label{sec:errorsanderrorprop}Errors and error propagation}

As was emphasized in the original description of GKP codes in Ref.~\cite{Gottesman01}, the discussion of fault tolerance is somewhat different for bosonic codes than for conventional qubit stabilizer codes. For bosonic codes, it is entirely acceptable for \emph{every} mode involved in the quantum computation to have small errors.  In fact, this is unavoidable in practice. In the context of number-phase codes, ``smallness" of errors is relative to a code's number and rotational distance, c.f.~\cref{eq:codedistances}, as we explain in more detail below. Therefore, propagation of small errors is not a major issue. What is catastrophic, however, is when 
initially small and approximately correctable errors are amplified to the point where
the probability of \emph{logical} errors become likely. Gates that turn small errors into large errors should therefore be avoided.

As we show in the following, the gates introduced in the previous section behave nicely in terms of error amplification. They do not amplify the magnitude of a number-shift error of the form $\sim \hat a^k$ ($\sim \hat a^{\dagger k}$), although they might map a number-shift error onto a number-shift error plus a rotation error, $\e^{i\theta \hat n}$. The upshot is that
the new rotation error is proportional to the size of the number-shift error relative to the Fock-space distance, \emph{i.e.}, $\theta \sim k/d_N$. In this sense, the new error is proportional in size to the initial error, and importantly, small errors remain small. Restricting to gates with such well-behaved error propagation properties is likely to be a pre-requisite for bosonic codes to have an advantage over unencoded qubits. We return to this point in~\cref{sec:concat} where we discuss a fault-tolerant concatenated quantum computing scheme.

\subsection{\label{sec:errors}Error bases and large vs. small errors}

There are two single-mode operator bases that are particularly useful in the context of rotation codes. The first is the set~\cite{Wunsche1999}
\begin{equation}\label{eq:errormodel0}
    \left\{\hat n^\ell \hat a^k, (\hat a\dg)^k \hat n^\ell \right\},
\end{equation}
where $k, \ell \ge 0$ run over all non-negative integers.
A straightforward way to show that this is an operator basis is to first expand the displacement operator in a normal ordered form,
\begin{equation}
  \begin{aligned}
    \hat D(\alpha) ={}& \e^{\alpha \hat a\dg - \alpha^*\hat a}
    = \e^{\alpha \hat a\dg}\e^{-\alpha^* \hat a}\e^{-\frac{1}{2} |\alpha|^2 }\\
    ={}& \e^{-\frac{1}{2} |\alpha|^2} \sum_{m,n=0}^\infty \frac{\alpha^m\left(\alpha^*\right)^n}{m!n!} (\hat a\dg)^m \hat a^n.
  \end{aligned}
\end{equation}
Since we can write $\left(\hat a\dg\right)^k \hat a^k = \sum_{\ell=0}^k c_{k\ell} \hat n^\ell$ by reordering operators, it follows that $\hat D(\alpha)$ can be expanded in terms of the set~\cref{eq:errormodel0}.\footnote{Explicitly, $c_{k \ell} = (-1)^{k-\ell}s(k,\ell)$ where $s(k,\ell)$ is a Stirling number of the first kind~\cite{Wunsche1999}.}
And, since the displacements form an operator basis, so does the set in~\cref{eq:errormodel0}, and the Kraus operators of any single-mode channel can therefore be expanded in terms of such operators.
    
A second useful operator basis is
\begin{equation}\label{eq:errormodel}
  \left\{\e^{i\theta\hat n}\hat a^k,\, \left(\hat a\dg\right)^k\e^{-i\theta \hat n} \right\},
\end{equation}
where $k\ge 0$ is an integer and $\theta \in[0,2\pi)$.
That this is a basis follows from~\cref{eq:errormodel0} since
$\hat n^\ell = \int_0^{2\pi} \frac{\dd\theta}{2\pi} c_\ell(\theta) \e^{i\theta\hat n}$ where $c_\ell(\theta) = \sum_{m=0}^\infty m^\ell \e^{-im\theta}$.\footnote{We can recognize that $c_\ell(\theta) = L(-\theta/2\pi,0,-\ell)$ with $L(\lambda,\alpha,s)$ the Lerch zeta function.} We will make use of this basis heavily in the following, and it is therefore convenient to introduce a short-hand notation:
\begin{equation}\label{eq:E}
  \hat E_k(\theta) \coloneqq \left\{ \begin{array}{cc}
      \e^{i\theta\hat n}\hat a^{|k|} \text{ for } k < 0 \\
      \left(\hat a\dg\right)\phantom{}^{\! |k|} \e^{i\theta\hat n} \text{ for } k \ge 0 
  \end{array}\right. .
\end{equation}
A negative $k$ thus denotes a downwards shift in number and a positive $k$ an upwards shift.

An error $\hat E_k(\theta)$ with $0 <  |k| < N$ is detectable for \emph{any} rotation code. This follows from the Fock-space separation of $N$ between $\ket{0_N}$ and $\ket{1_N}$. In general, we can not make a similarly sharp distinction between detectable and un-detectable rotations, but for the ideal number-phase codes introduced in~\cref{sec:CyclicStabilizerCodes}, a pure rotation error $E_0(\theta)$ with $0 < \theta < \pi/N$ is formally detectable.

Intuitively, an error with a small $|k|$ compared to $d_n/2 = N/2$ and small $|\theta|$ compared to $d_\theta/2 = \pi/(2N)$ is a ``small'' error and should be correctable to a good approximation for a number-phase code with $N$-fold rotation symmetry.
Typically, the codes discussed in this paper are only approximate error-correcting codes for physically relevant noise channels~\cite{Beny2010}.

\subsection{\label{sec:errorprop}Error propagation}

To show that our gates do not amplify errors too badly and to introduce the error-correction schemes in the next section, we need to understand how the errors propagate through the gates introduced in~\cref{sec:gates}. Here we describe errors using $\hat E_k(\theta)$ given in \cref{eq:E}.
The $\{\SN$, $\CROT_{NM} \}$ gates are generated by quadratic powers of the number operator, and although they are non-Gaussian, we show that they do not amplify errors too badly.
In contrast, the gate $\TN = \exp(i\pi \hat n^4/4N^4)$ is highly nonlinear, and as we show below this gate amplifies errors in a more damaging way, which is why we do not rely on using this gate in the scheme introduced in~\cref{sec:gates}.

All of the unitary gates in our scheme commute with pure rotation errors, $\e^{i\theta \hat n}$. For number-shift errors we can use that the commutation relation of an operator $\e^{c\hat n^\ell}$ with a power of $\hat a$ is given by the general formula
\begin{equation}\label{eq:error_propagate}
    \e^{c \hat n^\ell}  \hat{a}^k = \e^{c k f_{\ell-1}(\hat{n})} \hat{a}^k \e^{c \hat n^\ell},
\end{equation}
where $f_\ell(n) = n^{\ell+1} - (n+1)^{\ell+1}$ is a polynomial of order $\ell$.\footnote{Since $[\hat{n}^\ell, \hat{a}] = f_{\ell-1}(\hat{n}) \hat{a} $, we have $e^{c \hat{n}^\ell} \hat{a} e^{-c \hat{n}^\ell} =  e^{c f_{\ell-1}(\hat{n})} \hat{a}$.
} A similar commutation relation for $(\hat a\dg)^k$ follows straightforwardly by Hermitian conjugation.

Applying this to  $\ZN$ ($\ell = 1$), we have $f_0(\hat n) = -1$, and the prefactor in~\cref{eq:error_propagate} is a global phase. Explicitly,
\begin{equation}
  \ZN \, \hat E_k(\theta) = \e^{i \frac{\pi k}{N}}\hat E_k(\theta) \, \ZN,
\end{equation}
for a general error $\hat E_k(\theta)$.
Recall that we are labeling the errors with an index $k < 0$ for an error $\sim \hat a^{|k|}$ and $k>0$ for an error $\sim (\hat a\dg)^{|k|}$. We see that propagating an $\hat E_k(\theta)$-error through the Gaussian $\ZN$-gate only leads to a phase factor and the error is not amplified at all.

Gates with $\ell = 2$, such as $\hat{S}_N$ in~\cref{eq:S_N}, on the other hand, introduce new rotation errors:
\begin{equation}
    \hat{S}_N  \hat E_k(\theta)
    = \e^{i \frac{k \pi}{N^2} } \hat{E}_k \left(\theta + \frac{\pi k}{N^2} \right) \hat{S}_N.
\end{equation}
The initial rotation error $\theta$ is amplified by an amount proportional to the initial number-shift error $k$. Recall that the angular distance of the code is $d_\theta = \pi/N$ and the number distance is $d_n = N$. If $|k| < d_n/2$ is a small, approximately correctable number-shift error, then the additional rotation error $\pi|k|/N^2 < d_\theta/2$ is small as well. Thus if $k$ and $\theta$ are sufficiently small and approximately correctable, we expect the error after the gate to be approximately correctable as well. This is akin to a gate implemented by a constant depth circuit on a qubit stabilizer code, where an initial error can spread at most within a constant-size light cone~\cite{Bravyi2013,Pastawski2015}.\footnote{Of course, one also has to take into account that the \emph{type} of error introduced by the gate is different from the initial error, and it is not clearcut whether a rotation error of magnitude $\pi k/N^2$ is better or worse than a number-shift error of magnitude $\hat a^k$ or $(\hat a\dg)^k$, even though they are both in some sense small when $k$ is small compared to $N/2$. In general this will depend on the code's ability to deal with phase and number-shift errors.}

\begin{figure}
  \centering
  \includegraphics{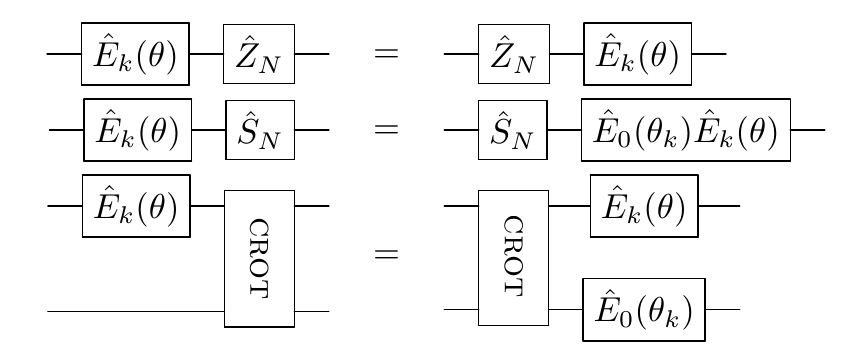}
  \caption{\label{fig:errorprop}Propagation of errors through gates. An error $\hat E_k(\theta)$ at the input of a gate (left column) is propagated to new errors at the output (right column). The circuit identities hold up to an overall phase factor (see the text for the exact relations). The new error $\hat E_0(\theta_k)$ for $\SN$ and $\CROT$ is a pure rotation error with $\theta_k = \frac{\pi k}{NM}$, with $N=M$ for $\SN$.
  }
\end{figure}

Commuting an error through the $\CROT$ gate spreads errors between the two modes involved in the gate. Label the two modes $a$ and $b$, where $a$ is encoded in an order-$N$ rotation code and $b$ in an order-$M$ rotation code. Commuting $\CROT$ through an error on mode $a$ gives
\begin{equation}\label{eq:croterror}
  \CROT_{NM}\, \hat E^a_k(\theta)
  = \hat E^a_k(\theta) \, \hat E^b_0\left(\frac{\pi k}{NM}\right) \CROT_{NM},
\end{equation}
where the superscript $a/b$ indicates the mode that an operator acts on.
Here, the initial number-shift error on mode $a$ spreads to a rotation error on the mode $b$.
Again, if $|k|<N/2$ is small with respect to the number distance of mode $a$, mode $b$ is rotated by an angle $\pi |k|/ (NM) < \pi/(2M)$ which is small compared to its angular distance. In other words, even though the error has spread to a new rotation error on mode $b$, the error remains small relative to the rotational distance of this mode. In the context of $[[n, 1]]$ qubit codes with a single logical qubit encoded in $n$ physical qubits~\cite{Gottesman09}, the $\CROT$ gate is analogous to a logical $\lbar{C}_Z$ gate enacted transversally between two code blocks. How errors propagate through the gate set $\{\ZN, \SN, \CROT_{NM} \}$ is summarized in~\cref{fig:errorprop}.

Finally, we show that gates generated by $\hat{n}^\ell$ for $\ell > 2$, such as the $\TN$ gate in~\cref{eq:T_N}, are unlikely to be useful for fault-tolerant quantum computing, because errors are amplified and spread in a potentially damaging way. 
Propagating an error through $\TN$ produces additional linear and nonlinear rotation errors,
	\begin{align} \label{eq:T_N}
		\hat{T}_N E_{k}(\theta)=& \left\{ \begin{array}{cc}
      e^{i \frac{ k\pi }{ 4 N^4} } \hat{F}_N^k \hat E_{k} \left(\theta + \frac{k \pi}{2 N^4} \right) \hat{T}_N \text{ } k < 0 \\
      e^{i \frac{ k\pi }{ 4 N^4} } \hat{E}_k\left(\theta + \frac{k \pi}{2 N^4} \right) \hat{F}_N^k \hat{T}_N \text{ } k \ge 0 
  \end{array}\right. ,
	\end{align}
where $\hat F_N^k= e^{i \frac{k \pi  }{4 N^4}  ( 4\hat{n}^3 + 6\hat{n}^2)}$ is a nonlinear rotation error.
The precise consequences of such nonlinear errors on relevant codes requires further study; however, we expect it to be rather damaging, see, \emph{e.g.}, Ref.~\cite{Sanders92}.
As already discussed in~\cref{sec:gates} we avoid using this gate in our scheme, instead relying on gate teleportation for a logical $\bar{T}$.

  \section{\label{sec:EC}Error correction}

Number-phase codes, introduced in \cref{sec:CyclicStabilizerCodes},  are naturally robust to shifts in number with higher-$N$ codes tolerating larger shifts. On the other hand, their robustness to phase errors is directly related to the rotational distance $d_\theta = \pi/N$, and rotations reasonably small compared to $d_\theta/2$ can be corrected. Here, we introduce a constructive and practical error correction scheme for number-phase codes and study its performance numerically.

In a well designed error-correction scheme one needs to (i) use gates that do not amplify errors such that they become uncorrectable, (ii) use measurements that are robust to noise, and (iii) carefully construct ancilla interactions such that errors do not spread from ancilla qubits to data qubits in a catastrophic way.
In the context of $[[n, k]]$ qubit stabilizer codes, where $k$ logical qubits are encoded in $n$ physical qubits, two common approaches to fault-tolerant error correction are Steane (Steane-EC) and Knill (Knill-EC) error correction~\cite{Gottesman09}.

In an adaptation of these schemes to single-mode codes (\emph{i.e.}, codes with each logical qubit encoded in a single bosonic mode), $[[n,1]]$ code blocks are replaced by bosonic modes, and gates and measurement are replaced by fault-tolerant bosonic counterparts. For example, in Ref.~\cite{Gottesman01}, Steane-EC was adapted to GKP codes by replacing transversal \CNOT s by the fault-tolerant (linear) \SUM{} gate and logical $\X$ and $\Z$ measurements by quadrature measurements.

Rotation codes, on the other hand, do not in general have a fault-tolerant \CNOT{} (or Hadamard), due to the highly non-linear nature of the $\X$-operator. Steane-EC can therefore not be used directly.
We get around this issue by adapting a version of Knill-EC to number-phase codes. 
This approach turns out to have many salient features. As an alternative, we also present a hybrid Steane-Knill scheme in~\cref{sec:hybrid}.

\subsection{\label{sec:Knill}Error correction by teleportation}

\begin{figure}
\centering
\includegraphics[width=0.95\columnwidth]{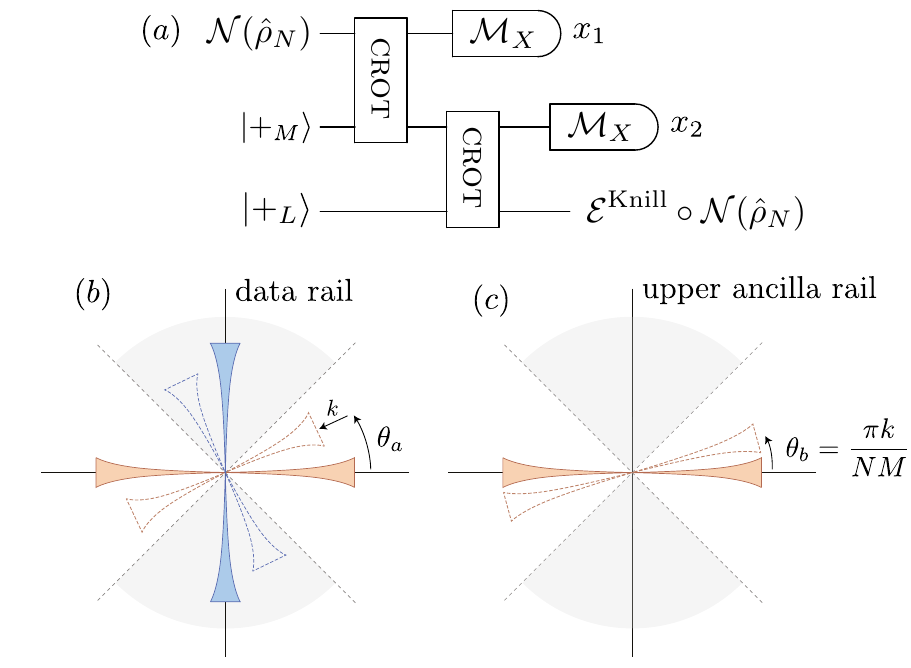}
\caption{\label{fig:ecschemes}
$(a)$ Schematic illustration of telecorrection using number-phase codes, where $\MX$ are phase measurements with outcomes $x_i$. The scheme is based on how errors spread through the \CROT{} gate. An arbitrary error $\hat E_k(\theta_a)$ on the order $N$ data rail $(b)$ induces a rotation of the order $M$ upper ancilla rail by $\theta_b = \frac{\pi k}{NM}$ $(c)$. Phase measurements on the data rail and the upper ancilla rail extract information about $\theta_a$ and $\theta_b\sim k$. By the end of the circuit the logical information has been teleported to the bottom order-$L$ ancilla rail (typically $L=N$). 
}
\end{figure}

We present a bosonic version of Knill-EC~\cite{Knill05,Knill05b,Dawson06} for number-phase codes. Its implementation requires $\CROT$ gates, phase measurements, and preparation of $\ket{+_N}$ states, as illustrated in~\cref{fig:ecschemes}$(a)$.
The measurements attempt to distinguish between damaged logical codewords in the dual basis.\footnote{For rotation codes that are not number-phase codes, \emph{i.e.} codes with poor phase resolution $\Delta_N(\theta)$, it is possible that other measurements might be devised.} Similar schemes have been considered previously in the context of measurement-based quantum computing~\cite{Dawson06}, where it is occasionally referred to as \emph{telecorrection} due to the teleportation-based error-correction circuit: the circuit in~\cref{fig:ecschemes}$(a)$ can be recognized as two consecutive one-bit teleportations.

A particular feature of the teleportation-based error-correction scheme makes it desirable for number-phase codes. Natural errors for bosonic modes include loss and gain errors that shift the codewords off the Fock grid $\ket{kN}$. Actively correcting such errors generally requires difficult, non-linear shift operations up or down the Fock ladder, see \emph{e.g.} Ref.~\cite{Radtke:2017aa}. In contrast, for the circuit in~\cref{fig:ecschemes}$(a)$, the damaged state is teleported into a fresh ancilla, thus restoring the codespace, with only a \emph{logical} error channel remaining (assuming, of course, error-free ancillae). Whether one can correct this remaining logical error depends on the magnitude of the initial error and the error-correcting properties of the code.

The ability to correct errors using teleportation is based on the following circuit identity (c.f., \cref{fig:errorprop})
\begin{equation}\label{eq:Knillidentity}
  \begin{aligned}
  \Qcircuit @C=1em @R=1.5em {
    \lstick{N} & \gate{\hat E} & \multigate{1}{\rotatebox{270}{\CROT}} & \qw                                       & \qw \\
    \lstick{M} & \qw           & \ghost{\rotatebox{270}{\CROT}}        & \multigate{1}{\rotatebox{270}{\,\CROT\,}} & \qw \\
    \lstick{L} & \qw           & \qw                                   & \ghost{\rotatebox{270}{\,\CROT\,}}        & \qw
 }
 \qquad
 \Qcircuit @C=1em @R=1.2em {
               & \multigate{1}{\rotatebox{270}{\CROT}} & \gate{\hat E}                         & \qw           & \qw \\
    \lstick{=} & \ghost{\rotatebox{270}{\CROT}}        & \multigate{1}{\rotatebox{270}{\CROT}} & \gate{\hat R} & \qw \\
               & \qw                                   & \ghost{\rotatebox{270}{\CROT}}        & \qw           & \qw
 }
  \end{aligned}
\end{equation}
where the labels $N$, $M$, and $L$ indicate the order of rotation symmetry for each encoded rail,
$\hat E = \hat E_k(\theta)$ is an arbitrary error as defined in~\cref{eq:E}, and $\hat R = \hat E_0\left(\frac{\pi k}{NM}\right)$ is a pure rotation error. Crucially, although the initial error $\hat E$ on the top rail spreads to a rotation error $\hat R$ on the second rail, this does not spread to an error on the third rail because $\hat R$ commutes with $\CROT$. By measuring the first and the second rail, we can teleport the state from the first to the third rail [c.f. \cref{eq:logicalteleport}] and remove the error in the process. The teleportation is successful as long as one is able to correctly distinguish between codewords in the dual basis in the presence of the errors.

We consider the action of the error-correction circuit in more detail for error-free ancillae.
Consider an encoded logical state $\hat{\rho}_N = \sum_{i,j=0}^1 \rho_{ij} \ket{i_N}\bra{j_N}$ that is sent through an arbitrary single-mode noise channel $\mathcal N(\hat{\rho}_N)$ followed by the error correction circuit in~\cref{fig:ecschemes}$(a)$.
One can show that the corresponding quantum channel can be written
\begin{equation}\label{eq:ec:Knill}
    \begin{aligned}
    \mathcal E^\text{Knill} \circ \mathcal N(\hat \rho_N)
    ={}& \frac{1}{4} \sum_{i,j=0}^3 \sum_{\vec x} 
    c_{ij}(\vec x)
    \Pauli_i \hat{\rho}_{L} \Pauli_j\dg,
    \end{aligned}
\end{equation}
where 
$\hat{\rho}_{L} = \sum_{i,j=0}^1 \rho_{ij} \ket{i_L}\bra{j_L}$ on the right-hand side represents the same logical state as $\hat{\rho}_N$ on the left hand side, but now encoded in the order-$L$ number-phase code of the bottom ancilla rail (in general this can be a different code). 
The operators $\Pauli_{i} \in \{\I, \Z, \X, \X\Z\}$ are \emph{logical} Paulis acting on the encoded order-$L$ output state.

The weights $c_{ij}(\vec x) \coloneqq \tr[\hat M_{x_1}\otimes\hat M_{x_2} \hat{\sigma}_{ij}]$ in~\cref{eq:ec:Knill} are set by the measuremnt scheme and the noise channel:
Here $\hat M_{x_i}$ are POVM elements for the two measurements in~\cref{fig:ecschemes}$(a)$, and the sum over outcomes, $\vec x = (x_1,x_2)$, should be understood as an integral for continuous measurement outcomes.
We have also introduced $\hat{\sigma}_{ij} \coloneqq \mathcal U_\CROT \circ \mathcal N \circ \mathcal U_\CROT\dg\big(\ket{i}\bra{j}\big)$ with
$\ket{i} = \Hd\ket{a_N}\otimes\Hd\ket{b_M}$ with $ab$ being the binary representation of $i$ (\emph{i.e.} $\ket i$ runs over $\ket{k_N}\otimes\ket{\ell_M}$ with $k,\ell=\pm$). 
Here $\mathcal U_\CROT \bullet = \CROT \bullet \CROT\dg$ and $\mathcal U_\CROT\dg \bullet = \CROT\dg \bullet \CROT$.
The operators $\hat{\sigma}_{ij}$ thus represents a damaged two-mode logical dual basis, where the noise channel $\mathcal N$ has been commuted through the $\CROT$ gate.

Crucially, \cref{eq:ec:Knill} shows that after error correction the remaining error channel acts entirely in the logical subspace of the output mode, for \emph{any} noise channel $\mathcal N$ acting on the input mode.
In principle, a measurement-dependent Pauli recovery $\Pauli_x$ could be applied, but the simplest solution is to track the \emph{most likely} Pauli correction $\Pauli_{i*}$ in a Pauli frame. The right Pauli correction can be guessed using, for example, a maximum likelihood decoder
\begin{equation}\label{eq:KnillML}
    i^*(\vec x) = \text{argmax}_i \tr[\hat M_{x_1}\otimes \hat M_{x_2} \hat \sigma_{ii}].
\end{equation}

\begin{figure*}
\centering
\includegraphics{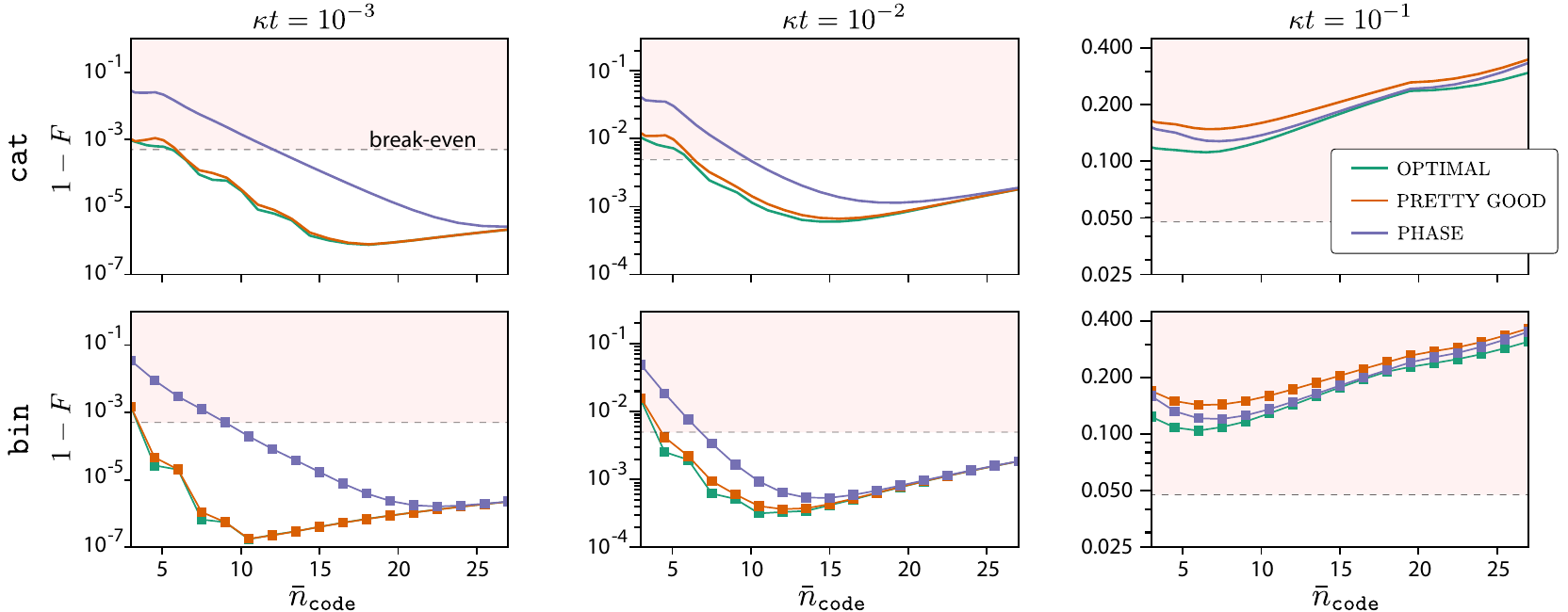}
\caption{\label{fig:numerics1} Average gate infidelity as a function of average excitation number in the code, $\ncode$ [\cref{eq:code_energy}], for an $N=3$ cat code (top row) and binomial code (bottom row). 
We compare the theoretically optimal error-correction scheme found numerically (\optimal) to our telecorrection scheme using Pretty Good Measurements (\pg) and canonical phase measurements (\phase).
The encoded data rail is subject to noise through evolution under the master equation in \cref{eq:lossdephasingme} with equal loss and dephasing strength, $\kappa_\phi t =\kappa t$. 
Each column shows results for a different amount of total noise $\kappa t$ before error correction is performed.
 A code performs better than break even (the uncorrected trivial encoding) whenever the gate infidelity is below dashed line (falls outside the shaded region).}
\end{figure*}

\subsection{Numerical results for loss and dephasing with error-free ancillae} \label{sec:ECnumerics}

In a recent work, Albert et al.~\cite{Albert17} investigated the error-correction performance of several bosonic codes including cat and binomial under pure loss. In that work, a recovery map was found by numerical optimization, and it was shown that bosonic codes can heavily suppress loss errors, in principle. While it sets a useful bound, numerical optimization does not provide any information about how to implement the recovery map in practice. An important step forward is therefore to compare an \emph{explicit} error-correction scheme to the numerically optimal recovery map for relevant noise channels. To make this comparison we focus in this section on a setting with noise-free ancillae and idealized measurements. Dealing with faulty ancillae and measurement noise is a non-trivial problem which we return to in a qualitative discussion in~\cref{sec:concat}. One of our main results is that the teleportation-based error correction scheme presented above is close to optimal in the relevant region of small noise, with a gap from the optimal scheme arising primarily from inherent noise in the phase measurements.

In this section we numerically compute the average gate fidelity~\cite{Nielsen02} for a channel composed of an ideal encoding of a qubit into a number-phase code, followed by a single-mode noise channel, and finally error correction using noise-free ancillae and ideal gates.\footnote{The average gate fidelity $F$ for a $d$ level system is related to the entanglement fidelity $F_\text{ent}$ as $F = (dF_\text{ent}+1)/(d+1)$~\cite{Nielsen02}.} 
We have confirmed that the Knill-EC scheme from the previous section and the hybrid Steane-Knill scheme presented in~\cref{sec:hybrid} perform identically for the codes and parameter ranges we investigated.

In the idealized situation where the ancillae are noise free, one can use, \emph{e.g.}, a simple $M=1$ cat code with $\ket{0_\cat} \propto \ket\alpha + \ket{-\alpha}$, $\ket{1_\cat} \propto \ket{\alpha} - \ket{-\alpha}$ for the middle ancilla rail in~\cref{fig:ecschemes}$(a)$.
For large enough $\alpha$, rotations of the state $\ket{+_\cat}$ can be detected arbitrarily well with phase measurements, giving essentially perfect syndrome measurements for this rail.
Moreover, we set $L=1$ for the bottom (output) rail and use the trivial encoding, $\ket{0_\triv} = \ket 0$ and $\ket{1_\triv} = \ket 1$, for this mode. The error-correction circuit therefore also decodes the encoded information.

The total quantum channel we consider is
\begin{equation}\label{eq:ec:totalchannel}
    \mathcal E^\text{Knill} \circ \mathcal N \circ \mathcal S,
\end{equation}
where $\mathcal S \bullet = \hat S \bullet \hat S\dg$, with $\hat S = \ket{0_N}\bra 0 + \ket{1_N}\bra 1$, is the ideal encoding map for a given code,
$\mathcal N$ is a noise map, and $\mathcal E^\text{Knill}$ is the error-correction circuit.
This is a logical channel with qubit input and qubit output, due to the use of a trivial encoding for the final output mode of the error-correction circuit.
Note that $\mathcal E^\text{Knill} \circ {\mathcal N}$ is given by~\cref{eq:ec:Knill} in the case when recoveries are tracked in software rather than applied explicitly. To compute the average gate fidelity, we explicitly apply 
a correction $\Pauli_{i^*}\dg$ to undo the most likely Pauli error using the decoder in~\cref{eq:KnillML}.

The noise channel we consider consists of simultaneous loss and dephasing. Specifically, $\mathcal N(\hat\rho)$ is the solution to the master equation
\begin{equation}\label{eq:lossdephasingme}
    \dot {\hat{\rho}} = \kappa \mathcal D[\hat a] \hat\rho + \kappa_\phi \mathcal D[\hat n] \hat\rho,
\end{equation}
integrated up to some unitless time $\kappa t$,
with $\mathcal D[\hat L] \hat\rho = \hat L \hat\rho \hat L\dg - \half \hat L\dg \hat L \hat\rho - \half \hat\rho \hat L\dg \hat L$. A Kraus-operator decomposition of $\mathcal N$ is given in~\cref{app:kraus}.

We wish to quantify the performance of the telecorrection scheme from~\cref{sec:Knill}. To this end, we compare the average gate fidelity for the channel in~\cref{eq:ec:totalchannel} with a channel where the error-correction map is replaced by the completely positive trace-preserving map $\mathcal E^\text{opt}$ that maximizes the fidelity. The latter can be found by solving a semi-definite program~\cite{Albert17}, and we refer to this recovery as \optimal.
The discrepancy between the teleportation-based scheme and the optimal is due in large part to the inherent uncertainty of the phase measurements, c.f.~\cref{sec:phaseest}, and is especially prominent for small average excitation numbers. We obtain further insight into the origin of this gap by comparing canonical phase measurements to Pretty Good Measurements (see~\cref{sec:prettygood}) for the data rail in the circuit~\cref{fig:ecschemes}$(a)$. The measurement on the middle ancilla rail is in both cases a canonical phase measurement as defined in~\cref{eq:POVM_Holevo}.\footnote{Numerically we implement phase measurements by discretizing the phase $\theta_j = \theta_0 + j2\pi/J$, $j=1,\dots,J$, and define POVMs $\hat M_j = \int_{\theta_j}^{\theta_{j+1}} \dd \phi \ket{\phi,s}{\bra{\phi,s}}$ where $\ket{\phi,s} = \frac{1}{\sqrt s} \sum_{n=0}^{s-1} \e^{in\phi}\ket n$ is a truncated Pegg-Barnett phase state.} We refer to these two error-correction schemes as \phase{} and \pg, respectively, depending on which measurement is performed on the data rail.
We emphasize that the \pg{} scheme is used in our numerics purely to gain insight into the origin of the gap between \phase{} and \optimal.

We focus here on cat and binomial codes as examples of number-phase codes. As expected, these codes have similar performance for large average excitation number, $\ncode$ [\cref{eq:code_energy}], but can show significant differences for smaller $\ncode$~\cite{Albert18}.
Remarkably, we find that the \phase{} error-correction scheme approaches \optimal{} for large $\ncode$ and small noise strength, and that the \pg{} scheme is \emph{near} \optimal{} for almost all $\ncode$ and small noise strengths.

In~\cref{fig:numerics1} we show examples of the average gate infidelity $1-F$~\cite{Horodecki1999,Nielsen02} for an $N=3$ cat code (top row) and $N=3$ binomial code (bottom row) as a function $\ncode$.
We fix the dephasing rate to be equal to the loss rate $\kappa_\phi = \kappa$ and compare three different noise strengths parameterized by $\kappa t$. The dephasing rate $\kappa_\phi$ models both natural dephasing and additional uncertainty in the phase measurements, which motivates choosing a fairly large $\kappa_\phi$. The dashed line in the figures shows the average gate infidelity using the trivial encoding $\ket{0_\triv} = \ket 0$ and $\ket{1_\triv} = \ket 1$ on the data rail with no error correction. This marks the \emph{break-even point} above which, in the pink region, encoding and error correction perform worse and provide no advantage.

There are several takeaway messages from the results in~\cref{fig:numerics1}. First, as advertised, both the \pg{} and \phase{} error-correction schemes are near optimal for large $\ncode$. While for \phase{} there is a significant gap at small to moderate $\ncode$, the \pg{} scheme performs very close to optimal except for very small $\ncode$ and/or large noise strengths. Second, for all schemes the codes exhibit an optimal $\ncode$ where the infidelity reaches a minimum. Under small loss rates the optimal $\ncode$ for \phase{} is much larger than for \pg{} and \optimal, due to the poor ability of phase measurements to distinguish states with small $\ncode$. Third, we see that the binomial code generally outperforms cat for small $\ncode$ and low noise strength, while the performance is identical for large $\ncode$. We note that for smaller dephasing rates, binomial codes would have increasing advantage over cat codes since a more loss-dominated noise channel generically favors smaller $\ncode$~\cite{Albert17}.

\begin{figure*}
\centering
\includegraphics{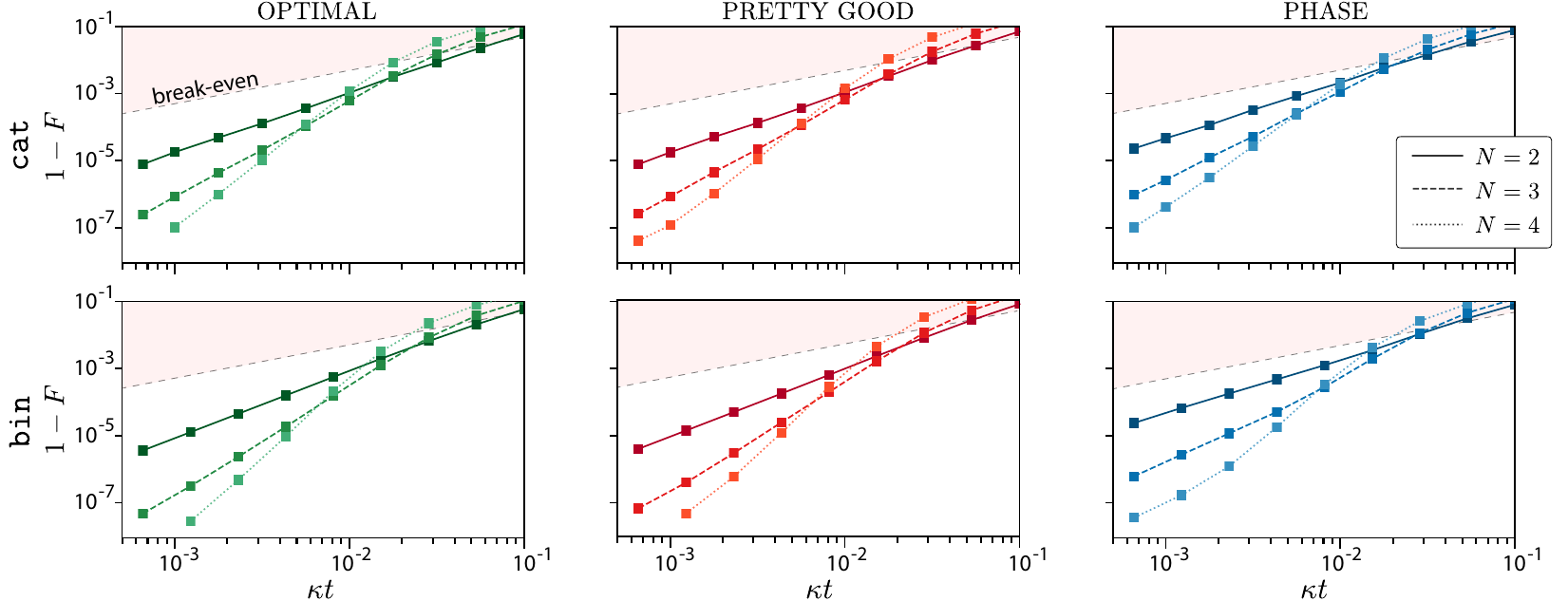}
\caption{\label{fig:numerics2}Average gate infidelity as a function of noise strength $\kappa t$ for cat codes (top row) and binomial codes (bottom row). For each $N$ and each $\kappa_\phi t = \kappa t$, the optimal average excitation number $\ncode$ is used for each code. 
A code performs better than break-even whenever the gate infidelity is below the dashed line (outside the shaded region). We only show results for $1-F\ge 2.5 \times 10^{-8}$, $\kappa t \ge 0.5 \times 10^{-3}$ and $N \le 4$ due to prohibitively large Fock-space truncation needed for numerical simulations and numerical accuracy issues for very small infidelity.}
\end{figure*}

\Cref{fig:numerics1} also shows that there is a significant potential for going beyond break-even, at least within the idealized error model considered here. To investigate this further we compare cat and binomial codes with different $N$ in~\cref{fig:numerics2}. For each $N$ and each code, we choose the optimal $\ncode$. We then plot the logical infidelity $1-F$ as a function of noise strength for the three different error-correction schemes \optimal, \pg{} and \phase.

There are two main observations from~\cref{fig:numerics2}. First, both cat and binomial codes break even by orders of magnitude for all error-correction schemes under the simplified noise model considered here. We can identify break-even pseudo-thresholds, defined as the noise strength $(\kappa t)_\text{be}$, where the infidelity is equal to that of the trivial encoding with no error correction (\emph{i.e.}, where the lines in~\cref{fig:numerics2} cross the pink/white boundary). The break-even thresholds are fairly high, falling in the $1$--$10$\% range, for the range of $N$ considered, but they decrease with larger $N$. The second observation is that the logical infidelity $1-F$ falls off more rapidly with higher $N$, over a range of $\kappa t$. However, the gain in performance is diminishing with increasing $N$. We do not expect the infidelity to become arbitrarily small with increasing $N$, since the protection against phase noise decreases. The relatively large reductions in infidelity with increasing $N$ seen in~\cref{fig:numerics2} suggest that the performance is not limited by dephasing for the noise parameters and codes considered here.

The results presented in this section have consequences for the prospects of using number-phase codes for fault-tolerant quantum computing. On one hand, the large break-even potential motivates further study of these codes under realistic noise models including noisy ancillae, noisy encoding, more realistic measurement models, and noise during gates. A pertinent question is how sensitive the performance is to nonlinear unitary errors such as those from unwanted self-Kerr interactions. On the other hand, the expectation that arbitrarily low infidelities cannot be reached means that cat and binomial codes must ultimately be concatenated with a second code to suppress errors further.

\section{\label{sec:concat}Roadmap to Fault Tolerance}

In this section we outline a scheme that achieves a long-standing goal for number-phase codes such as cat and binomial: A fault-tolerant universal set of operations. 
First, let us summarize the challenges that need to be overcome for fault-tolerance with number-phase codes (and rotation codes more generally):
\begin{itemize}
\item We touched upon some of the challenges with state preparation in~\cref{sec:stateprep}. A modular excitation-number measurement (\cref{sec:numbermodN}) can be performed to check a state for loss and gain errors, but not without risk of introducing new rotation errors. Unfortunately, we have not been able to find a practical, nondestructive measurement to check for rotation errors. If only number-shift errors are checked for, state-preparation noise will be biased towards rotation errors.\footnote{This is a noticeable difference with GKP codes, where the two respective syndrome measurements for quadrature shifts are equally easy~\cite{Gottesman01}. Fundamentally the issue stems from the highly non-linear number-translation ``stabilizer'' $\TransN$, c.f.~\cref{eq:transstabilizer} and~\cref{fig:rotGKP_comparison}.}

\item Realistic syndrome measurements are going to be rather noisy. A standard approach, especially in topological stabilizer codes, is to repeat the syndrome measurements multiple times and decode based on a record of measurement outcomes~\cite{terhal2015quantum}. However, due to the destructive nature of the measurements in~\cref{fig:ecschemes}, they cannot straightforwardly be repeated.

\item We expect that logical error rates can only be suppressed up to a point for realistic error channels. In particular, there is a tradeoff between resilience to number-shift and rotation errors, c.f.~\cref{eq:codedistances}.\footnote{GKP codes suffer a similar tradeoff between resilience to position and momentum shifts~\cite{Gottesman01}.}
\end{itemize}

One way to overcome these issues is to concatenate the bosonic code with a conventional qubit code so that the qubit code can deal with errors not handled by the bosonic code. This is broadly the approach taken recently for GKP codes, \emph{e.g.}, in Refs.~\cite{Menicucci14,Vuillot2018,Noh:2019aa}. Tailoring fault-tolerant gadgets to the respective strengths and weaknesses of the bosonic code is, however, not a trivial task. If this is not done carefully, the advantage of using a bosonic code at the ground level is diminished, and any performance benefits over using bare qubits might be lost. Simply, we want to optimize the concatenated scheme to maximally exploit the error-correction properties and noise resilience of the ground-level bosonic code.

In the following we outline how this goal can be achieved using concatenation with a Bacon-Shor subsystem code~\cite{Bacon2006,Poulin2005,Aliferis2007}. This specific choice serves to illustrate the broader point of how a fault-tolerant concatenated scheme can be tailored to exploit the strengths of the underlying bosonic code.

A single Bacon-Shor code can have a high fault-tolerant pseudo-threshold at intermediate code size, even though the threshold vanishes in the limit of infinitely large system size~\cite{cross2009comparative,Napp2013,brooks2013fault}. Bacon-Shor codes are attractive candidates for near- and intermediate term logical qubits due to their geometrically local layout in two dimensions and weight-two gauge operator checks~\cite{Aliferis2007,brooks2013fault,Gottesman2016}. Specifically, we will make use of a teleportation-based scheme developed by Aliferis and Preskill for repetition codes in Ref.~\cite{Aliferis08} and generalized to Bacon-Shor codes by Brooks and Preskill in Ref.~\cite{brooks2013fault}.
We refer to these two works collectively as the Aliferis-Brooks-Preskill (ABP) scheme.

The key to our fault-tolerant gadgets is to use only bosonic operations that do not amplify errors within a mode and do not spread small errors to larger errors at the bosonic level. This allows the error correction properties of the bosonic code to be exploited fully. The bosonic code operations we make use of are
\begin{equation}\label{eq:Gbos}
    \mathcal G_\text{bosonic} = \{ \mathcal P_\ket{+_N}, \MX, \CROT \}.
\end{equation}
We also assume universal control over two-level ancillae, or equivalently, bosonic modes in the trivial encoding $\ket{0_\triv} = \ket 0$, $\ket{1_\triv} = \ket 1$, and $\CROT_{N1}$ gates between the two-level ancillae and order $N$ rotation codes.
Using the ABP scheme, this suffices for universal quantum computing.
Remarkably, the bosonic error correction is seamlessly integrated into the Bacon-Shor error correction with our approach, such that minimal overhead is incurred. The bosonic error correction is in this sense highly efficient.

\subsection{Parallel teleportation}

\begin{figure}
\centering
\includegraphics{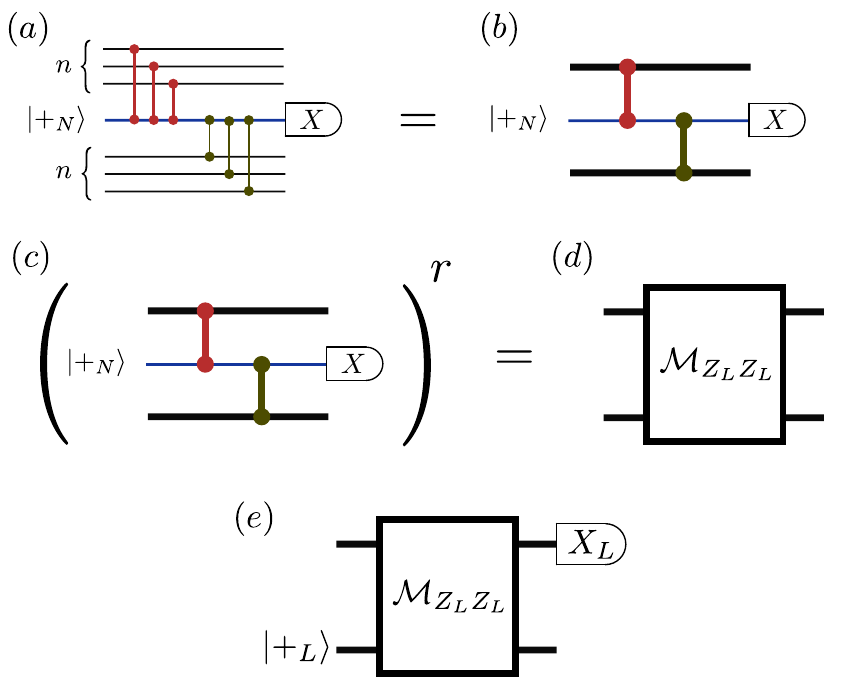}
\caption{\label{fig:aliferis}Building blocks of the error-correction scheme from Ref.~\cite{Aliferis08}. $(a)$ The rails represent number-phase codes and the $\hat C_Z$ gates are $\CROT$ gates. A $\hat C_Z^{\otimes n}$ gate enacted between a repetition code block and a single ancilla acts as a logical $\hat C_Z$. $(b)$ The same circuit as in $(a)$ where the thick lines represent length $n$ repetition code blocks. $(c,d)$ By repeating a nondestructive measurement of $\hat Z_L \otimes \hat Z_L$ $r$ times, we get a robust $\MZZL$ gadget. $(e)$ The error-correction gadget from Ref.~\cite{Aliferis08}. Here, $\ket{+_L} = \ket{+_N}^{\otimes n}$, and the $\hat X_L$-measurement is independent, destructive phase measurement of each mode in the repetition code block.}
\end{figure}

Let us start with the issue of measurement errors in the error-correction scheme in~\cref{fig:ecschemes}, as this 
gives an entry point into the ABP scheme. As already stated, because the syndrome information is extracted using destructive measurements, the measurements can not simply be repeated. However, since the measurements and state preparation in~\cref{fig:ecschemes} are all in the $\X$-basis, a simple remedy is to concatenate with a length $n$ repetition code (for $n$ odd) in the dual basis,
\begin{equation}\label{eq:rep1}
\ket{\pm_L} ={} \ket{\pm_N}^{\otimes n}.
\end{equation}
Logical $\hat X_L$ measurements and state preparation for the repetition code are then simply independent phase measurements and independent preparation of the $n$ bosonic modes using a number-phase encoding. Measurement errors are suppressed by performing a majority vote on the $n$ outcomes.\footnote{Digitizing each bosonic mode measurement result as in~\cref{fig:phasestructure} and performing a majority vote is unlikely to be the optimal scheme. Recent work on GKP codes has shown that decoders that explicitly take advantage of continuous variable measurements have better performance~\cite{Fukui:2017aa,Vuillot2018,Fukui:2018aa}.} This means that, if we can generalize the teleportation circuit in~\cref{fig:ecschemes} to teleportations of repetition code blocks, we can use the simple encoding in~\cref{eq:rep1} to suppress measurement errors. This is essentially a classical encoding, and the repetition code plays a role analogous to repeated measurements in topological codes~\cite{terhal2015quantum}. 

A useful observation is that a $\hat C_Z$-gate enacted between each qubit in a repetition code block and a single ancilla qubit acts as a \emph{logical} $\hat C_Z$ between the repetition \emph{code block} and the one ancilla. So, for example, the red $\CROT$ gates between the upper block and the ancilla in~\cref{fig:aliferis}$(a)$ gives a logical $\hat C_Z$ between this block and the ancilla, leading to the identity with~\cref{fig:aliferis}$(b)$ where the thick lines represent repetition code blocks of length $n$. The circuit in~\cref{fig:aliferis}$(a,b)$ represents a nondestructive measurement of $\hat Z_L \otimes \hat Z_L$, where $\hat Z_L$ is a logical Pauli-$\hat Z$ for the repetition code block. To increase robustness to ancilla measurement errors, we can repeat this circuit $r$ times, as illustrated in~\cref{fig:aliferis}$(c,d)$ to get a robust $\MZZL$ gadget. Error correction is performed, following Ref.~\cite{Aliferis08}, by the logical one-bit teleportation circuit in~\cref{fig:aliferis}$(e)$. Note that with $n=r=1$, the error-correction gadget in~\cref{fig:aliferis}$(e)$ is identical to that in~\cref{fig:ecschemes}$(a)$. This is thus a direct generalization of the scheme we introduced above in~\cref{sec:EC}, and the error correction circuit performs \emph{both} bosonic code and repetition code error correction in one step.

Let us pause at this point to clarify what is achieved by concatenating with a repetition code.
First of all, the repetition code is used to suppress measurement errors, but it also protects against rotation errors more generally. This is especially useful if state-preparation noise is biased towards rotations, as will be the case if number-syndrome checks are used in this step.

To further understand the concatenated bosonic-repetition code's robustness to number-shift errors,
consider for the sake of concreteness
an $N=2$ number-phase code and a noise channel biased towards loss with negligible gain. 
An $N=2$ number-phase code can on its own be robust against a single loss error when gain is negligible.
Say, for example, a single loss error occurs on one of the ancilla rails in the $\MZZL$ gadget in~\cref{fig:aliferis}. It will spread to a clockwise rotation error of magnitude $\pi/4$ on \emph{all} of the upper and lower data rails. This is acceptable, because the number-phase code with angular distance $d_\theta=\pi/2$ can deal with rotation errors of this magnitude (note that under the assumption of no gain, one should bias the decoding procedure in~\cref{fig:phasestructure} towards clockwise rotations). Confidence in detecting this loss error is increased by the fact it is present in all $n$ phase measurements on the upper data rails.

Despite using only a repetition code, the scheme is thus robust to both rotation and loss errors, because the bosonic code at the ground level can handle both. This is what we mean when we say that the bosonic-code error correction is integrated in the qubit-code error correction.

On the other hand if, say, two of the $r$ ancillae in the $\MZZL$ have a single loss error, this will propagate to a rotation error of magnitude $\pi/2$, which is too large for an $N=2$ number-phase code to handle. Consequently, this leads to a \emph{logical} error at the repetition-code level with high likelihood. The number of ancillae $r$ and the repetition-code length $n$ thus have to be chosen dependent on the order of rotation symmetry of the number-phase code, $N$, as well as the physical loss rate.

We note that concatenation with a repetition code has recently been proposed in the context of $N=1$ cat codes~\cite{Puri:2019aa,Guillaud:2019aa}. An $N=1$ cat code does not provide any protection against loss or gain errors, but gives rise to a highly biased effective noise model because transitions $\ket{\alpha} \to \ket{-\alpha}$ are exponentially suppressed in $|\alpha|^2$. In this situation, no active error correction is performed at the bosonic-code level, and the purpose of the repetition code is to protect against loss and gain errors, which is orthogonal to what we propose here.

\subsection{Cats of cats and repetitions of repetitions}

The above concatenated error-correction scheme can be improved following Ref.~\cite{brooks2013fault}. The techniques presented there can be adopted straightforwardly to the present context, so we only briefly comment on the main ideas here.

Despite the scheme already described above being robust to both loss and rotation errors, it may be impractical to prepare number-phase codes with very large $N$. This limits the number of ancilla measurements, as explained above, because a single ancilla loss or gain error spreads to \emph{all} of the data rails, and vice versa. This problem can be overcome by replacing each of the $r$ $\ket{+_N}$ middle-rail ancillae in~\cref{fig:aliferis} by logical $\ket{+_L'}$ for a repetition code which is \emph{dual} to~\cref{eq:rep1}:
\begin{equation}\label{eq:rep2}
\ket{\pm_L'} = \smallfrac{1}{\sqrt 2}\big( \ket{0_N}^{\otimes n} \pm \ket{1_N}^{\otimes n} \big).
\end{equation}
The logical Pauli-$X$ for this code is $\hat X^{\otimes n}$. However, since we only require destructive measurements of the ancillae, one can perform destructive phase measurement to measure each of the $n$ ancilla modes in the $\ket{\pm_N}$ basis, as in~\cref{fig:phasestructure}, and take the parity of the digitized outcomes.

\begin{figure}
\centering
\includegraphics{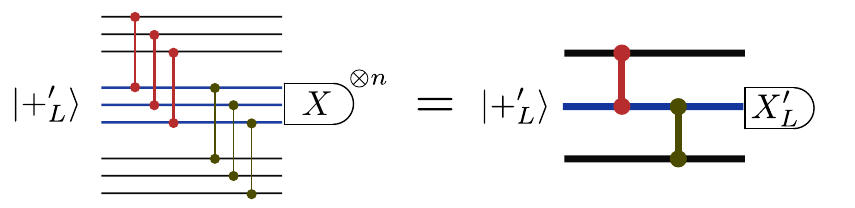}
\caption{\label{fig:aliferiscats}An extension of the error correction scheme from Ref.~\cite{Aliferis08} where each ancilla is encoded in the state $\ket{+_L'}$, \cref{eq:rep2}. On the left, the rails represent number-phase codes, the $\hat C_Z$ gates are $\CROT$ gates, and the measurement is independent destructive phase measurement of each mode in a block. The parity of the $n$ digitized measurement outcomes is computed to perform a measurement in the $\ket{\pm_L'}$ basis. On the right, the thick lines represent reptition code blocks and the gates are logical $\hat C_Z$ between the code blocks. Error correction is otherwise performed exactly as in~\cref{fig:aliferis}: The circuit is repeated $r$ times and finally the upper data block is measured $\MX^{\otimes n}$, with the lower data block prepared in $\ket{+_N}^{\otimes n}$.}
\end{figure}

The purpose of using the dual repetition code~\cref{eq:rep2} for the middle-rail ancillae is that a fully transversal $\hat C_Z^{\otimes n}$ gate enacted between the two codes,~\cref{eq:rep1,eq:rep2}, performs a \emph{logical} $\hat C_Z$ gate between the two code blocks.\footnote{As explained in Ref.~\cite{brooks2013fault}, one can also use dual repetition codewords $\ket{+_L'}$ of any length $1 \le p \le n$. For $p<n$, some ancilla modes interact with multiple data modes. The two extremes $p=1$ and $p=n$ are shown in~\cref{fig:aliferis,fig:aliferiscats}, respectively.} The basic building block of the error correction gadget then simply becomes an encoded version of the scheme from~\cref{fig:ecschemes}$(a)$, as illustrated in~\cref{fig:aliferiscats}. The error correction scheme is otherwise identical to that in~\cref{fig:aliferis}. The crucial difference is that since the $\hat C_Z$ gates are now fully transversal, an error on any mode in the circuit spreads to at most two other modes.

The state $\frac{1}{\sqrt 2}\ket{0}^{\otimes n} + \frac{1}{\sqrt 2}\ket{1}^{\otimes n}$ is sometimes referred to as a ``cat state.'' Since any bosonic rotation code is also a generalized cat code, according to~\cref{eq:01codewords}, we can refer to the state $\ket{+_L'}$ as a ``cat-of-cats.'' Of course, preparing these cat states also incurs overhead. As explained in Ref.~\cite{brooks2013fault}, this can be done fault-tolerantly using only the basic operations $\mathcal G_\text{bosonic}$ (with about $2n$ $\CROT$ gates and $n$ extra ancillae for a length $n$ cat-of-cats.).

The last extension from Ref.~\cite{brooks2013fault} is to replace the repetition code with a Bacon-Shor subsystem code. A Bacon-Shor code can be thought of as a concatenation of a length $n$ repetition code with a second, length $m$, repetition code. While the code in~\cref{eq:rep1} 
only gives protection against rotation errors (beyond the protection provided by the bosonic code), a Bacon-Shor code with $m>1$ will provide additional protection against loss and gain errors. In general, one can optimize the ratio $m/n$ to the noise model and the capabilities of the bosonic code at the ground level. Error correction is still performed with a gadget as in~\cref{fig:aliferis}$(e)$ using a relatively straightforward generalization of the $\MZZL$ measurement, and where logical $\ket{+_L}$ for the bosonic-Bacon-Shor code is now a product of cat-of-cats states~\cite{brooks2013fault}. Crucially, the only operations ever needed are $\CROT$ gates between pairs of modes, preparation of $\ket{+_N}$ states, and destructive phase measurements to measure each mode independently in the $\ket{\pm_N}$ basis.

\subsection{Universality}

In the ABP scheme, a logical $\CNOT$ gate is built out of the fundamental operations in~\cref{eq:Gbos}~\cite{Aliferis08,brooks2013fault}. Universality is furthermore achieved by injecting noisy states $\ket{+i} = (\ket 0 + i\ket 1)/\sqrt 2$ and $\ket{T} = (\ket 0 + \e^{i\pi/4}\ket 1)/\sqrt 2$ directly into the Bacon-Shor (or repetition) code block. In the present context, state injection can be performed by preparing the two lowest levels of an ancilla, \emph{e.g.} a transmon, in the desired state $\ket\psi$, and using a cross-Kerr interaction between the ancilla and a bosonic mode to perform a $\CROT_{N1}$ gate. The one-bit teleportation gadget in~\cref{fig:aliferis}$(e)$ is then used to teleport the ancilla state into a code block. In this case, the upper rail in~\cref{fig:aliferis}$(e)$ is the unencoded two-level ancilla in the state $\ket\psi$, and the lower rail is the Bacon-Shor code block. This circuit prepares the code block in an encoded state $\ket{\psi_L}$ up to a known measurement-dependent logical Pauli. Several noisy $\ket{+i_L}$ or $\ket{T_L}$ states can be distilled to higher fidelity using standard state distillation protocols, as analyzed in Ref.~\cite{brooks2013fault}.

As a final remark, we emphasize that although the full-blown scheme using Bacon-Shor codes 
and cat-of-cats ancillae is quite complex, the simplest repetition code with single-mode ancillae should already be able to demonstrate all the key ingredients of a fault-tolerant scheme. It is therefore a promising candidate to realize small-scale logical qubits in near-to-medium term experiments~\cite{Gottesman2016}.

\section{\label{sec:conclusion} Conclusions and discussion}

We have presented a broad class of bosonic error-correcting codes characterized by discrete rotation symmetry, which we call bosonic rotation codes, or simply rotation codes for short.
Several well known codes, including cat, binomial and \ON-codes, are rotation codes.
Cat and binomial codes moreover belong to a subset of rotation codes, the number-phase codes, which are characterized by vanishing phase uncertainty for large excitation numbers.  Another member of this subset are Pegg-Barnett codes, which can be interpreted as the shift-resistant qudit codes introduced in Ref.~\cite{Gottesman01} embedded in the infinite-dimensional Hilbert space of a bosonic mode.

\begin{figure}
\centering
\includegraphics[width=1\columnwidth]{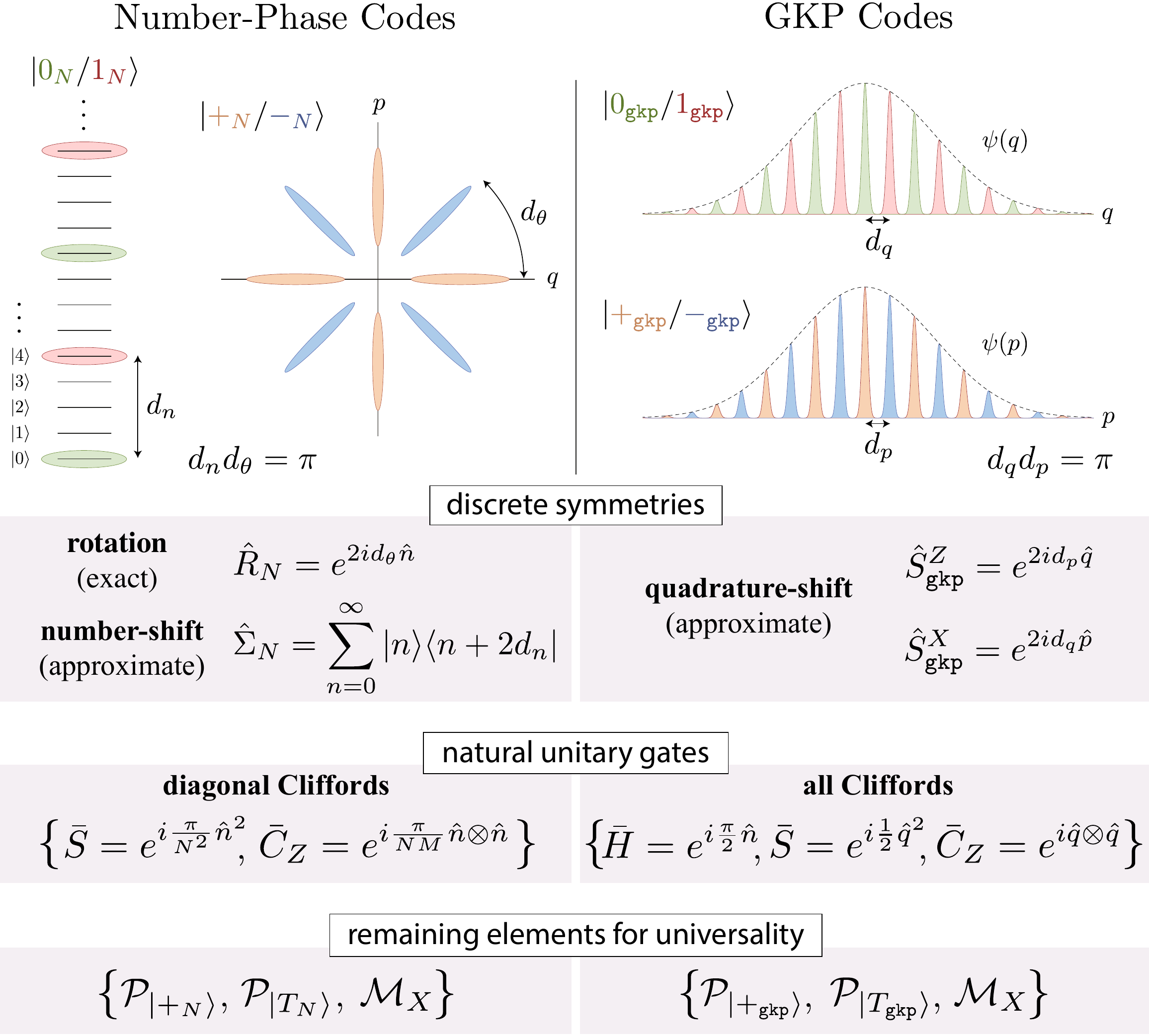}
\caption{\label{fig:rotGKP_comparison} Comparison between the number-phase codes introduced in this work and  GKP codes~\cite{Gottesman01}. For simplicity we consider the square-lattice GKP code. The cartoon shows codewords for an $N=4$ number-phase code, with $d_n=4$, $d_\theta=\pi/4$, and a GKP code with $d_q=d_p=\sqrt\pi$. For each code type there are unitary gates that are ``natural'' in the sense that they are generated by physically natural Hamiltonians and that they map small errors to small errors. To achieve universality, the unitary gates are supplemented with state preparation and Pauli measurements for both code types. Note that it was recently shown that the GKP magic state $\ket{T_\gkp}$ can be distilled using only Gaussian operations~\cite{Baragiola:2019aa}. The number-phase-code analog to the envelopes in the approximate GKP codewords are number-amplitude envelopes such as those in \cref{fig:FockDist}. }
\end{figure}

The theoretical framework we present here for number-phase codes allows a natural comparison with GKP codes, summarized in~\cref{fig:rotGKP_comparison}.
First, number-phase codes are analogous to approximate GKP codes, with number and phase playing dual roles in place of position and momentum.\footnote{Recently it was shown that GKP codes can be decomposed into two subsystems---a logical qubit and gauge mode---according to their discrete translation symmetry \cite{Pantaleoni:2019aa}, and a similar decomposition for rotation codes based on discrete rotation symmetry is likely. Note that the translation symmetry of any GKP code also implies two-fold rotation symmetry (see~\cref{sec:GKP}).}
Second, the fact that the GKP stabilizers are generated by Hamiltonians linear in annihilation and creation operators leads to Gaussian Clifford gates. In contrast, number-phase codes have one symmetry generated by a quadratic Hamiltonian ($\CN$) and one non-Gaussian symmetry ($\TransN$). As a consequence, a subset of the Clifford group---the diagonal Cliffords---can be generated by Hamiltonians that are quartic in annihilation and creation operators, while the full Clifford group contains highly non-linear gates. The smaller set of natural unitary gates and the higher degree of non-linearity is a disadvantage compared to GKP codes. However, the issue of non-linearity is not entirely clear cut. Self- and cross-Kerr interactions are very natural in some platforms~\cite{Wallraff2004,Nigg2012}, and 
some of the GKP Cliffords will require underlying physical nonlinearities as well---for example, to realize the shear gate, $\bar{S}$, and the two-mode gate $\bar{C}_Z$.

A feature of number-phase codes is that phase measurement is a natural candidate for a robust logical measurement in the dual basis. In practice, phase measurements are realized by heterodyne and homodyne measurements as well as adaptive homodyne schemes~\cite{Wiseman1995,Berry:2009aa}. The existence of a robust dual-basis measurement is crucial for the quantum computing and error correction schemes presented in~\cref{sec:gates,sec:EC}.

The entangling gate ($\CROT$) in our scheme is a controlled-rotation gate based on a cross-Kerr interaction between two modes. The $\CROT$-gate plays a central role in many pieces of our quantum computing scheme including gate teleportation, modular number measurement, and error correction. An attractive feature of the $\CROT$ gate is that it can be used to interface any two rotation codes. 
In fact, the $\CROT$ gate can even be used as an entangling gate between any bosonic rotation code and a square- or hexagonal-lattice GKP code, due to the underlying rotation symmetry of the GKP code lattices, see~\cref{sec:GKP}.
This could be used in hybrid schemes employing multiple encodings and for switching between codes using teleportation.
The generality of the $\CROT$ gate is similar in spirit to that of the exponential-SWAP (eSWAP) gate~\cite{Lau:2016aa, Gao:2019aa}, which is agnostic to code type but makes no guarantees about error propagation. 
In contrast, the Kerr-based interactions that underlie $\CROT$ gates amplify errors only in a very limited way, and 
are thus attractive candidates for use in a fault-tolerant scheme. 

There are two particularly salient features of the teleportation-based error-correction scheme we introduced in~\cref{sec:EC}.
The first is that for large average excitation number in the code, the performance (as measured by average gate fidelity) is near that of the optimal recovery map allowed by the laws of quantum mechanics. For small average excitation number, the gap from optimal stems mainly from the inability of phase measurements to accurately distinguish the logical codewords in this limit. The second is that teleportation-based error correction does away with the need for explicit recovery operations to restore the codespace, and logical recoveries can be tracked entirely in software.
Since explicit recovery operations would require highly nonlinear implementations, this is a major simplification.

Finally, we outlined an approach to universal fault-tolerant quantum computing with number-phase codes in~\cref{sec:concat}. 
Given the highly non-linear nature of the number-translation symmetry $\TransN$, \cref{eq:transstabilizer}, one can wonder if fault-tolerant error correction is possible. We showed that fault-tolerance can indeed be achieved, through concatenation with a Bacon-Shor code and exploiting schemes that were originally developed to deal with highly asymmetric noise~\cite{Aliferis08,Napp2013,brooks2013fault}.
This illustrates the broader point of how a concatenated scheme can be tailored to the strengths and weaknesses of the underlying bosonic code.
We showed how bosonic-code error correction can be seamlessly integrated with the qubit-code error correction and further how fault-tolerant gadgets can be constructed using bosonic operations with minimal error spread or amplification. In this way, one can hope to maximize the advantage of using bosonically encoded qubits over bare qubits at the ground level. 

It is worth reiterating that no additional syndrome measurements are required to perform error correction at the bosonic level beyond those needed for the qubit subsystem code itself. In other words, error correction at both levels is performed in one step, such that no additional resources are used compared to if bare qubits were used at the ground level.
This is an interesting contrast to other recent work on concatenation of GKP codes with topological qubit codes~\cite{Vuillot2018, Noh:2019aa}. In these latter works, dedicated ancillae and syndrome-measurement circuits are introduced to perform bosonic-code error correction, thus incurring more overhead.

It is still largely an open problem to quantify the performance of bosonic codes for fault-tolerant error correction and computing, and thus their potential to give an improvement over bare qubits. An important part of future research on bosonic codes is therefore to quantitatively study different schemes in a fault-tolerant setting. 
Our scheme based on concatenation with Bacon-Shor codes is the first example of a fault-tolerant scheme for bosonic rotation codes, and an interesting direction for future research is to explore this path further with other concatenated codes and to optimize fault-tolerant protocols. In particular, decoders tailored to bosonic codes at the ground level have already shown promise in theoretical work on GKP codes~\cite{Fukui:2017aa,Vuillot2018,Fukui:2018aa}.

\begin{acknowledgements}
We thank Victor Albert, Rafael Alexander, Stephen Bartlett, Andrew Doherty, Giulia Ferrini, Jonathan Gross, Anirudh Krishna, Nicolas Menicucci, Stefanus Edgar Tanuarta, Barbara Terhal, Matthew Scott Winnel and Pei Zeng for valuable discussions and feedback during the preparation of this manuscript and Dumpling King for providing the environment where it was conceived.
This work was supported by the Australian Research Council (ARC) via the Centre of Excellence in Engineered Quantum Systems (EQUS) project number CE170100009, the Centre of Excellence in Quantum Computation and Communication Technology (CQC$^2$T) project number CE170100012, and a Discovery Early Career Research Award (DECRA) project number DE160100356.
\end{acknowledgements}

\appendix

\section{Dual-basis primitives} \label{sec:conjugateprimitive}
	
The dual-basis codewords $\ket{\pm_N}$ can themselves be described by a set of rotated dual primitives $e^{i \theta \hat{n}} \ket{\Theta'}$ parameterized by $N$ equally spaced angles $\theta = 2 m \pi/N $ for $m=0,\dots,N-1$:
	\begin{subequations} \label{eq:pmstates}
	\begin{align}
		\ket{+_{N,\Theta} }
		= & \frac{1}{\sqrt{ \mathcal{N}_+}} \sum_{m=0}^{N-1} e^{i  \frac{2m \pi}{N}  \hat{n} }\ket{  \Theta' } \label{eq:plusstate}, \\
		\ket{-_{N,\Theta} }
		= & \frac{1}{\sqrt{ \mathcal{N}_-}} \sum_{m=0}^{N-1} e^{i  \frac{(2m+1)\pi}{N} \hat{n} }\ket{\Theta' }.
	\end{align}
	\end{subequations}
The dual primitive $\ket{\Theta'}$ is a weighted superposition of the original primitive $\ket{\Theta}$ with itself rotated by $\pi/N$,
	\begin{align} \label{eq:primitive_conj}
		\ket{\Theta'}
		= & \frac{1}{\sqrt{\mathcal{N}_{\Theta'} } } \left( c_+ + c_- e^{i \frac{\pi}{N} \hat{n}} \right)  \ket{\Theta},
	\end{align}
where $c_\pm \coloneqq \sqrt{ \mathcal{N}_1} \pm \sqrt{ \mathcal{N}_0} $, and the normalization  is
	\begin{equation}
		\mathcal{N}_{\Theta'} = 2(\mathcal{N}_0 + \mathcal{N}_1) + 2(\mathcal{N}_1 - \mathcal{N}_0) \bra{\Theta} \cos \left( \tfrac{2\pi}{N} \hat{n} \right) \ket{\Theta}. 
	\end{equation}
Note that for a fixed primitive $\ket{\Theta}$, the dual primitive $\ket{\Theta'}$ is in general different for each $N$.

When the normalization constants satisfy $\mathcal N_0 = \mathcal N_1$,
the computational- and dual-basis primitives coincide. 
This holds for the number-phase codes in \cref{code_table} in the large average excitation-number limit.
An example are cat codes in the limit $\alpha \rightarrow \infty$. 
In this case, the code states in \cref{eq:pmstates} can be expressed simply as
	\begin{subequations} \label{eq:pmcodewordslimit}
	\begin{align}
		\ket{+_N} &\rightarrow \frac{1}{\sqrt{N}} \sum_{m=0}^{N-1} e^{i  \frac{2m \pi}{N} \hat{n}} \ket{\Theta}, \\ 
		\ket{-_N} &\rightarrow \frac{1}{\sqrt{N}} \sum_{m=0}^{N-1} e^{i  \frac{(2m+1)\pi}{N} \hat{n}} \ket{\Theta}, 
	\end{align}
	\end{subequations}
with a set of angles, $m \frac{2\pi}{N}$, that are twice the size of those in the computation-basis codewords. Moreover, we have that codes of different order are related via
\begin{equation}
\ket{ 0_{N,\Theta}} = \ket{+_{2N,\Theta} } \quad (\mathcal N_0 = \mathcal N_1).
\end{equation}

\section{\label{sec:CodeZoo}Examples of rotation codes}

The angular and Fock-space structures that characterize rotation codes leave freedom to define different codes through the coefficients $\{ \fc_{kN} \}$, \cref{eq:01logical}, that arise from a choice of primitive $\ket{\Theta}$. Several rotation codes have been presented in the literature, such as cat codes \cite{Mirrahimi14} and binomial codes \cite{Michael16} and modified versions of these codes that improve their error-correcting capability by changing the relative phases of their Fock-grid coefficients \cite{Linshu-Li:2019aa}. Their structure, properties, and error-resilience have been studied recently in Ref.~\cite{Albert17}. We summarize cat and binomial codes here and introduce two other examples. The first is a generalization of cat codes to include squeezing. The second is a new class of code based on the phase states introduced by Pegg and Barnett \cite{BarnPegg89}.  For each we give a primitive $\ket{\Theta}$ and the Fock-grid coefficients, $\{ \fc_{kN} \}$, associated with an order-$N$ code. 
For binomial codes the primitive itself depends on $N$, while the other code families use the same primitive for all $N$. Note that in the main text we are primarily interested in the subset of rotation codes that have approximate number-translation symmetry---the number-phase codes introduced in \cref{sec:CyclicStabilizerCodes}. Each of the codes introduced here can serve as a number-phase code in an appropriate limit.

\subsection{Squeezed cat codes} \label{Appendix:squeezedcat}

Cat codes consist of coherent states superposed in phase space. This can be generalized in a straightforward to include squeezing. A displaced, squeezed vacuum state is given by
    \begin{equation}
        \ket{\alpha, r, \phi} \coloneqq \hat D(\alpha) \hat S(\zeta) \ket{0},
    \end{equation}
where $\hat D(\alpha) = e^{\alpha a\dg - \alpha^* a}$ is the displacement operator and $\hat S(\zeta)= e^{ \frac{1}{2} (\zeta a\dg{}^2 - \zeta^* a^2)}$ is the squeeze operator with squeezing parameter $\zeta = r e^{-2i \phi}$, where $r$ is the squeezing amplitude and $\phi$ the squeezing angle. 
The Fock-space representation for a single displaced squeezed vacuum state, 
	$	\ket{\alpha, r, \phi} = \sum_{n=0}^\infty \tilde{c}_n \ket{n}$
is given by the coefficients
	\begin{equation} \label{eq:disqcoef}
		\tilde{c}_n
			 = \sum_{\ell=0}^\infty D_{n,\ell}(\alpha) S_{\ell,0}(r,\phi) .
	\end{equation}
where the squeezed vacuum coefficients are $S_{\ell,0}(r,\phi) = 0$ for $\ell$ odd and 
	\begin{equation} \label{eq:sqzVac}
		S_{\ell,0}(r, \phi) = \sqrt{\mbox{sech } r } \left( -\frac{1}{2} e^{2i \phi} \tanh r \right)^{\frac{\ell}{2}} \frac{\sqrt{\ell!}}{(\ell/2)!},
	\end{equation}
for $\ell$ even. In the limit of no squeezing, $S_{\ell,0}(r\rightarrow 0,\phi) = \delta_{\ell,0}$. 
The coefficients for a displaced number state
are
	\begin{equation} \label{eq:disFock}
		D_{n,\ell}(\alpha) = e^{-\frac{1}{2}|\alpha|^2} \left( \frac{\ell!}{n!} \right)^{\frac{1}{2}} \alpha^{n-\ell} \mathcal{L}_\ell^{n-\ell}(|\alpha|^2) ,
	\end{equation}
with associated Laguerre polynomials $\mathcal{L}_\ell^{k}(x)$. 

The $\ket{0_N}$ and $\ket{1_N}$ codewords are constructed using a displaced squeezed state as the primitive $\ket{\Theta_\scat} = \ket{\alpha, r, \phi=0}$ with $\alpha$ real and positive. 
From \cref{eq:comp0_rot} the computational basis states are given by\footnote{In Ref. \cite{LiZouAlbe17} and related works, the coherent-state phases are given by $\omega^k = e^{ik \pi/d}$. In our notation $d=N$ is the Fock-space code distance. }
	\begin{subequations}
    \begin{align}
        \ket{0_{N,\scat}}
            &= \frac{1}{\sqrt{\mathcal{N}_0}}\sum_{m=0}^{2N-1} \ket{\alpha e^{i m \frac{\pi}{N} }, r, m \tfrac{\pi}{N}}, \\
        \ket{1_{N,\scat}} 
            &= \frac{1}{\sqrt{\mathcal{N}_1}}\sum_{m=0}^{2N-1} (-1)^m \ket{\alpha e^{i m \frac{\pi}{N} }, r, m \tfrac{\pi}{N}} .
    \end{align} 
	\end{subequations}    
The Fock-grid coefficients can be found readily from the $\tilde{c}_n$ in \cref{eq:disqcoef}, see \cref{sec:FockStructure} for details. 

Two code subclasses can be identified by specific parameters: Cat codes emerge in the limit that squeezing vanishes, $r=0$, and squeezed vacuum codes arise when the displacement vanishes, $\alpha = 0$.\footnote{Note that $\alpha$ and $r$ cannot both vanish, in which case the state is simply vacuum and two codewords cannot be defined.}
For the smallest cat code ($\alpha \rightarrow 0, r=0$), the trivial encoding in \cref{eq:trivialencoding} is realized.
The Fock-grid coefficients for cat codes are
	\begin{align}
		\fc_{kN} = \sqrt{ \frac{ 2 }{ \mathcal{N}^*_i } }  e^{-\frac{1}{2} |\alpha|^2} \frac{\alpha^{kN}}{\sqrt{(kN)!}},
	\end{align}
where $\mathcal{N}^*_i$ indicates the Fock-space normalization factor $\mathcal{N}^*_0$ ($\mathcal{N}^*_1$) (see \cref{sec:FockStructure}) for $k$ even (odd).

Note that the cat codes we consider in this work are distinct from the two-mode cat codes in Ref. \cite{Albert18} and the single-mode cat codes in Refs. \cite{RalpGilcMilb03,LundRalpHase08}, where the codewords are manifestly nonorthogonal while still exhibiting discrete rotational symmetry. 

\subsection{Binomial codes} \label{appendix:binomial}

Binomial codes were introduced in Ref. \cite{Michael16} as a class of codes that can exactly correct loss, gain, and dephasing errors up to a certain order. The codewords are most straightforwardly defined in the conjugate basis\footnote{We use a different notation than previous references. Specifically, $N$ was denoted $S+1$ and $K$ was denoted $N+1$, where $S$ and $N$ are referred to as ``spacing" and the ``order" in Ref. \cite{Albert17}.} :
\begin{subequations}
    \begin{align}
  \ket{+_{N,\bin}} ={}& 
   \frac{1}{\sqrt{2}}  \sum_{k=0}^{K} \sqrt{\frac{1}{2^{K-1} } \binom{K}{k}} \ket{kN}, \\
  \ket{-_{N,\bin}} ={}&  \frac{1}{\sqrt{2}} \sum_{k=0}^{K} (-1)^k  \sqrt{ \frac{1}{2^{K-1}}  \binom{K}{k}} \ket{kN}.
    \end{align}
\end{subequations}
The Fock-grid coefficients $\{ \fc_{kN} \}$ can be read off directly, c.f.~\cref{eq:pmlogical}.
There are many binomial codes with $N$-fold rotation symmetry, one for each value of $K = 1,2\dots$, which sets the truncation level in Fock space. The mean excitation number is $\bar{n}_\bin = \frac{1}{2} N K$ \cite{Albert17}, and the mean modular phase can be simplified to
	\begin{align}
		\expt{e^{i \theta N}} = \frac{1}{2^K} \sum_{k=0}^{K-1} \sqrt{ \frac{K-k}{k+1} } \binom{K}{k}.
	\end{align}

Binomial codes are defined explicitly at the codeword level, leaving freedom to describe associated primitives. However, each binomial code, specified by its rotational order $N$ and truncation parameter $K$, has a different set of primitives. An example primitive can be defined as 
	\begin{equation}
    \big| \Theta_{\bin}^{N,K} \big\rangle = \frac{1}{ \sqrt{\mathcal{N}_{\bin}} }\sum_{n=0}^{KN} 
		\sqrt{\frac{1}{2^{K-1} } \binom{K}{ \lfloor n/N \rfloor }},
	\end{equation}
with $\mathcal{N}_\text{\bin}$ a normalization constant,
and $\lfloor x \rfloor$ is the floor function that gives the largest integer less than or equal to the real number $x$.
The smallest binomial codes ($K=1$) yield the \ON{} code, $\ket{+_N} ={}\tfrac{1}{\sqrt{2}} \left( \ket{0} + \ket{N} \right),$
which is the trivial encoding, \cref{eq:trivialencoding}, for $N=1$.

\subsection{Pegg-Barnett codes} \label{appendix:PB}

A Pegg-Barnett phase state is given by~\cite{BarnPegg89}
    \begin{equation}
        \ket{\phi,s} \coloneqq \frac{1}{\sqrt{s}} \sum_{n=0}^{s-1} e^{i n \phi} \ket{n},
    \end{equation}
where the parameter $s$ sets the truncation level. The set of states $\{ \ket{\phi_m = 2\pi m/s,s} \}$ with $m=0,\dots,s-1$ forms an orthonormal basis for the $s$-dimensional truncated Fock space. 

We use a specific phase state as a primitive, $| \Theta_\PB \rangle = \ket{\phi = 0,s}$, to define \emph{Pegg-Barnett codes}. To ensure adequate Fock space for the codewords, the truncation $s-1$ is required to be at least $N$. 
Given that $e^{i\theta \hat{n}} \ket{\phi,s} = \ket{\phi + \theta,s}$, the codewords can be expressed simply as
	\begin{subequations}
    \begin{align}
        \ket{0_{N,\PB}}
            &= \frac{1}{\sqrt{\mathcal{N}_0}} \sum_{m=0}^{2N-1} \ket{\phi = m \tfrac{\pi}{N},s},\\
        \ket{1_{N,\PB}} 
            &= \frac{1}{\sqrt{\mathcal{N}_1}} \sum_{m=0}^{2N-1} (-1)^m \ket{\phi = m \tfrac{\pi}{N},s}.
    \end{align}    
	\end{subequations} 
Note that if the truncation of the Pegg-Barnett phase state used as a primitive is commensurate with the order of the code, $s = p\times 2N$ for $p = 1,2,3 \dots$, then the rotated primitives are automatically orthogonal, $\ip{\phi = m \frac{\pi}{N},s}{\phi = m' \frac{\pi}{N},s} = \delta_{m,m'}$. In this case the conjugate-basis codewords and the dual-basis codewords are simple superpositions of the primitive, c.f., \cref{eq:pmcodewordslimit}. 
For $s=p\times 2N$ we also have that the Pegg-Barnett codes can be recognized as the shift-resitant qudit codes from Ref.~\cite{Gottesman01}, with $d=s$, $n=2$, $r_1=N$ and $r_2=p$, using the notation from Ref.~\cite{Gottesman01}.

The Fock-grid coefficients in general are $\fc_{kN} = \sqrt{2/\lceil s/N \rceil }$, where $\lceil x \rceil$ is the ceiling function that gives the least integer greater than or equal to the real number $x$.
The mean excitation number is $\bar{n}_\PB = \frac{N}{2} \big( \lceil s/N \rceil - 1\big)$, and the mean modular phase is
	\begin{equation}
		\expt{e^{iN\theta}} = 1 - \frac{1}{ \lceil s/N \rceil }. 
	\end{equation}
The Pegg-Barnett code $[N; s=N+1]$ is the \ON{} code 
and the smallest Pegg-Barnett code $[N=1; s=2]$ is the trivial encoding, \cref{eq:trivialencoding}.

\section{\label{sec:GKP}Rotation symmetry for Gottesman-Kitaev-Preskill codes}

Logical Pauli gates for GKP codes are realized by discrete translations in phase space. 
As a consequence of translation symmetry, all GKP codes also exhibit $N=2$ discrete rotation symmetry, as we show below.
However, GKP codes are not rotation codes as we define them in~\cref{sec:gcb}, because the operator $\hat Z_2$ does not act as a Pauli operator.
Depending on the underlying lattice defining the GKP codewords, $\ZN$ may nevertheless act as a logical Clifford operator for some values of $N$.
Two notable GKP codes where this is the case are the square- and hexagonal-lattice GKP codes, which means that these codes exhibit some properties of rotation codes.
We briefly elucidate this point. 

The codespace of a single-mode ideal GKP qubit is defined to be the $+1$ eigenspace of the stabilizers
\begin{equation} \label{eq:GKPstabs}
    \hat S_\gkp^X = \hat D(2\alpha),\qquad \hat S_\gkp^Z = \hat D(2\beta),
\end{equation}
where $\hat D(\zeta) = \e^{\zeta \hat a\dg - \zeta^* \hat a}$ is the displacement operator,\footnote{Note that $\hat D(\zeta)$ generates shifts in $qp$-phase space of magnitude $\sqrt{2}|\zeta|$. For example, $\hat D \big(\sqrt{\pi/2} \big) = e^{-i \sqrt{\pi} \hat{p}}$, which acts as a logical $\X$ for square-lattice GKP.} and $\alpha$ and $\beta$ are  complex numbers satisfying
\begin{equation}\label{eq:GKP_parallelogram}
    \alpha^*\beta - \alpha \beta^* = i\pi.
\end{equation}
The stabilizer group consists of all powers $(\hat S^X_\gkp )^{n_1} (\hat S^Z_\gkp)^{n_2}$ for $n_1,n_2\in\mathbb{Z}$.
The shift operators that act as logical Paulis for the GKP qubit are
\begin{subequations}
\begin{align} \label{eq:GKPPaulilike}
    \hat X_\gkp ={}& \hat D(\alpha),\\
    \hat Z_\gkp ={}& \hat D(\beta),\\
    \hat Y_\gkp ={}& i\hat Z_\gkp \hat X_\gkp = \hat D(\alpha + \beta),
\end{align}
\end{subequations}
where the form of $\hat Y_\gkp$ follows from the general identity $\hat D(\alpha) \hat D(\beta) = \e^{\frac{1}{2} (\alpha\beta^*-\alpha^*\beta )}\hat D(\alpha + \beta)$.
The constraint~\cref{eq:GKP_parallelogram} ensures that the stabilizers in \cref{eq:GKPstabs} commute, and that $\hat X_\gkp$ and $\hat Z_\gkp$ commute with the stabilizers and anti-commute with each other.

From the identity $\e^{-i\theta \hat n}\hat D(\zeta) \e^{i\theta \hat n} = \hat D(\e^{-i\theta}\zeta)$ we see that the discrete rotation operator $\CN$ for $N=2$ has the following action on the Pauli shift operators:
\begin{subequations} \label{eq:GKPcodesymmetry}
\begin{align} 
\hat R_2\dg \hat X_\gkp \hat R_2 ={}& \hat D(-\alpha) = \hat X_\gkp\dg,\\
\hat R_2\dg \hat Z_\gkp \hat R_2 ={}& \hat D(-\beta) = \hat Z_\gkp\dg.
\end{align}
\end{subequations}
Since $\hat X_\gkp\dg = \hat X_\gkp(\hat S_\gkp^X)^{-1}$ and $\hat Z_\gkp\dg = \hat Z_\gkp(\hat S_\gkp^Z)^{-1}$, they are equivalent up to stabilizers and have the same action on the codespace.
Thus, $\hat R_2$ acts as the identity operator on the codespace of an ideal GKP code. In other words, all GKP codes have 2-fold discrete rotation symmetry.

The computational-basis codewords can be written as a superposition of coherent states:
\begin{equation}
    \begin{aligned}
    &\ket{j_\gkp} = \sum_{n_1,n_2\in\mathbb{Z}} \hat D[(2n_1+j)\alpha]\hat D(n_2\beta)\ket 0 \\
    ={}& \sum_{n_1,n_2\in\mathbb{Z}} \e^{-i\pi(n_1 n_2 + j n_2/2)} \ket{(2n_1 + j)\alpha + n_2\beta},
    \end{aligned}
\end{equation}
with $j=0,1$.
It is natural to define state lattices in the complex plane for the computational-basis codewords,
\begin{equation}
\mathcal L_j = \left\{(2n_1 + j)\alpha + n_2\beta\mid n_1, n_2 \in \mathbb{Z} \right\},
\end{equation}
as well as the codespace lattice, $\mathcal L = \mathcal L_0 \cup \mathcal L_1$, consisting of all points $n_1\alpha + n_2\beta$. As described in Ref. \cite{Albert17}, the codeword lattices are basis-dependent, while the codespace lattice $\mathcal{L}$ is not. Note that~\cref{eq:GKP_parallelogram} means that the unit cell, or fundamental parallelogram, of $\mathcal L$ has area $\pi/2$.\footnote{Using standard quadrature operators $\hat{q}$ and $\hat{p}$ satisfying $[\hat{q}, \hat{p}] = i$, the lattice distances are expanded to $\sqrt{2} \alpha$ and $\sqrt{2} \beta$, and the area of the parellelogram in $qp$-phase space is $\pi$. }

\begin{figure}
\centering
\includegraphics{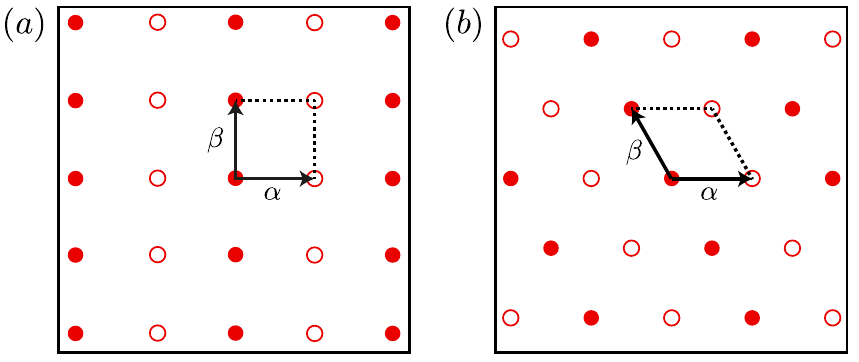} 
\caption{\label{fig:GKPlattice}Code lattices $\mathcal{L}_0$ (filled circles) and $\mathcal{L}_1$ (open circles) in the complex plane for (a) square-lattice GKP and (b) hexagonal-lattice GKP. The area of the lattice-parallelogram is $\pi/2$. The codespace lattice $\mathcal{L}$ is also apparent in the Wigner function $W_{\gkp}(\alpha)$ of the completely mixed state $\frac{1}{2} \hat{ \Pi}_\gkp$. Using quadrature axes, $Q$ and $I$, as defined in \cref{fig:summary}, $W_{\gkp}(\alpha)$ is identical to the above plots with positive $\delta$-functions at each point in $\mathcal{L}$. For the Wigner functions, arrows indicate direction of displacement for $\hat{X}_\gkp$ and $\hat{Z}_\gkp$. 
}
\end{figure}

We consider two canonical examples of ideal GKP codes defined on square (\gkps) and hexagonal (\gkph) lattices. The codes can be defined in terms of their respective lattice-basis translations $\alpha$ and $\beta$:
\begin{subequations}
\begin{align}
\text{\gkps:} &\quad \alpha = \sqrt{\frac{\pi}{2}}, \quad \beta = i\sqrt{\frac{\pi}{2}},\\
\text{\gkph:} &\quad \alpha = \sqrt{\frac{\pi}{\sqrt 3}}, \quad \beta = \e^{i \frac{ 2\pi}{3}}\sqrt{\frac{\pi}{\sqrt 3}}.
\end{align}
\end{subequations}
 Codespace lattices for $\gkps$ and $\gkph$ are shown in \cref{fig:GKPlattice}.

Depending on the lattice, specific discrete rotations can act as non-trivial logical Clifford gates. 
In particular, for square-lattice GKP, $N=4$ rotations transform the Paulis:
\begin{subequations}
\begin{align}
    \hat R_4\dg \hat X_\gkps \hat R_4
             &= \hat Z_\gkps\dg,\\
    \hat R_4\dg \hat Z_\gkps \hat R_4  
        &= \hat X_\gkps.
\end{align}
\end{subequations}
Thus, $\hat{R}_4 = \hat Z_2 = e^{i \frac{\pi}{2} \hat{n}}$ (the Fourier transform operator) acts as a logical Hadamard $\Had$ for \gkps.
It follows that the eigenstates of $\Had$ for this code can be written in the form~\cref{eq:01logical}, a fact that was recognized in Ref.~\cite{Gottesman01}. It was also suggested there that measurement of number mod 4 to can be used to prepare Hadamard eigenstates, which are equivalent to the magic state $\ket{T_\gkps}$ up to Cliffords.
Since mapping logical $\Had$ to $\Z$ is a non-Clifford operation, this is not a trivial change of basis for a GKP code, and we therefore do not classify the square-lattice GKP code as a rotation code.

For hexagonal-lattice GKP, $N=6$ rotations perform logical $\pi/3$-rotations around the $X Y Z$-axis: 
\begin{subequations}
\begin{align}
    \hat R_6\dg \hat X_\gkph \hat R_6 &= \hat Z_\gkph\dg,\\
    \hat R_6\dg \hat Z_\gkph \hat R_6 &= \hat Y_\gkph,
\end{align}
\end{subequations}
giving a cyclic permutation of the logical Pauli operators. Thus $\hat{R}_6 = \hat Z_3$ acts as the logical Clifford gate $\Had \Sgate\dg$.\footnote{Interestingly, $\Had \Sgate\dg = (\Sgate \Had)\dg$, and $\Sgate \Had$ has $T$-type magic states as its eigenstates, where we use the original terminology from~\cite{Bravyi2005}.}
An $N=3$ rotation, given by $\hat{R}_3 = \hat{R}_6^2$, acts as the logical gate $(\Had \Sgate\dg)^2 =e^{ -i \frac{\pi}{4}}\Sgate \Had$, which performs a cyclic permutation of the Paulis in the opposite order.

The fact that the square- and hexagonal-lattice GKP codes have non-trivial Cliffords implementable as discrete rotations also implies that the $\CROT$ gate can be used between a bosonic rotation code and these two GKP codes. In particular $\CROT_{N4}$ ($\CROT_{N6}$) implements a controlled-$\Had$ (controlled-$\Had \Sgate\dg$) between an order-$N$ rotation code and a square (hexagonal) GKP code, with the rotation code as control and the GKP code as target.

In contrast to square- and hexagonal-, the rectangular-lattice GKP code does not benefit from additional Clifford gates performed via discrete rotations. Lastly, we note that the crystallographic restriction theorem implies that no lattice can have higher than 6-fold rotation symmetry, and $\hat R_6$ is thus the highest degree rotation operator that can preserve the codespace of a GKP code.

\section{\label{sec:prettygood}Pretty Good Measurements}

As we alluded to in~\cref{sec:phaseest}, the canonical phase measurement might not be the optimal measurement to distinguish the codewords $\ket{\pm_{N}}$. Even in the absence of noise, two orthogonal codewords cannot in general be perfectly distinguished by a canonical phase measurement due to embedded phase uncertainty $\dtheta$. Moreover, in a realistic scenario where the codewords may be damaged, defining a phase estimation problem with respect to the ideal codewords is sub-optimal. We do not attempt to find an optimal scheme in general, as this must likely be done on a case-by-case basis for different codes and noise models. Instead, we introduce a measurement that perform well for all the codes and noise models we study in this paper. This is the Pretty Good Measurement introduced in Refs.~\cite{Belavkin1975aa, Belavkin1975bb, Hausladen94}. 
The Pretty Good Measurement perfectly distinguishes orthogonal states and is designed to also distinguish fairly well states with small non-zero overlaps. The measurement is defined through the POVM elements
\begin{equation}\label{eq:PGM}
    \hat M_i^\text{Pretty Good} = \hat \sigma^{-1/2} \mathcal N\left( \ket{i}\bra{i} \right) \hat \sigma^{-1/2},
\end{equation}
where $\ket i$ runs over the states to be distinguished, and $\mathcal N$ represents an arbitrary noise channel.
The operator $\hat \sigma = \mathcal N( \hat P )$ with $\hat P = \sum_i \ket{i}\bra{i}$ represents the projector onto the subspace in question sent through the noise channel.
 The POVM elements satisfy $\sum_{i} \hat M_{i} = \hat P_\sigma$, where $\hat P_\sigma$ is a projector onto the support of $\hat \sigma$. To have a complete measurement, one can add a POVM element $\hat{I} - \hat P_\sigma$ projecting onto the complement of $\hat P_\sigma$. Note that in contrast to the canonical phase measurement, the Pretty Good Measurement explicitly exploits knowledge about the noise $\mathcal N$.

\section{\label{sec:hybrid}Hybrid Steane/Knill error correction}

\begin{figure}
\centering
\includegraphics{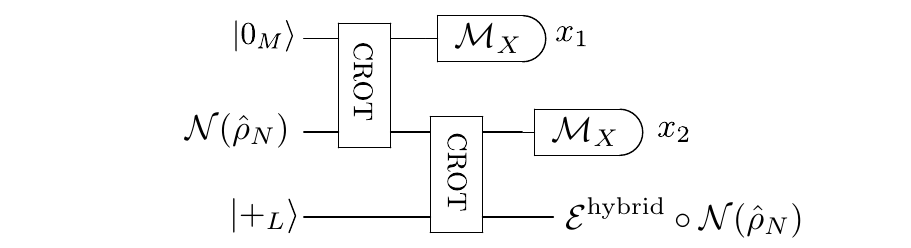}
\caption{\label{fig:hybrid}A hybrid Steane-Knill error correction scheme for rotation codes.}
\end{figure}

Here we introduce an error-correction circuit, depicted in~\cref{fig:hybrid}, as an alternative to the circuit in~\cref{fig:ecschemes}. We call this error-correction method ``hybrid'' (hybrid-EC), because the first $\CROT$ gate and the measurement of the top ancilla rail can be recognized as identical to one of the two steps of Steane-EC, while the second $\CROT$ and the measurement on the data rail is a one-bit teleportation, identical to one of the steps in Knill-EC. Essentially, this circuit uses Steane-EC for one of the syndromes (number-shift errors) and Knill-EC for the other (dephasing errors).

This error-correction scheme is based on the circuit identity
\begin{equation}\label{eq:hybrididentity}
  \begin{aligned}
  \Qcircuit @C=1em @R=1.5em {
    \lstick{M} & \qw           & \multigate{1}{\rotatebox{270}{\CROT}} & \qw                                       & \qw \\
    \lstick{N} & \gate{\hat E} & \ghost{\rotatebox{270}{\CROT}}        & \multigate{1}{\rotatebox{270}{\,\CROT\,}} & \qw \\
    \lstick{L} & \qw           & \qw                                   & \ghost{\rotatebox{270}{\,\CROT\,}}        & \qw
 }
 \qquad
 \Qcircuit @C=1em @R=1.2em {
               & \multigate{1}{\rotatebox{270}{\CROT}} & \qw                                   & \gate{\hat R} & \qw \\
    \lstick{=} & \ghost{\rotatebox{270}{\CROT}}        & \multigate{1}{\rotatebox{270}{\CROT}} & \gate{\hat E}  & \qw \\
               & \qw                                   & \ghost{\rotatebox{270}{\CROT}}        & \gate{\hat R'} & \qw
 }
  \end{aligned}
\end{equation}
where $\hat E = \hat E_k(\theta)$ is again an arbitrary initial error and the two rotation errors, $\hat R = \hat E_0\left(\frac{\pi k}{N M}\right)$ and $\hat R' = \hat E_0\left(\frac{\pi k}{N L}\right)$, are both proportional to $k$. A phase measurement on the top rail thus estimates $k$, and a recovery operation can be applied to the bottom rail to undo the leftover error $\hat R'$.

The first step of the circuit in~\cref{fig:ecschemes}$(b)$, involving only the top ancilla,
\begin{equation}\label{eq:hybrid:numbersyndrome}
  \begin{aligned}
  \Qcircuit @C=1em @R=1.3em {
    \lstick{\ket{0_M}} & \multigate{1}{\rotatebox{270}{\CROT}} & \measureD{\MX} \\
                       & \ghost{1}{\rotatebox{270}{\CROT}}     & \qw & \lstick{,}
 }
  \end{aligned}
\end{equation}
performs a nondestructive measurement of the excitation number mod $N$ on the data rail, as explained in~\cref{sec:numbermodN}.
The second step of the circuit,
\begin{equation}\label{eq:hybrid:phasesyndrome}
  \begin{aligned}
  \Qcircuit @C=1em @R=1.3em {
                           & \multigate{1}{\rotatebox{270}{\CROT}} & \measureD{\MX} \\
     \lstick{\ket{+_{L}}} & \ghost{1}{\rotatebox{270}{\CROT}}     & \qw & \lstick{,}
 }
  \end{aligned}
\end{equation}
teleports the data to a fresh new ancilla, thus restoring the codespace up to a known rotation error as per~\cref{eq:hybrididentity}.

More explicitly, we consider an encoded state $\hat \rho_N$ in an order $N$ rotation code sent through a noise noise channel $\mathcal N\bullet = \sum_n \hat A_n \bullet \hat A_n\dg$, followed by the error-correction circuit in~\cref{fig:hybrid}. In this case, it is convenient to expand the noise Kraus operators as $\hat A_n = \sum_k \int\dd\theta c_{nk}(\theta) \hat E_k(\theta) = \sum_k \hat E_{nk}$ where we define $\hat E_{nk} = \int \dd\theta c_{nk}(\theta) \hat E_k(\theta)$. The state after the error-correction circuit can be written
\begin{equation}\label{eq:hybridchannel}
    \begin{aligned}
    \mathcal E^\text{hybrid} \circ \mathcal N(\hat \rho_N) &=
    \half \sum_{a,b=\pm} \sum_{\vec x} \sum_{n,k,k'} \\
    &\times \tr \left[\hat M_{x_1} \hat \sigma_{ab}^{nkk'} \right]
    \tr\left[\hat M_{x_2} \hat \tau^{kk'} \right] \\
    &\times \hat E_0\left(\smallfrac{\pi k}{NL}\right) \Clifford_a\, \hat \rho_{L} \, \Clifford_b\dg \hat E_0\dg\big(\smallfrac{\pi k'}{NL}\big),
    \end{aligned}
\end{equation}
where
\begin{align}
    \hat \sigma_{ab}^{nkk'} \coloneqq {}& \hat E_{nk} \ket{a_N}\bra{b_N} \hat E_{nk'}\dg,\\
    \hat \tau^{kk'} \coloneqq{}& \hat E_0\left(\smallfrac{\pi k}{NM}\right) \ket{0_M}\bra{0_M} \hat E_0\dg \big(\smallfrac{\pi k'}{NM}\big),
\end{align}
represent damaged codewords for the data and ancilla, respectively,
and $\Clifford_+ = \Hd, \Clifford_- = \X \Hd$ are logical Cliffords. We again use a notation where $\hat \rho_{L}$ on the right-hand side represents the same logical state as $\hat \rho_{N}$ on the left-hand side, but where the latter is encoded in the order-$L$ number-phase code of the bottom ancilla in~\cref{fig:hybrid}.

The most likely rotation error on the output, $\hat E_0 \big(\frac{\pi k^*}{NL} \big)$, as well as the most likely logical Clifford error $\Clifford_{a^*}$ can be deduced using, \emph{e.g.}, a maximum likelihood decoder. The rotation error is straightforward to correct, or alternatively can be tracked by updating later phase measurements. The error $\Clifford_{a^*}$ is not a Pauli error; however, by performing the error-correction circuit twice, the most likely error will be a Pauli from the set $\Pauli_i \in \{\I, \Z, \X, \X\Z\}$ which can be tracked in a Pauli frame.

We have verified in numerical simulations that the hybrid-EC scheme performs identically to the Knill-EC scheme presented in~\cref{sec:Knill}, \emph{i.e.}, the results in~\cref{sec:ECnumerics} are identical for both schemes. In fact, there is arguably a slight advantage of the hybrid-EC scheme in the case of noiseless ancillae, because we can use an unencoded state $\ket\alpha$ for the upper ancilla rail in~\cref{fig:hybrid}. This state give slightly better phase resolution compared to an $M=1$ cat code, as used in~\cref{sec:ECnumerics}, at the same value of $\alpha$.

\subsubsection{Relation to recent experiments}
The hybrid-EC scheme generalizes the experimental protocol used in the two experiments in Refs.~\cite{Ofek16,Hu:2019aa}, where error correction for $N=2$ cat and binomial codes was implemented, respectively. The scheme employed the error-syndrome circuit~\cref{eq:hybrid:numbersyndrome} with the ancilla replaced by
an unencoded qubit in the state $\ket + = (\ket 0 + \ket 1)/\sqrt 2$ and the $\CROT$ gate in their case was $\CROT = \CROT_{N1/2} = \e^{i \frac{2\pi}{N} \hat n \otimes \hat n}.$
An error $E_k(\theta)$ on the data qubit was thus imprinted as a phase on the ancilla, 
\emph{i.e.}, $\ket + \to \ket 0 + \e^{i\frac{2\pi k}{N}}\ket 1$, and the syndrome measurement attempted to estimate this phase.
Note that with $N=2$, an initial error \emph{prior} to the gate with $|k| \mod 2 =1$ on the data qubit flips the ancilla to $\ket -$, while an error with $|k| \mod 2 = 0$ leaves the ancilla in the state $\ket +$.
This syndrome does not detect dephasing errors, nor does it restore the state back to the Fock grid $\ket{kN}$.
For cat codes, loss events can be tracked in software so that recoveries for this particular error are not necessary (this does not include the no-jump part of the bosonic loss channel, however)~\cite{Ofek16}. For binomial codes, optimal control was employed in Ref.~\cite{Hu:2019aa} to find a unitary that approximately restores the codespace.
In the hybrid-EC scheme we propose, both detection of dephasing errors and restoration of the codespace is performed by the \emph{second} step of the error correction circuit, \cref{eq:hybrid:phasesyndrome}.
The difficulty of restoring the codespace has thus been replaced by the ability to prepare copies of encoded $\ket{+_{N}}$ states.

\section{\label{app:kraus}Loss and dephasing channel}

We seek a convenient representation of the channel that describes simultaneous loss and dephasing:
\begin{equation}\label{eq:app:lossdephasing:channel}
\mathcal N = \e^{t(\mathcal L_\kappa + \mathcal L_{\kappa_\phi})},
\end{equation}
where $\mathcal L_\kappa = \kappa \mathcal D[\hat a]$ and $\mathcal L_{\kappa_\phi} = \kappa_\phi \mathcal D[\hat n]$, c.f.~\cref{eq:lossdephasingme}. First, we note that the pure dephasing channel can be written
\begin{equation}\label{eq:app:lossdephasing:dephasingchannel}
    \e^{t \mathcal L_{\kappa_\phi}}\hat \rho = \int_{-\infty}^{\infty} \dd\theta \, p_{\kappa_\phi t}(\theta) \e^{i\theta \hat n}\hat \rho \e^{-i\theta \hat n},
\end{equation}
where $p_{\kappa_\phi t} = (2\pi\kappa_\phi t)^{-1/2}\e^{-\theta^2/(2\kappa_\phi t)}$ is a Gaussian with zero mean and variance $\kappa_\phi t$. This can be shown by first considering the small $\kappa_\phi t$ limit:
\begin{equation}
    \begin{aligned}
    \e^{\mathcal L_{\kappa_\phi} t}&\hat \rho
    \simeq{} \int_{-\infty}^{\infty} \dd\theta \, p_{\kappa_\phi t}(\theta) (1+i\theta \hat n)\hat \rho (1-i\theta \hat n) \\
    \simeq{}& \int_{-\infty}^\infty \dd\theta \, p(\theta) \left[ \e^{i\theta \hat n} \hat \rho + \hat \rho \e^{-i\theta \hat n} + \theta^2 \hat n\hat \rho \hat n - \hat \rho\right] \\
    ={}& \e^{-\half \kappa_\phi t \hat n^2}\hat \rho + \hat \rho \e^{-\half \kappa_\phi t \hat n^2} + \kappa_\phi t \hat n \hat \rho \hat n - \hat \rho \\
    \simeq{}& \hat \rho - \half \kappa_\phi t \hat n^2\hat \rho  - \half \hat \rho \kappa_\phi t \hat n^2 + \kappa_\phi t \hat n \rho \hat n,
    \end{aligned}
\end{equation}
and thus
\begin{equation}
  \dot{\hat\rho} = \lim_{t\to 0} \frac{\e^{t \mathcal L_{\kappa_\phi}}\hat \rho -\hat \rho}{t} = \mathcal L_{\kappa_\phi}\hat\rho,
\end{equation}
as required.

Returning to~\cref{eq:app:lossdephasing:channel} we can write the solution $\hat \rho(t) = \mathcal N(\hat \rho_0)$ in terms of a generalized Dyson expansion~\cite{Carmichael1989,Wiseman1993}:
\begin{equation}\label{eq:app:lossdephasing:dyson}
  \begin{aligned}
  \hat \rho(t) ={}& \sum_{k=0}^\infty \int_0^t \dd t_k \int_{0}^{t_k} \dd t_{k-1} \dots \int_{0}^{t_2} \dd t_1 \\
  &\times \e^{(t-t_{k})\mathcal S} \mathcal J \dots \mathcal J \e^{(t_2-t_{1})\mathcal S}\mathcal J \e^{t_1\mathcal S} \hat \rho_0,
  \end{aligned}
\end{equation}
where $\mathcal J\hat\rho = \kappa \hat a\hat \rho \hat a\dg$, and $\mathcal S = \mathcal L_{\kappa_\phi} + \mathcal L_{\kappa} - \mathcal J$. Next, note that the no-jump part of the evolution factorizes as
\begin{equation}
    \e^{t \mathcal S} = \e^{t (\mathcal L_{\kappa_\phi} + \mathcal L_{\kappa} - \mathcal J)} = \e^{t \mathcal L_{\kappa_\phi}}\e^{t(\mathcal L_{\kappa} - \mathcal J)},
\end{equation}
where $\e^{t \mathcal L_{\kappa_\phi}}$ is given by~\cref{eq:app:lossdephasing:dephasingchannel} and
\begin{equation}
    \e^{t(\mathcal L_{\kappa} - \mathcal J)} \hat\rho = \e^{-\kappa \hat n t/2}\hat\rho\e^{-\kappa \hat n t/2}.
\end{equation}
Using this, it is straightforward to pull the jump superoperators $\mathcal J$ through all the no-jump superoperators in~\cref{eq:app:lossdephasing:dyson}:
\begin{align}
\mathcal J \e^{t \mathcal L_{\kappa_\phi}} ={}& \e^{t \mathcal L_{\kappa_\phi}} \mathcal J \\
\mathcal J \e^{t(\mathcal L_{\kappa} - \mathcal C)} ={}& \e^{-\kappa t} \e^{t(\mathcal L_{\kappa} - \mathcal C)} \mathcal J.
\end{align}
After pushing all the $\mathcal J$'s to the right, rewriting the $k$ time-ordered integrals as $k$ integrals over $[0,t]$ and performing the integrals, one readily finds
\begin{equation}
    \hat\rho(t) = \e^{t\mathcal L_{\kappa_\phi}} \sum_{k=0}^\infty \hat A_k \hat \rho_0 \hat A_k\dg,
\end{equation}
where $\hat A_k$ is given by
\begin{equation}\label{eq:kraus:loss}
    \hat A_k = \frac{\left(1-\e^{-\kappa t}\right)^{k/2}}{\sqrt{k!}} \e^{-\kappa \hat n t/2} \hat a^k.
\end{equation}
One can, of course, also find a Kraus-operator representation of $\e^{t\mathcal L_{\kappa_\phi}}$ using exactly the same method,
\begin{equation}
    \e^{t\mathcal L_{\kappa_\phi}}\hat \rho = \sum_{\ell=0}^\infty \hat B_\ell \hat \rho \hat B_\ell\dg,
\end{equation}
with
\begin{equation}
    \hat B_\ell = \frac{(\kappa_\phi t)^{\ell/2}}{\sqrt{\ell!}} \e^{-\half \kappa_\phi \hat n^2 t} \hat n^\ell.
\end{equation}
We, however, do not need such a Kraus-operator representation of the dephasing channel for our purposes in~\cref{sec:EC}, since the dephasing channel commutes with all the $\CROT$ gates. We found it more convenient to solve the master equation
\begin{equation}
  \dot{\mathcal N}_{\kappa_\phi} = \mathcal L_{\kappa_\phi}\mathcal N_{\kappa_\phi}
\end{equation}
numerically, with $\mathcal N_{\kappa_\phi}(0) = \mathcal I$ the identity superoperator as initial condition,
 to find the pure dephasing channel $\mathcal N_{\kappa_\phi}(t) = \e^{t\mathcal L_{\kappa_\phi}}$.


\begin{thebibliography}{101}%
\makeatletter
\providecommand \@ifxundefined [1]{%
 \@ifx{#1\undefined}
}%
\providecommand \@ifnum [1]{%
 \ifnum #1\expandafter \@firstoftwo
 \else \expandafter \@secondoftwo
 \fi
}%
\providecommand \@ifx [1]{%
 \ifx #1\expandafter \@firstoftwo
 \else \expandafter \@secondoftwo
 \fi
}%
\providecommand \natexlab [1]{#1}%
\providecommand \enquote  [1]{``#1''}%
\providecommand \bibnamefont  [1]{#1}%
\providecommand \bibfnamefont [1]{#1}%
\providecommand \citenamefont [1]{#1}%
\providecommand \href@noop [0]{\@secondoftwo}%
\providecommand \href [0]{\begingroup \@sanitize@url \@href}%
\providecommand \@href[1]{\@@startlink{#1}\@@href}%
\providecommand \@@href[1]{\endgroup#1\@@endlink}%
\providecommand \@sanitize@url [0]{\catcode `\\12\catcode `\$12\catcode
  `\&12\catcode `\#12\catcode `\^12\catcode `\_12\catcode `\%12\relax}%
\providecommand \@@startlink[1]{}%
\providecommand \@@endlink[0]{}%
\providecommand \url  [0]{\begingroup\@sanitize@url \@url }%
\providecommand \@url [1]{\endgroup\@href {#1}{\urlprefix }}%
\providecommand \urlprefix  [0]{URL }%
\providecommand \Eprint [0]{\href }%
\providecommand \doibase [0]{http://dx.doi.org/}%
\providecommand \selectlanguage [0]{\@gobble}%
\providecommand \bibinfo  [0]{\@secondoftwo}%
\providecommand \bibfield  [0]{\@secondoftwo}%
\providecommand \translation [1]{[#1]}%
\providecommand \BibitemOpen [0]{}%
\providecommand \bibitemStop [0]{}%
\providecommand \bibitemNoStop [0]{.\EOS\space}%
\providecommand \EOS [0]{\spacefactor3000\relax}%
\providecommand \BibitemShut  [1]{\csname bibitem#1\endcsname}%
\let\auto@bib@innerbib\@empty
\bibitem [{\citenamefont {Chuang}\ \emph {et~al.}(1997)\citenamefont {Chuang},
  \citenamefont {Leung},\ and\ \citenamefont {Yamamoto}}]{ChuaLeunYama97}%
  \BibitemOpen
  \bibfield  {author} {\bibinfo {author} {\bibfnamefont {I.~L.}\ \bibnamefont
  {Chuang}}, \bibinfo {author} {\bibfnamefont {D.~W.}\ \bibnamefont {Leung}}, \
  and\ \bibinfo {author} {\bibfnamefont {Y.}~\bibnamefont {Yamamoto}},\
  }\bibfield  {title} {\enquote {\bibinfo {title} {Bosonic quantum codes for
  amplitude damping},}\ }\href {https://dx.doi.org/10.1103/PhysRevA.56.1114}
  {\bibfield  {journal} {\bibinfo  {journal} {Phys. Rev. A}\ }\textbf {\bibinfo
  {volume} {56}},\ \bibinfo {pages} {1114} (\bibinfo {year}
  {1997})}\BibitemShut {NoStop}%
\bibitem [{\citenamefont {Cochrane}\ \emph {et~al.}(1999)\citenamefont
  {Cochrane}, \citenamefont {Milburn},\ and\ \citenamefont
  {Munro}}]{CochMilbMunr99}%
  \BibitemOpen
  \bibfield  {author} {\bibinfo {author} {\bibfnamefont {P.~T.}\ \bibnamefont
  {Cochrane}}, \bibinfo {author} {\bibfnamefont {G.~J.}\ \bibnamefont
  {Milburn}}, \ and\ \bibinfo {author} {\bibfnamefont {W.~J.}\ \bibnamefont
  {Munro}},\ }\bibfield  {title} {\enquote {\bibinfo {title} {Macroscopically
  distinct quantum-superposition states as a bosonic code for amplitude
  damping},}\ }\href {\doibase 10.1103/PhysRevA.59.2631} {\bibfield  {journal}
  {\bibinfo  {journal} {Phys. Rev. A}\ }\textbf {\bibinfo {volume} {59}},\
  \bibinfo {pages} {2631} (\bibinfo {year} {1999})}\BibitemShut {NoStop}%
\bibitem [{\citenamefont {Gottesman}\ \emph {et~al.}(2001)\citenamefont
  {Gottesman}, \citenamefont {Kitaev},\ and\ \citenamefont
  {Preskill}}]{Gottesman01}%
  \BibitemOpen
  \bibfield  {author} {\bibinfo {author} {\bibfnamefont {D.}~\bibnamefont
  {Gottesman}}, \bibinfo {author} {\bibfnamefont {A.}~\bibnamefont {Kitaev}}, \
  and\ \bibinfo {author} {\bibfnamefont {J.}~\bibnamefont {Preskill}},\
  }\bibfield  {title} {\enquote {\bibinfo {title} {Encoding a qubit in an
  oscillator},}\ }\href {\doibase 10.1103/PhysRevA.64.012310} {\bibfield
  {journal} {\bibinfo  {journal} {Phys. Rev. A}\ }\textbf {\bibinfo {volume}
  {64}},\ \bibinfo {pages} {012310} (\bibinfo {year} {2001})}\BibitemShut
  {NoStop}%
\bibitem [{\citenamefont {Lund}\ \emph {et~al.}(2008)\citenamefont {Lund},
  \citenamefont {Ralph},\ and\ \citenamefont {Haselgrove}}]{LundRalpHase08}%
  \BibitemOpen
  \bibfield  {author} {\bibinfo {author} {\bibfnamefont {A.~P.}\ \bibnamefont
  {Lund}}, \bibinfo {author} {\bibfnamefont {T.~C.}\ \bibnamefont {Ralph}}, \
  and\ \bibinfo {author} {\bibfnamefont {H.~L.}\ \bibnamefont {Haselgrove}},\
  }\bibfield  {title} {\enquote {\bibinfo {title} {Fault-tolerant linear
  optical quantum computing with small-amplitude coherent states},}\ }\href
  {\doibase 10.1103/PhysRevLett.100.030503} {\bibfield  {journal} {\bibinfo
  {journal} {Phys. Rev. Lett.}\ }\textbf {\bibinfo {volume} {100}},\ \bibinfo
  {pages} {030503} (\bibinfo {year} {2008})}\BibitemShut {NoStop}%
\bibitem [{\citenamefont {Mirrahimi}\ \emph {et~al.}(2014)\citenamefont
  {Mirrahimi}, \citenamefont {Leghtas}, \citenamefont {Albert}, \citenamefont
  {Touzard}, \citenamefont {Schoelkopf}, \citenamefont {Jiang},\ and\
  \citenamefont {Devoret}}]{Mirrahimi14}%
  \BibitemOpen
  \bibfield  {author} {\bibinfo {author} {\bibfnamefont {M.}~\bibnamefont
  {Mirrahimi}}, \bibinfo {author} {\bibfnamefont {Z.}~\bibnamefont {Leghtas}},
  \bibinfo {author} {\bibfnamefont {V.~V.}\ \bibnamefont {Albert}}, \bibinfo
  {author} {\bibfnamefont {S.}~\bibnamefont {Touzard}}, \bibinfo {author}
  {\bibfnamefont {R.~J.}\ \bibnamefont {Schoelkopf}}, \bibinfo {author}
  {\bibfnamefont {L.}~\bibnamefont {Jiang}}, \ and\ \bibinfo {author}
  {\bibfnamefont {M.~H.}\ \bibnamefont {Devoret}},\ }\bibfield  {title}
  {\enquote {\bibinfo {title} {Dynamically protected cat-qubits: a new paradigm
  for universal quantum computation},}\ }\href {\doibase
  10.1088/1367-2630/16/4/045014} {\bibfield  {journal} {\bibinfo  {journal}
  {New J. Phys.}\ }\textbf {\bibinfo {volume} {16}},\ \bibinfo {pages} {045014}
  (\bibinfo {year} {2014})}\BibitemShut {NoStop}%
\bibitem [{\citenamefont {Michael}\ \emph {et~al.}(2016)\citenamefont
  {Michael}, \citenamefont {Silveri}, \citenamefont {Brierley}, \citenamefont
  {Albert}, \citenamefont {Salmilehto}, \citenamefont {Jiang},\ and\
  \citenamefont {Girvin}}]{Michael16}%
  \BibitemOpen
  \bibfield  {author} {\bibinfo {author} {\bibfnamefont {M.~H.}\ \bibnamefont
  {Michael}}, \bibinfo {author} {\bibfnamefont {M.}~\bibnamefont {Silveri}},
  \bibinfo {author} {\bibfnamefont {R.~T.}\ \bibnamefont {Brierley}}, \bibinfo
  {author} {\bibfnamefont {V.~V.}\ \bibnamefont {Albert}}, \bibinfo {author}
  {\bibfnamefont {J.}~\bibnamefont {Salmilehto}}, \bibinfo {author}
  {\bibfnamefont {L.}~\bibnamefont {Jiang}}, \ and\ \bibinfo {author}
  {\bibfnamefont {S.~M.}\ \bibnamefont {Girvin}},\ }\bibfield  {title}
  {\enquote {\bibinfo {title} {New class of quantum error-correcting codes for
  a bosonic mode},}\ }\href {\doibase 10.1103/PhysRevX.6.031006} {\bibfield
  {journal} {\bibinfo  {journal} {Phys. Rev. X}\ }\textbf {\bibinfo {volume}
  {6}},\ \bibinfo {pages} {031006} (\bibinfo {year} {2016})}\BibitemShut
  {NoStop}%
\bibitem [{\citenamefont {Puri}\ \emph {et~al.}(2017)\citenamefont {Puri},
  \citenamefont {Andersen}, \citenamefont {Grimsmo},\ and\ \citenamefont
  {Blais}}]{Puri:2017aa}%
  \BibitemOpen
  \bibfield  {author} {\bibinfo {author} {\bibfnamefont {S.}~\bibnamefont
  {Puri}}, \bibinfo {author} {\bibfnamefont {C.~K.}\ \bibnamefont {Andersen}},
  \bibinfo {author} {\bibfnamefont {A.~L.}\ \bibnamefont {Grimsmo}}, \ and\
  \bibinfo {author} {\bibfnamefont {A.}~\bibnamefont {Blais}},\ }\bibfield
  {title} {\enquote {\bibinfo {title} {Quantum annealing with all-to-all
  connected nonlinear oscillators},}\ }\href
  {https://doi.org/10.1038/ncomms15785} {\bibfield  {journal} {\bibinfo
  {journal} {Nature Comm.}\ }\textbf {\bibinfo {volume} {8}},\ \bibinfo {pages}
  {15785 EP } (\bibinfo {year} {2017})}\BibitemShut {NoStop}%
\bibitem [{\citenamefont {Puri}\ \emph {et~al.}(2018)\citenamefont {Puri},
  \citenamefont {Grimm}, \citenamefont {Campagne-Ibarcq}, \citenamefont
  {Eickbusch}, \citenamefont {Noh}, \citenamefont {Roberts}, \citenamefont
  {Jiang}, \citenamefont {Mirrahimi}, \citenamefont {Devoret},\ and\
  \citenamefont {Girvin}}]{Puri:2018aa}%
  \BibitemOpen
  \bibfield  {author} {\bibinfo {author} {\bibfnamefont {S.}~\bibnamefont
  {Puri}}, \bibinfo {author} {\bibfnamefont {A.}~\bibnamefont {Grimm}},
  \bibinfo {author} {\bibfnamefont {P.}~\bibnamefont {Campagne-Ibarcq}},
  \bibinfo {author} {\bibfnamefont {A.}~\bibnamefont {Eickbusch}}, \bibinfo
  {author} {\bibfnamefont {K.}~\bibnamefont {Noh}}, \bibinfo {author}
  {\bibfnamefont {G.}~\bibnamefont {Roberts}}, \bibinfo {author} {\bibfnamefont
  {L.}~\bibnamefont {Jiang}}, \bibinfo {author} {\bibfnamefont
  {M.}~\bibnamefont {Mirrahimi}}, \bibinfo {author} {\bibfnamefont {M.~H.}\
  \bibnamefont {Devoret}}, \ and\ \bibinfo {author} {\bibfnamefont {S.~M.}\
  \bibnamefont {Girvin}},\ }\bibfield  {title} {\enquote {\bibinfo {title}
  {Stabilized cat in driven nonlinear cavity: A fault-tolerant error syndrome
  detector},}\ }\href {https://arxiv.org/abs/1807.09334} {\bibfield  {journal}
  {\bibinfo  {journal} {arXiv:1807.09334}\ } (\bibinfo {year}
  {2018})}\BibitemShut {NoStop}%
\bibitem [{\citenamefont {Noh}\ \emph {et~al.}(2018)\citenamefont {Noh},
  \citenamefont {Albert},\ and\ \citenamefont {Jiang}}]{Noh:2018aa}%
  \BibitemOpen
  \bibfield  {author} {\bibinfo {author} {\bibfnamefont {K.}~\bibnamefont
  {Noh}}, \bibinfo {author} {\bibfnamefont {V.~V.}\ \bibnamefont {Albert}}, \
  and\ \bibinfo {author} {\bibfnamefont {L.}~\bibnamefont {Jiang}},\ }\bibfield
   {title} {\enquote {\bibinfo {title} {Quantum capacity bounds of gaussian
  thermal loss channels and achievable rates with gottesman-kitaev-preskill
  codes},}\ }\href@noop {} {\bibfield  {journal} {\bibinfo  {journal} {IEEE
  Transactions on Information Theory}\ } (\bibinfo {year} {2018})}\BibitemShut
  {NoStop}%
\bibitem [{\citenamefont {Li}\ \emph {et~al.}(2019)\citenamefont {Li},
  \citenamefont {Young}, \citenamefont {Albert}, \citenamefont {Noh},
  \citenamefont {Zou},\ and\ \citenamefont {Jiang}}]{Linshu-Li:2019aa}%
  \BibitemOpen
  \bibfield  {author} {\bibinfo {author} {\bibfnamefont {L.}~\bibnamefont
  {Li}}, \bibinfo {author} {\bibfnamefont {D.~J.}\ \bibnamefont {Young}},
  \bibinfo {author} {\bibfnamefont {V.~V.}\ \bibnamefont {Albert}}, \bibinfo
  {author} {\bibfnamefont {K.}~\bibnamefont {Noh}}, \bibinfo {author}
  {\bibfnamefont {C.-L.}\ \bibnamefont {Zou}}, \ and\ \bibinfo {author}
  {\bibfnamefont {L.}~\bibnamefont {Jiang}},\ }\bibfield  {title} {\enquote
  {\bibinfo {title} {Designing good bosonic quantum codes via creating
  destructive interference},}\ }\href {https://arxiv.org/abs/1901.05358}
  {\bibfield  {journal} {\bibinfo  {journal} {arXiv:1901.05358}\ } (\bibinfo
  {year} {2019})}\BibitemShut {NoStop}%
\bibitem [{\citenamefont {Menicucci}(2014)}]{Menicucci14}%
  \BibitemOpen
  \bibfield  {author} {\bibinfo {author} {\bibfnamefont {N.~C.}\ \bibnamefont
  {Menicucci}},\ }\bibfield  {title} {\enquote {\bibinfo {title}
  {Fault-tolerant measurement-based quantum computing with continuous-variable
  cluster states},}\ }\href {\doibase 10.1103/PhysRevLett.112.120504}
  {\bibfield  {journal} {\bibinfo  {journal} {Phys. Rev. Lett.}\ }\textbf
  {\bibinfo {volume} {112}},\ \bibinfo {pages} {120504} (\bibinfo {year}
  {2014})}\BibitemShut {NoStop}%
\bibitem [{\citenamefont {Fukui}\ \emph {et~al.}(2017)\citenamefont {Fukui},
  \citenamefont {Tomita},\ and\ \citenamefont {Okamoto}}]{Fukui:2017aa}%
  \BibitemOpen
  \bibfield  {author} {\bibinfo {author} {\bibfnamefont {K.}~\bibnamefont
  {Fukui}}, \bibinfo {author} {\bibfnamefont {A.}~\bibnamefont {Tomita}}, \
  and\ \bibinfo {author} {\bibfnamefont {A.}~\bibnamefont {Okamoto}},\
  }\bibfield  {title} {\enquote {\bibinfo {title} {Analog quantum error
  correction with encoding a qubit into an oscillator},}\ }\href {\doibase
  10.1103/PhysRevLett.119.180507} {\bibfield  {journal} {\bibinfo  {journal}
  {Phys. Rev. Lett.}\ }\textbf {\bibinfo {volume} {119}},\ \bibinfo {pages}
  {180507} (\bibinfo {year} {2017})}\BibitemShut {NoStop}%
\bibitem [{\citenamefont {Vuillot}\ \emph {et~al.}(2019)\citenamefont
  {Vuillot}, \citenamefont {Asasi}, \citenamefont {Wang}, \citenamefont
  {Pryadko},\ and\ \citenamefont {Terhal}}]{Vuillot2018}%
  \BibitemOpen
  \bibfield  {author} {\bibinfo {author} {\bibfnamefont {C.}~\bibnamefont
  {Vuillot}}, \bibinfo {author} {\bibfnamefont {H.}~\bibnamefont {Asasi}},
  \bibinfo {author} {\bibfnamefont {Y.}~\bibnamefont {Wang}}, \bibinfo {author}
  {\bibfnamefont {L.~P.}\ \bibnamefont {Pryadko}}, \ and\ \bibinfo {author}
  {\bibfnamefont {B.~M.}\ \bibnamefont {Terhal}},\ }\bibfield  {title}
  {\enquote {\bibinfo {title} {Quantum error correction with the toric
  gottesman-kitaev-preskill code},}\ }\href {\doibase
  10.1103/PhysRevA.99.032344} {\bibfield  {journal} {\bibinfo  {journal} {Phys.
  Rev. A}\ }\textbf {\bibinfo {volume} {99}},\ \bibinfo {pages} {032344}
  (\bibinfo {year} {2019})}\BibitemShut {NoStop}%
\bibitem [{\citenamefont {Noh}\ and\ \citenamefont
  {Chamberland}(2019)}]{Noh:2019aa}%
  \BibitemOpen
  \bibfield  {author} {\bibinfo {author} {\bibfnamefont {K.}~\bibnamefont
  {Noh}}\ and\ \bibinfo {author} {\bibfnamefont {C.}~\bibnamefont
  {Chamberland}},\ }\bibfield  {title} {\enquote {\bibinfo {title}
  {Fault-tolerant bosonic quantum error correction with the surface-gkp
  code},}\ }\href {https://arxiv.org/abs/1908.03579} {\bibfield  {journal}
  {\bibinfo  {journal} {arXiv:1908.03579}\ } (\bibinfo {year}
  {2019})}\BibitemShut {NoStop}%
\bibitem [{\citenamefont {Fukui}\ \emph {et~al.}(2018)\citenamefont {Fukui},
  \citenamefont {Tomita},\ and\ \citenamefont {Okamoto}}]{Fukui:2018aa}%
  \BibitemOpen
  \bibfield  {author} {\bibinfo {author} {\bibfnamefont {K.}~\bibnamefont
  {Fukui}}, \bibinfo {author} {\bibfnamefont {A.}~\bibnamefont {Tomita}}, \
  and\ \bibinfo {author} {\bibfnamefont {A.}~\bibnamefont {Okamoto}},\
  }\bibfield  {title} {\enquote {\bibinfo {title} {Tracking quantum error
  correction},}\ }\href {\doibase 10.1103/PhysRevA.98.022326} {\bibfield
  {journal} {\bibinfo  {journal} {Phys. Rev. A}\ }\textbf {\bibinfo {volume}
  {98}},\ \bibinfo {pages} {022326} (\bibinfo {year} {2018})}\BibitemShut
  {NoStop}%
\bibitem [{\citenamefont {Kimble}(1998)}]{Kimble:1998aa}%
  \BibitemOpen
  \bibfield  {author} {\bibinfo {author} {\bibfnamefont {H.~J.}\ \bibnamefont
  {Kimble}},\ }\bibfield  {title} {\enquote {\bibinfo {title} {Strong
  interactions of single atoms and photons in cavity {QED}},}\ }\href {\doibase
  10.1238/physica.topical.076a00127} {\bibfield  {journal} {\bibinfo  {journal}
  {Physica Scripta}\ }\textbf {\bibinfo {volume} {T76}},\ \bibinfo {pages}
  {127} (\bibinfo {year} {1998})}\BibitemShut {NoStop}%
\bibitem [{\citenamefont {Mabuchi}\ and\ \citenamefont
  {Doherty}(2002)}]{Mabuchi2002}%
  \BibitemOpen
  \bibfield  {author} {\bibinfo {author} {\bibfnamefont {H.}~\bibnamefont
  {Mabuchi}}\ and\ \bibinfo {author} {\bibfnamefont {A.}~\bibnamefont
  {Doherty}},\ }\bibfield  {title} {\enquote {\bibinfo {title} {Cavity quantum
  electrodynamics: coherence in context},}\ }\href@noop {} {\bibfield
  {journal} {\bibinfo  {journal} {Science}\ }\textbf {\bibinfo {volume}
  {298}},\ \bibinfo {pages} {1372} (\bibinfo {year} {2002})}\BibitemShut
  {NoStop}%
\bibitem [{\citenamefont {Yokoyama}\ \emph {et~al.}(2013)\citenamefont
  {Yokoyama}, \citenamefont {Ukai}, \citenamefont {Armstrong}, \citenamefont
  {Sornphiphatphong}, \citenamefont {Kaji}, \citenamefont {Suzuki},
  \citenamefont {Yoshikawa}, \citenamefont {Yonezawa}, \citenamefont
  {Menicucci},\ and\ \citenamefont {Furusawa}}]{Yokoyama:2013aa}%
  \BibitemOpen
  \bibfield  {author} {\bibinfo {author} {\bibfnamefont {S.}~\bibnamefont
  {Yokoyama}}, \bibinfo {author} {\bibfnamefont {R.}~\bibnamefont {Ukai}},
  \bibinfo {author} {\bibfnamefont {S.~C.}\ \bibnamefont {Armstrong}}, \bibinfo
  {author} {\bibfnamefont {C.}~\bibnamefont {Sornphiphatphong}}, \bibinfo
  {author} {\bibfnamefont {T.}~\bibnamefont {Kaji}}, \bibinfo {author}
  {\bibfnamefont {S.}~\bibnamefont {Suzuki}}, \bibinfo {author} {\bibfnamefont
  {J.-i.}\ \bibnamefont {Yoshikawa}}, \bibinfo {author} {\bibfnamefont
  {H.}~\bibnamefont {Yonezawa}}, \bibinfo {author} {\bibfnamefont {N.~C.}\
  \bibnamefont {Menicucci}}, \ and\ \bibinfo {author} {\bibfnamefont
  {A.}~\bibnamefont {Furusawa}},\ }\bibfield  {title} {\enquote {\bibinfo
  {title} {Ultra-large-scale continuous-variable cluster states multiplexed in
  the time domain},}\ }\href {https://doi.org/10.1038/nphoton.2013.287}
  {\bibfield  {journal} {\bibinfo  {journal} {Nat. Photonics}\ }\textbf
  {\bibinfo {volume} {7}},\ \bibinfo {pages} {982 EP } (\bibinfo {year}
  {2013})}\BibitemShut {NoStop}%
\bibitem [{\citenamefont {Wallraff}\ \emph {et~al.}(2004)\citenamefont
  {Wallraff}, \citenamefont {Schuster}, \citenamefont {Blais}, \citenamefont
  {Frunzio}, \citenamefont {Huang}, \citenamefont {Majer}, \citenamefont
  {Kumar}, \citenamefont {Girvin},\ and\ \citenamefont
  {Schoelkopf}}]{Wallraff2004}%
  \BibitemOpen
  \bibfield  {author} {\bibinfo {author} {\bibfnamefont {A.}~\bibnamefont
  {Wallraff}}, \bibinfo {author} {\bibfnamefont {D.~I.}\ \bibnamefont
  {Schuster}}, \bibinfo {author} {\bibfnamefont {A.}~\bibnamefont {Blais}},
  \bibinfo {author} {\bibfnamefont {L.}~\bibnamefont {Frunzio}}, \bibinfo
  {author} {\bibfnamefont {R.-S.}\ \bibnamefont {Huang}}, \bibinfo {author}
  {\bibfnamefont {J.}~\bibnamefont {Majer}}, \bibinfo {author} {\bibfnamefont
  {S.}~\bibnamefont {Kumar}}, \bibinfo {author} {\bibfnamefont {S.~M.}\
  \bibnamefont {Girvin}}, \ and\ \bibinfo {author} {\bibfnamefont {R.~J.}\
  \bibnamefont {Schoelkopf}},\ }\bibfield  {title} {\enquote {\bibinfo {title}
  {Strong coupling of a single photon to a superconducting qubit using circuit
  quantum electrodynamics},}\ }\href@noop {} {\bibfield  {journal} {\bibinfo
  {journal} {Nature}\ }\textbf {\bibinfo {volume} {431}},\ \bibinfo {pages}
  {162} (\bibinfo {year} {2004})}\BibitemShut {NoStop}%
\bibitem [{\citenamefont {Devoret}\ and\ \citenamefont
  {Schoelkopf}(2013)}]{Devoret2013}%
  \BibitemOpen
  \bibfield  {author} {\bibinfo {author} {\bibfnamefont {M.~H.}\ \bibnamefont
  {Devoret}}\ and\ \bibinfo {author} {\bibfnamefont {R.~J.}\ \bibnamefont
  {Schoelkopf}},\ }\bibfield  {title} {\enquote {\bibinfo {title}
  {Superconducting circuits for quantum information: an outlook},}\ }\href@noop
  {} {\bibfield  {journal} {\bibinfo  {journal} {Science}\ }\textbf {\bibinfo
  {volume} {339}},\ \bibinfo {pages} {1169} (\bibinfo {year}
  {2013})}\BibitemShut {NoStop}%
\bibitem [{\citenamefont {Barends}\ \emph {et~al.}(2014)\citenamefont
  {Barends}, \citenamefont {Kelly}, \citenamefont {Megrant}, \citenamefont
  {Veitia}, \citenamefont {Sank}, \citenamefont {Jeffrey}, \citenamefont
  {White}, \citenamefont {Mutus}, \citenamefont {Fowler}, \citenamefont
  {Campbell} \emph {et~al.}}]{Barends2014}%
  \BibitemOpen
  \bibfield  {author} {\bibinfo {author} {\bibfnamefont {R.}~\bibnamefont
  {Barends}}, \bibinfo {author} {\bibfnamefont {J.}~\bibnamefont {Kelly}},
  \bibinfo {author} {\bibfnamefont {A.}~\bibnamefont {Megrant}}, \bibinfo
  {author} {\bibfnamefont {A.}~\bibnamefont {Veitia}}, \bibinfo {author}
  {\bibfnamefont {D.}~\bibnamefont {Sank}}, \bibinfo {author} {\bibfnamefont
  {E.}~\bibnamefont {Jeffrey}}, \bibinfo {author} {\bibfnamefont {T.~C.}\
  \bibnamefont {White}}, \bibinfo {author} {\bibfnamefont {J.}~\bibnamefont
  {Mutus}}, \bibinfo {author} {\bibfnamefont {A.~G.}\ \bibnamefont {Fowler}},
  \bibinfo {author} {\bibfnamefont {B.}~\bibnamefont {Campbell}},  \emph
  {et~al.},\ }\bibfield  {title} {\enquote {\bibinfo {title} {Superconducting
  quantum circuits at the surface code threshold for fault tolerance},}\
  }\href@noop {} {\bibfield  {journal} {\bibinfo  {journal} {Nature}\ }\textbf
  {\bibinfo {volume} {508}},\ \bibinfo {pages} {500} (\bibinfo {year}
  {2014})}\BibitemShut {NoStop}%
\bibitem [{\citenamefont {Haroche}\ and\ \citenamefont
  {Raimond}(2006)}]{Haroche2006}%
  \BibitemOpen
  \bibfield  {author} {\bibinfo {author} {\bibfnamefont {S.}~\bibnamefont
  {Haroche}}\ and\ \bibinfo {author} {\bibfnamefont {J.-M.}\ \bibnamefont
  {Raimond}},\ }\href@noop {} {\emph {\bibinfo {title} {Exploring the quantum:
  atoms, cavities, and photons}}}\ (\bibinfo  {publisher} {Oxford university
  press},\ \bibinfo {year} {2006})\BibitemShut {NoStop}%
\bibitem [{\citenamefont {Reagor}\ \emph {et~al.}(2013)\citenamefont {Reagor},
  \citenamefont {Paik}, \citenamefont {Catelani}, \citenamefont {Sun},
  \citenamefont {Axline}, \citenamefont {Holland}, \citenamefont {Pop},
  \citenamefont {Masluk}, \citenamefont {Brecht}, \citenamefont {Frunzio} \emph
  {et~al.}}]{Reagor2013}%
  \BibitemOpen
  \bibfield  {author} {\bibinfo {author} {\bibfnamefont {M.}~\bibnamefont
  {Reagor}}, \bibinfo {author} {\bibfnamefont {H.}~\bibnamefont {Paik}},
  \bibinfo {author} {\bibfnamefont {G.}~\bibnamefont {Catelani}}, \bibinfo
  {author} {\bibfnamefont {L.}~\bibnamefont {Sun}}, \bibinfo {author}
  {\bibfnamefont {C.}~\bibnamefont {Axline}}, \bibinfo {author} {\bibfnamefont
  {E.}~\bibnamefont {Holland}}, \bibinfo {author} {\bibfnamefont {I.~M.}\
  \bibnamefont {Pop}}, \bibinfo {author} {\bibfnamefont {N.~A.}\ \bibnamefont
  {Masluk}}, \bibinfo {author} {\bibfnamefont {T.}~\bibnamefont {Brecht}},
  \bibinfo {author} {\bibfnamefont {L.}~\bibnamefont {Frunzio}},  \emph
  {et~al.},\ }\bibfield  {title} {\enquote {\bibinfo {title} {Reaching 10 ms
  single photon lifetimes for superconducting aluminum cavities},}\ }\href@noop
  {} {\bibfield  {journal} {\bibinfo  {journal} {Appl. Phys. Lett.}\ }\textbf
  {\bibinfo {volume} {102}},\ \bibinfo {pages} {192604} (\bibinfo {year}
  {2013})}\BibitemShut {NoStop}%
\bibitem [{\citenamefont {Leibfried}\ \emph {et~al.}(2003)\citenamefont
  {Leibfried}, \citenamefont {Blatt}, \citenamefont {Monroe},\ and\
  \citenamefont {Wineland}}]{Leibfried:2003aa}%
  \BibitemOpen
  \bibfield  {author} {\bibinfo {author} {\bibfnamefont {D.}~\bibnamefont
  {Leibfried}}, \bibinfo {author} {\bibfnamefont {R.}~\bibnamefont {Blatt}},
  \bibinfo {author} {\bibfnamefont {C.}~\bibnamefont {Monroe}}, \ and\ \bibinfo
  {author} {\bibfnamefont {D.}~\bibnamefont {Wineland}},\ }\bibfield  {title}
  {\enquote {\bibinfo {title} {Quantum dynamics of single trapped ions},}\
  }\href {\doibase 10.1103/RevModPhys.75.281} {\bibfield  {journal} {\bibinfo
  {journal} {Rev. Mod. Phys.}\ }\textbf {\bibinfo {volume} {75}},\ \bibinfo
  {pages} {281} (\bibinfo {year} {2003})}\BibitemShut {NoStop}%
\bibitem [{\citenamefont {Monroe}\ and\ \citenamefont
  {Kim}(2013)}]{Monroe1164}%
  \BibitemOpen
  \bibfield  {author} {\bibinfo {author} {\bibfnamefont {C.}~\bibnamefont
  {Monroe}}\ and\ \bibinfo {author} {\bibfnamefont {J.}~\bibnamefont {Kim}},\
  }\bibfield  {title} {\enquote {\bibinfo {title} {Scaling the ion trap quantum
  processor},}\ }\href {\doibase 10.1126/science.1231298} {\bibfield  {journal}
  {\bibinfo  {journal} {Science}\ }\textbf {\bibinfo {volume} {339}},\ \bibinfo
  {pages} {1164} (\bibinfo {year} {2013})}\BibitemShut {NoStop}%
\bibitem [{\citenamefont {Fl{\"u}hmann}\ \emph {et~al.}(2019)\citenamefont
  {Fl{\"u}hmann}, \citenamefont {Nguyen}, \citenamefont {Marinelli},
  \citenamefont {Negnevitsky}, \citenamefont {Mehta},\ and\ \citenamefont
  {Home}}]{Fluhmann:2019aa}%
  \BibitemOpen
  \bibfield  {author} {\bibinfo {author} {\bibfnamefont {C.}~\bibnamefont
  {Fl{\"u}hmann}}, \bibinfo {author} {\bibfnamefont {T.~L.}\ \bibnamefont
  {Nguyen}}, \bibinfo {author} {\bibfnamefont {M.}~\bibnamefont {Marinelli}},
  \bibinfo {author} {\bibfnamefont {V.}~\bibnamefont {Negnevitsky}}, \bibinfo
  {author} {\bibfnamefont {K.}~\bibnamefont {Mehta}}, \ and\ \bibinfo {author}
  {\bibfnamefont {J.~P.}\ \bibnamefont {Home}},\ }\bibfield  {title} {\enquote
  {\bibinfo {title} {Encoding a qubit in a trapped-ion mechanical
  oscillator},}\ }\href {\doibase 10.1038/s41586-019-0960-6} {\bibfield
  {journal} {\bibinfo  {journal} {Nature}\ }\textbf {\bibinfo {volume} {566}},\
  \bibinfo {pages} {513} (\bibinfo {year} {2019})}\BibitemShut {NoStop}%
\bibitem [{\citenamefont {Leghtas}\ \emph {et~al.}(2013)\citenamefont
  {Leghtas}, \citenamefont {Kirchmair}, \citenamefont {Vlastakis},
  \citenamefont {Schoelkopf}, \citenamefont {Devoret},\ and\ \citenamefont
  {Mirrahimi}}]{LeghKircVlas13}%
  \BibitemOpen
  \bibfield  {author} {\bibinfo {author} {\bibfnamefont {Z.}~\bibnamefont
  {Leghtas}}, \bibinfo {author} {\bibfnamefont {G.}~\bibnamefont {Kirchmair}},
  \bibinfo {author} {\bibfnamefont {B.}~\bibnamefont {Vlastakis}}, \bibinfo
  {author} {\bibfnamefont {R.~J.}\ \bibnamefont {Schoelkopf}}, \bibinfo
  {author} {\bibfnamefont {M.~H.}\ \bibnamefont {Devoret}}, \ and\ \bibinfo
  {author} {\bibfnamefont {M.}~\bibnamefont {Mirrahimi}},\ }\bibfield  {title}
  {\enquote {\bibinfo {title} {Hardware-efficient autonomous quantum memory
  protection},}\ }\href {\doibase 10.1103/PhysRevLett.111.120501} {\bibfield
  {journal} {\bibinfo  {journal} {Phys. Rev. Lett.}\ }\textbf {\bibinfo
  {volume} {111}},\ \bibinfo {pages} {120501} (\bibinfo {year}
  {2013})}\BibitemShut {NoStop}%
\bibitem [{\citenamefont {Bergmann}\ and\ \citenamefont {van
  Loock}(2016)}]{BergLooc16}%
  \BibitemOpen
  \bibfield  {author} {\bibinfo {author} {\bibfnamefont {M.}~\bibnamefont
  {Bergmann}}\ and\ \bibinfo {author} {\bibfnamefont {P.}~\bibnamefont {van
  Loock}},\ }\bibfield  {title} {\enquote {\bibinfo {title} {Quantum error
  correction against photon loss using multicomponent cat states},}\ }\href
  {\doibase 10.1103/PhysRevA.94.042332} {\bibfield  {journal} {\bibinfo
  {journal} {Phys. Rev. A}\ }\textbf {\bibinfo {volume} {94}},\ \bibinfo
  {pages} {042332} (\bibinfo {year} {2016})}\BibitemShut {NoStop}%
\bibitem [{\citenamefont {Ofek}\ \emph {et~al.}(2016)\citenamefont {Ofek},
  \citenamefont {Petrenko}, \citenamefont {Heeres}, \citenamefont {Reinhold},
  \citenamefont {Leghtas}, \citenamefont {Vlastakis}, \citenamefont {Liu},
  \citenamefont {Frunzio}, \citenamefont {Girvin}, \citenamefont {Jiang} \emph
  {et~al.}}]{Ofek16}%
  \BibitemOpen
  \bibfield  {author} {\bibinfo {author} {\bibfnamefont {N.}~\bibnamefont
  {Ofek}}, \bibinfo {author} {\bibfnamefont {A.}~\bibnamefont {Petrenko}},
  \bibinfo {author} {\bibfnamefont {R.}~\bibnamefont {Heeres}}, \bibinfo
  {author} {\bibfnamefont {P.}~\bibnamefont {Reinhold}}, \bibinfo {author}
  {\bibfnamefont {Z.}~\bibnamefont {Leghtas}}, \bibinfo {author} {\bibfnamefont
  {B.}~\bibnamefont {Vlastakis}}, \bibinfo {author} {\bibfnamefont
  {Y.}~\bibnamefont {Liu}}, \bibinfo {author} {\bibfnamefont {L.}~\bibnamefont
  {Frunzio}}, \bibinfo {author} {\bibfnamefont {S.}~\bibnamefont {Girvin}},
  \bibinfo {author} {\bibfnamefont {L.}~\bibnamefont {Jiang}},  \emph
  {et~al.},\ }\bibfield  {title} {\enquote {\bibinfo {title} {Extending the
  lifetime of a quantum bit with error correction in superconducting
  circuits},}\ }\href@noop {} {\bibfield  {journal} {\bibinfo  {journal}
  {Nature}\ }\textbf {\bibinfo {volume} {536}},\ \bibinfo {pages} {441}
  (\bibinfo {year} {2016})}\BibitemShut {NoStop}%
\bibitem [{\citenamefont {Hu}\ \emph {et~al.}(2019)\citenamefont {Hu},
  \citenamefont {Ma}, \citenamefont {Cai}, \citenamefont {Mu}, \citenamefont
  {Xu}, \citenamefont {Wang}, \citenamefont {Wu}, \citenamefont {Wang},
  \citenamefont {Song}, \citenamefont {Zou}, \citenamefont {Girvin},
  \citenamefont {Duan},\ and\ \citenamefont {Sun}}]{Hu:2019aa}%
  \BibitemOpen
  \bibfield  {author} {\bibinfo {author} {\bibfnamefont {L.}~\bibnamefont
  {Hu}}, \bibinfo {author} {\bibfnamefont {Y.}~\bibnamefont {Ma}}, \bibinfo
  {author} {\bibfnamefont {W.}~\bibnamefont {Cai}}, \bibinfo {author}
  {\bibfnamefont {X.}~\bibnamefont {Mu}}, \bibinfo {author} {\bibfnamefont
  {Y.}~\bibnamefont {Xu}}, \bibinfo {author} {\bibfnamefont {W.}~\bibnamefont
  {Wang}}, \bibinfo {author} {\bibfnamefont {Y.}~\bibnamefont {Wu}}, \bibinfo
  {author} {\bibfnamefont {H.}~\bibnamefont {Wang}}, \bibinfo {author}
  {\bibfnamefont {Y.~P.}\ \bibnamefont {Song}}, \bibinfo {author}
  {\bibfnamefont {C.~L.}\ \bibnamefont {Zou}}, \bibinfo {author} {\bibfnamefont
  {S.~M.}\ \bibnamefont {Girvin}}, \bibinfo {author} {\bibfnamefont {L.-M.}\
  \bibnamefont {Duan}}, \ and\ \bibinfo {author} {\bibfnamefont
  {L.}~\bibnamefont {Sun}},\ }\bibfield  {title} {\enquote {\bibinfo {title}
  {Quantum error correction and universal gate set operation on a binomial
  bosonic logical qubit},}\ }\href {\doibase 10.1038/s41567-018-0414-3}
  {\bibfield  {journal} {\bibinfo  {journal} {Nature Physics}\ } (\bibinfo
  {year} {2019}),\ 10.1038/s41567-018-0414-3}\BibitemShut {NoStop}%
\bibitem [{\citenamefont {Rosenblum}\ \emph {et~al.}(2018)\citenamefont
  {Rosenblum}, \citenamefont {Reinhold}, \citenamefont {Mirrahimi},
  \citenamefont {Jiang}, \citenamefont {Frunzio},\ and\ \citenamefont
  {Schoelkopf}}]{Rosenblum2018}%
  \BibitemOpen
  \bibfield  {author} {\bibinfo {author} {\bibfnamefont {S.}~\bibnamefont
  {Rosenblum}}, \bibinfo {author} {\bibfnamefont {P.}~\bibnamefont {Reinhold}},
  \bibinfo {author} {\bibfnamefont {M.}~\bibnamefont {Mirrahimi}}, \bibinfo
  {author} {\bibfnamefont {L.}~\bibnamefont {Jiang}}, \bibinfo {author}
  {\bibfnamefont {L.}~\bibnamefont {Frunzio}}, \ and\ \bibinfo {author}
  {\bibfnamefont {R.~J.}\ \bibnamefont {Schoelkopf}},\ }\bibfield  {title}
  {\enquote {\bibinfo {title} {Fault-tolerant detection of a quantum error},}\
  }\href {\doibase 10.1126/science.aat3996} {\bibfield  {journal} {\bibinfo
  {journal} {Science}\ }\textbf {\bibinfo {volume} {361}},\ \bibinfo {pages}
  {266} (\bibinfo {year} {2018})}\BibitemShut {NoStop}%
\bibitem [{\citenamefont {Holevo}(2011)}]{Holevo11}%
  \BibitemOpen
  \bibfield  {author} {\bibinfo {author} {\bibfnamefont {A.~S.}\ \bibnamefont
  {Holevo}},\ }\href@noop {} {\emph {\bibinfo {title} {Probabilistic and
  statistical aspects of quantum theory}}},\ Vol.~\bibinfo {volume} {1}\
  (\bibinfo  {publisher} {Springer Science \& Business Media},\ \bibinfo {year}
  {2011})\BibitemShut {NoStop}%
\bibitem [{\citenamefont {Knill}(2005{\natexlab{a}})}]{Knill05}%
  \BibitemOpen
  \bibfield  {author} {\bibinfo {author} {\bibfnamefont {E.}~\bibnamefont
  {Knill}},\ }\bibfield  {title} {\enquote {\bibinfo {title} {Quantum computing
  with realistically noisy devices},}\ }\href
  {http://dx.doi.org/10.1038/nature03350} {\bibfield  {journal} {\bibinfo
  {journal} {Nature}\ }\textbf {\bibinfo {volume} {434}},\ \bibinfo {pages}
  {39} (\bibinfo {year} {2005}{\natexlab{a}})}\BibitemShut {NoStop}%
\bibitem [{\citenamefont {Bacon}(2006)}]{Bacon2006}%
  \BibitemOpen
  \bibfield  {author} {\bibinfo {author} {\bibfnamefont {D.}~\bibnamefont
  {Bacon}},\ }\bibfield  {title} {\enquote {\bibinfo {title} {Operator quantum
  error-correcting subsystems for self-correcting quantum memories},}\ }\href
  {\doibase 10.1103/PhysRevA.73.012340} {\bibfield  {journal} {\bibinfo
  {journal} {Phys. Rev. A}\ }\textbf {\bibinfo {volume} {73}},\ \bibinfo
  {pages} {012340} (\bibinfo {year} {2006})}\BibitemShut {NoStop}%
\bibitem [{\citenamefont {Kribs}\ \emph {et~al.}(2005)\citenamefont {Kribs},
  \citenamefont {Laflamme},\ and\ \citenamefont {Poulin}}]{Poulin2005}%
  \BibitemOpen
  \bibfield  {author} {\bibinfo {author} {\bibfnamefont {D.}~\bibnamefont
  {Kribs}}, \bibinfo {author} {\bibfnamefont {R.}~\bibnamefont {Laflamme}}, \
  and\ \bibinfo {author} {\bibfnamefont {D.}~\bibnamefont {Poulin}},\
  }\bibfield  {title} {\enquote {\bibinfo {title} {Unified and generalized
  approach to quantum error correction},}\ }\href {\doibase
  10.1103/PhysRevLett.94.180501} {\bibfield  {journal} {\bibinfo  {journal}
  {Phys. Rev. Lett.}\ }\textbf {\bibinfo {volume} {94}},\ \bibinfo {pages}
  {180501} (\bibinfo {year} {2005})}\BibitemShut {NoStop}%
\bibitem [{\citenamefont {Aliferis}\ and\ \citenamefont
  {Preskill}(2008)}]{Aliferis08}%
  \BibitemOpen
  \bibfield  {author} {\bibinfo {author} {\bibfnamefont {P.}~\bibnamefont
  {Aliferis}}\ and\ \bibinfo {author} {\bibfnamefont {J.}~\bibnamefont
  {Preskill}},\ }\bibfield  {title} {\enquote {\bibinfo {title} {Fault-tolerant
  quantum computation against biased noise},}\ }\href {\doibase
  10.1103/PhysRevA.78.052331} {\bibfield  {journal} {\bibinfo  {journal} {Phys.
  Rev. A}\ }\textbf {\bibinfo {volume} {78}},\ \bibinfo {pages} {052331}
  (\bibinfo {year} {2008})}\BibitemShut {NoStop}%
\bibitem [{\citenamefont {Brooks}\ and\ \citenamefont
  {Preskill}(2013)}]{brooks2013fault}%
  \BibitemOpen
  \bibfield  {author} {\bibinfo {author} {\bibfnamefont {P.}~\bibnamefont
  {Brooks}}\ and\ \bibinfo {author} {\bibfnamefont {J.}~\bibnamefont
  {Preskill}},\ }\bibfield  {title} {\enquote {\bibinfo {title} {Fault-tolerant
  quantum computation with asymmetric bacon-shor codes},}\ }\href@noop {}
  {\bibfield  {journal} {\bibinfo  {journal} {Physical Review A}\ }\textbf
  {\bibinfo {volume} {87}},\ \bibinfo {pages} {032310} (\bibinfo {year}
  {2013})}\BibitemShut {NoStop}%
\bibitem [{\citenamefont {Sabapathy}\ and\ \citenamefont
  {Weedbrook}(2018)}]{Sabapathy:2018aa}%
  \BibitemOpen
  \bibfield  {author} {\bibinfo {author} {\bibfnamefont {K.~K.}\ \bibnamefont
  {Sabapathy}}\ and\ \bibinfo {author} {\bibfnamefont {C.}~\bibnamefont
  {Weedbrook}},\ }\bibfield  {title} {\enquote {\bibinfo {title} {On states as
  resource units for universal quantum computation with photonic
  architectures},}\ }\href {\doibase 10.1103/PhysRevA.97.062315} {\bibfield
  {journal} {\bibinfo  {journal} {Phys. Rev. A}\ }\textbf {\bibinfo {volume}
  {97}},\ \bibinfo {pages} {062315} (\bibinfo {year} {2018})}\BibitemShut
  {NoStop}%
\bibitem [{\citenamefont {Elder}\ \emph {et~al.}(2019)\citenamefont {Elder},
  \citenamefont {Wang}, \citenamefont {Reinhold}, \citenamefont {Hann},
  \citenamefont {Chou}, \citenamefont {Lester}, \citenamefont {Rosenblum},
  \citenamefont {Frunzio}, \citenamefont {Jiang},\ and\ \citenamefont
  {Schoelkopf}}]{Elder:2019aa}%
  \BibitemOpen
  \bibfield  {author} {\bibinfo {author} {\bibfnamefont {S.~S.}\ \bibnamefont
  {Elder}}, \bibinfo {author} {\bibfnamefont {C.~S.}\ \bibnamefont {Wang}},
  \bibinfo {author} {\bibfnamefont {P.}~\bibnamefont {Reinhold}}, \bibinfo
  {author} {\bibfnamefont {C.~T.}\ \bibnamefont {Hann}}, \bibinfo {author}
  {\bibfnamefont {K.~S.}\ \bibnamefont {Chou}}, \bibinfo {author}
  {\bibfnamefont {B.~J.}\ \bibnamefont {Lester}}, \bibinfo {author}
  {\bibfnamefont {S.}~\bibnamefont {Rosenblum}}, \bibinfo {author}
  {\bibfnamefont {L.}~\bibnamefont {Frunzio}}, \bibinfo {author} {\bibfnamefont
  {L.}~\bibnamefont {Jiang}}, \ and\ \bibinfo {author} {\bibfnamefont {R.~J.}\
  \bibnamefont {Schoelkopf}},\ }\bibfield  {title} {\enquote {\bibinfo {title}
  {High-fidelity measurement of qubits encoded in multilevel superconducting
  circuits},}\ }\href {\doibase https://arxiv.org/abs/1908.01869} {\bibfield
  {journal} {\bibinfo  {journal} {arXiv:1908.01869}\ } (\bibinfo {year}
  {2019}),\ https://arxiv.org/abs/1908.01869}\BibitemShut {NoStop}%
\bibitem [{\citenamefont {Albert}\ \emph
  {et~al.}(2018{\natexlab{a}})\citenamefont {Albert}, \citenamefont {Noh},
  \citenamefont {Duivenvoorden}, \citenamefont {Young}, \citenamefont
  {Brierley}, \citenamefont {Reinhold}, \citenamefont {Vuillot}, \citenamefont
  {Li}, \citenamefont {Shen}, \citenamefont {Girvin}, \citenamefont {Terhal},\
  and\ \citenamefont {Jiang}}]{Albert17}%
  \BibitemOpen
  \bibfield  {author} {\bibinfo {author} {\bibfnamefont {V.~V.}\ \bibnamefont
  {Albert}}, \bibinfo {author} {\bibfnamefont {K.}~\bibnamefont {Noh}},
  \bibinfo {author} {\bibfnamefont {K.}~\bibnamefont {Duivenvoorden}}, \bibinfo
  {author} {\bibfnamefont {D.~J.}\ \bibnamefont {Young}}, \bibinfo {author}
  {\bibfnamefont {R.~T.}\ \bibnamefont {Brierley}}, \bibinfo {author}
  {\bibfnamefont {P.}~\bibnamefont {Reinhold}}, \bibinfo {author}
  {\bibfnamefont {C.}~\bibnamefont {Vuillot}}, \bibinfo {author} {\bibfnamefont
  {L.}~\bibnamefont {Li}}, \bibinfo {author} {\bibfnamefont {C.}~\bibnamefont
  {Shen}}, \bibinfo {author} {\bibfnamefont {S.~M.}\ \bibnamefont {Girvin}},
  \bibinfo {author} {\bibfnamefont {B.~M.}\ \bibnamefont {Terhal}}, \ and\
  \bibinfo {author} {\bibfnamefont {L.}~\bibnamefont {Jiang}},\ }\bibfield
  {title} {\enquote {\bibinfo {title} {Performance and structure of single-mode
  bosonic codes},}\ }\href {\doibase 10.1103/PhysRevA.97.032346} {\bibfield
  {journal} {\bibinfo  {journal} {Phys. Rev. A}\ }\textbf {\bibinfo {volume}
  {97}},\ \bibinfo {pages} {032346} (\bibinfo {year}
  {2018}{\natexlab{a}})}\BibitemShut {NoStop}%
\bibitem [{\citenamefont {Helstrom}(1969)}]{Helstrom69}%
  \BibitemOpen
  \bibfield  {author} {\bibinfo {author} {\bibfnamefont {C.~W.}\ \bibnamefont
  {Helstrom}},\ }\bibfield  {title} {\enquote {\bibinfo {title} {Quantum
  detection and estimation theory},}\ }\href@noop {} {\bibfield  {journal}
  {\bibinfo  {journal} {J. Stat. Phys.}\ }\textbf {\bibinfo {volume} {1}},\
  \bibinfo {pages} {231} (\bibinfo {year} {1969})}\BibitemShut {NoStop}%
\bibitem [{\citenamefont {Motes}\ \emph {et~al.}(2017)\citenamefont {Motes},
  \citenamefont {Baragiola}, \citenamefont {Gilchrist},\ and\ \citenamefont
  {Menicucci}}]{MoteBaraGilc17}%
  \BibitemOpen
  \bibfield  {author} {\bibinfo {author} {\bibfnamefont {K.~R.}\ \bibnamefont
  {Motes}}, \bibinfo {author} {\bibfnamefont {B.~Q.}\ \bibnamefont
  {Baragiola}}, \bibinfo {author} {\bibfnamefont {A.}~\bibnamefont
  {Gilchrist}}, \ and\ \bibinfo {author} {\bibfnamefont {N.~C.}\ \bibnamefont
  {Menicucci}},\ }\bibfield  {title} {\enquote {\bibinfo {title} {Encoding
  qubits into oscillators with atomic ensembles and squeezed light},}\ }\href
  {\doibase 10.1103/PhysRevA.95.053819} {\bibfield  {journal} {\bibinfo
  {journal} {Phys. Rev. A}\ }\textbf {\bibinfo {volume} {95}},\ \bibinfo
  {pages} {053819} (\bibinfo {year} {2017})}\BibitemShut {NoStop}%
\bibitem [{\citenamefont {Belavkin}(1975{\natexlab{a}})}]{Belavkin1975aa}%
  \BibitemOpen
  \bibfield  {author} {\bibinfo {author} {\bibfnamefont {V.~P.}\ \bibnamefont
  {Belavkin}},\ }\bibfield  {title} {\enquote {\bibinfo {title} {Optimal
  distinction of non-orthogonal quantum signals},}\ }\href@noop {} {\bibfield
  {journal} {\bibinfo  {journal} {Radio Eng. Electron. Phys.}\ }\textbf
  {\bibinfo {volume} {20}},\ \bibinfo {pages} {39} (\bibinfo {year}
  {1975}{\natexlab{a}})}\BibitemShut {NoStop}%
\bibitem [{\citenamefont {Belavkin}(1975{\natexlab{b}})}]{Belavkin1975bb}%
  \BibitemOpen
  \bibfield  {author} {\bibinfo {author} {\bibfnamefont {V.~P.}\ \bibnamefont
  {Belavkin}},\ }\bibfield  {title} {\enquote {\bibinfo {title} {Optimal
  multiple quantum statistical hypothesis testing},}\ }\href {\doibase
  https://doi.org/10.1080/17442507508833114} {\bibfield  {journal} {\bibinfo
  {journal} {Stochastics}\ }\textbf {\bibinfo {volume} {1}},\ \bibinfo {pages}
  {315} (\bibinfo {year} {1975}{\natexlab{b}})}\BibitemShut {NoStop}%
\bibitem [{\citenamefont {Hausladen}\ and\ \citenamefont
  {Wootters}(1994)}]{Hausladen94}%
  \BibitemOpen
  \bibfield  {author} {\bibinfo {author} {\bibfnamefont {P.}~\bibnamefont
  {Hausladen}}\ and\ \bibinfo {author} {\bibfnamefont {W.~K.}\ \bibnamefont
  {Wootters}},\ }\bibfield  {title} {\enquote {\bibinfo {title} {A ``pretty
  good'' measurement for distinguishing quantum states},}\ }\href@noop {}
  {\bibfield  {journal} {\bibinfo  {journal} {J. Mod. Opt.}\ }\textbf {\bibinfo
  {volume} {41}},\ \bibinfo {pages} {2385} (\bibinfo {year}
  {1994})}\BibitemShut {NoStop}%
\bibitem [{\citenamefont {Wiseman}\ and\ \citenamefont
  {Killip}(1998)}]{Wiseman98}%
  \BibitemOpen
  \bibfield  {author} {\bibinfo {author} {\bibfnamefont {H.~M.}\ \bibnamefont
  {Wiseman}}\ and\ \bibinfo {author} {\bibfnamefont {R.~B.}\ \bibnamefont
  {Killip}},\ }\bibfield  {title} {\enquote {\bibinfo {title} {Adaptive
  single-shot phase measurements: The full quantum theory},}\ }\href {\doibase
  10.1103/PhysRevA.57.2169} {\bibfield  {journal} {\bibinfo  {journal} {Phys.
  Rev. A}\ }\textbf {\bibinfo {volume} {57}},\ \bibinfo {pages} {2169}
  (\bibinfo {year} {1998})}\BibitemShut {NoStop}%
\bibitem [{\citenamefont {Higgins}\ \emph {et~al.}(2007)\citenamefont
  {Higgins}, \citenamefont {Berry}, \citenamefont {Bartlett}, \citenamefont
  {Wiseman},\ and\ \citenamefont {Pryde}}]{Higgins07}%
  \BibitemOpen
  \bibfield  {author} {\bibinfo {author} {\bibfnamefont {B.~L.}\ \bibnamefont
  {Higgins}}, \bibinfo {author} {\bibfnamefont {D.~W.}\ \bibnamefont {Berry}},
  \bibinfo {author} {\bibfnamefont {S.~D.}\ \bibnamefont {Bartlett}}, \bibinfo
  {author} {\bibfnamefont {H.~M.}\ \bibnamefont {Wiseman}}, \ and\ \bibinfo
  {author} {\bibfnamefont {G.~J.}\ \bibnamefont {Pryde}},\ }\bibfield  {title}
  {\enquote {\bibinfo {title} {Entanglement-free heisenberg-limited phase
  estimation},}\ }\href@noop {} {\bibfield  {journal} {\bibinfo  {journal}
  {Nature}\ }\textbf {\bibinfo {volume} {450}},\ \bibinfo {pages} {393}
  (\bibinfo {year} {2007})}\BibitemShut {NoStop}%
\bibitem [{\citenamefont {Berni}\ \emph {et~al.}(2015)\citenamefont {Berni},
  \citenamefont {Gehring}, \citenamefont {Nielsen}, \citenamefont
  {H{\"a}ndchen}, \citenamefont {Paris},\ and\ \citenamefont
  {Andersen}}]{Berni15}%
  \BibitemOpen
  \bibfield  {author} {\bibinfo {author} {\bibfnamefont {A.~A.}\ \bibnamefont
  {Berni}}, \bibinfo {author} {\bibfnamefont {T.}~\bibnamefont {Gehring}},
  \bibinfo {author} {\bibfnamefont {B.~M.}\ \bibnamefont {Nielsen}}, \bibinfo
  {author} {\bibfnamefont {V.}~\bibnamefont {H{\"a}ndchen}}, \bibinfo {author}
  {\bibfnamefont {M.~G.}\ \bibnamefont {Paris}}, \ and\ \bibinfo {author}
  {\bibfnamefont {U.~L.}\ \bibnamefont {Andersen}},\ }\bibfield  {title}
  {\enquote {\bibinfo {title} {Ab initio quantum-enhanced optical phase
  estimation using real-time feedback control},}\ }\href@noop {} {\bibfield
  {journal} {\bibinfo  {journal} {Nat. Photonics}\ }\textbf {\bibinfo {volume}
  {9}},\ \bibinfo {pages} {577} (\bibinfo {year} {2015})}\BibitemShut {NoStop}%
\bibitem [{\citenamefont {Daryanoosh}\ \emph {et~al.}(2018)\citenamefont
  {Daryanoosh}, \citenamefont {Slussarenko}, \citenamefont {Berry},
  \citenamefont {Wiseman},\ and\ \citenamefont {Pryde}}]{Daryanoosh17}%
  \BibitemOpen
  \bibfield  {author} {\bibinfo {author} {\bibfnamefont {S.}~\bibnamefont
  {Daryanoosh}}, \bibinfo {author} {\bibfnamefont {S.}~\bibnamefont
  {Slussarenko}}, \bibinfo {author} {\bibfnamefont {D.~W.}\ \bibnamefont
  {Berry}}, \bibinfo {author} {\bibfnamefont {H.~M.}\ \bibnamefont {Wiseman}},
  \ and\ \bibinfo {author} {\bibfnamefont {G.~J.}\ \bibnamefont {Pryde}},\
  }\bibfield  {title} {\enquote {\bibinfo {title} {Experimental optical phase
  measurement approaching the exact heisenberg limit},}\ }\href {\doibase
  10.1038/s41467-018-06601-7} {\bibfield  {journal} {\bibinfo  {journal}
  {Nature Communications}\ }\textbf {\bibinfo {volume} {9}},\ \bibinfo {pages}
  {4606} (\bibinfo {year} {2018})}\BibitemShut {NoStop}%
\bibitem [{\citenamefont {Berry}\ and\ \citenamefont
  {Wiseman}(2000)}]{Berry00}%
  \BibitemOpen
  \bibfield  {author} {\bibinfo {author} {\bibfnamefont {D.~W.}\ \bibnamefont
  {Berry}}\ and\ \bibinfo {author} {\bibfnamefont {H.~M.}\ \bibnamefont
  {Wiseman}},\ }\bibfield  {title} {\enquote {\bibinfo {title} {Optimal states
  and almost optimal adaptive measurements for quantum interferometry},}\
  }\href {\doibase 10.1103/PhysRevLett.85.5098} {\bibfield  {journal} {\bibinfo
   {journal} {Phys. Rev. Lett.}\ }\textbf {\bibinfo {volume} {85}},\ \bibinfo
  {pages} {5098} (\bibinfo {year} {2000})}\BibitemShut {NoStop}%
\bibitem [{\citenamefont {Holevo}(1984)}]{Holevo84}%
  \BibitemOpen
  \bibfield  {author} {\bibinfo {author} {\bibfnamefont {A.}~\bibnamefont
  {Holevo}},\ }\bibfield  {title} {\enquote {\bibinfo {title} {Covariant
  measurements and imprimitivity systems},}\ }in\ \href@noop {} {\emph
  {\bibinfo {booktitle} {Quantum Probability and Applications to the Quantum
  Theory of Irreversible Processes}}}\ (\bibinfo  {publisher} {Springer},\
  \bibinfo {year} {1984})\ pp.\ \bibinfo {pages} {153--172}\BibitemShut
  {NoStop}%
\bibitem [{\citenamefont {Wiseman}\ \emph {et~al.}(2009)\citenamefont
  {Wiseman}, \citenamefont {Berry}, \citenamefont {Bartlett}, \citenamefont
  {Higgins},\ and\ \citenamefont {Pryde}}]{Wiseman09}%
  \BibitemOpen
  \bibfield  {author} {\bibinfo {author} {\bibfnamefont {H.~M.}\ \bibnamefont
  {Wiseman}}, \bibinfo {author} {\bibfnamefont {D.~W.}\ \bibnamefont {Berry}},
  \bibinfo {author} {\bibfnamefont {S.~D.}\ \bibnamefont {Bartlett}}, \bibinfo
  {author} {\bibfnamefont {B.~L.}\ \bibnamefont {Higgins}}, \ and\ \bibinfo
  {author} {\bibfnamefont {G.~J.}\ \bibnamefont {Pryde}},\ }\bibfield  {title}
  {\enquote {\bibinfo {title} {Adaptive measurements in the optical quantum
  information laboratory},}\ }\href@noop {} {\bibfield  {journal} {\bibinfo
  {journal} {IEEE J. Sel. Top. Quantum Electron.}\ }\textbf {\bibinfo {volume}
  {15}},\ \bibinfo {pages} {1661} (\bibinfo {year} {2009})}\BibitemShut
  {NoStop}%
\bibitem [{\citenamefont {Martin}\ \emph {et~al.}(2019)\citenamefont {Martin},
  \citenamefont {Livingston}, \citenamefont {Hacohen-Gourgy}, \citenamefont
  {Wiseman},\ and\ \citenamefont {Siddiqi}}]{LeighMartin:2019aa}%
  \BibitemOpen
  \bibfield  {author} {\bibinfo {author} {\bibfnamefont {L.~S.}\ \bibnamefont
  {Martin}}, \bibinfo {author} {\bibfnamefont {W.~P.}\ \bibnamefont
  {Livingston}}, \bibinfo {author} {\bibfnamefont {S.}~\bibnamefont
  {Hacohen-Gourgy}}, \bibinfo {author} {\bibfnamefont {H.~M.}\ \bibnamefont
  {Wiseman}}, \ and\ \bibinfo {author} {\bibfnamefont {I.}~\bibnamefont
  {Siddiqi}},\ }\bibfield  {title} {\enquote {\bibinfo {title} {Implementation
  of a canonical phase measurement with quantum feedback},}\ }\href {\doibase
  https://arxiv.org/abs/1906.07274} {\bibfield  {journal} {\bibinfo  {journal}
  {arXiv:1906.07274}\ } (\bibinfo {year} {2019}),\
  https://arxiv.org/abs/1906.07274}\BibitemShut {NoStop}%
\bibitem [{\citenamefont {Raynal}\ \emph {et~al.}(2010)\citenamefont {Raynal},
  \citenamefont {Kalev}, \citenamefont {Suzuki},\ and\ \citenamefont
  {Englert}}]{Raynal:2010aa}%
  \BibitemOpen
  \bibfield  {author} {\bibinfo {author} {\bibfnamefont {P.}~\bibnamefont
  {Raynal}}, \bibinfo {author} {\bibfnamefont {A.}~\bibnamefont {Kalev}},
  \bibinfo {author} {\bibfnamefont {J.}~\bibnamefont {Suzuki}}, \ and\ \bibinfo
  {author} {\bibfnamefont {B.-G.}\ \bibnamefont {Englert}},\ }\bibfield
  {title} {\enquote {\bibinfo {title} {Encoding many qubits in a rotor},}\
  }\href {https://dx.doi.org/10.1103/PhysRevA.81.052327} {\bibfield  {journal}
  {\bibinfo  {journal} {Phys. Rev. A}\ }\textbf {\bibinfo {volume} {81}},\
  \bibinfo {pages} {052327} (\bibinfo {year} {2010})}\BibitemShut {NoStop}%
\bibitem [{\citenamefont {Pegg}\ and\ \citenamefont
  {Barnett}(1989)}]{PeggBarn89}%
  \BibitemOpen
  \bibfield  {author} {\bibinfo {author} {\bibfnamefont {D.~T.}\ \bibnamefont
  {Pegg}}\ and\ \bibinfo {author} {\bibfnamefont {S.~M.}\ \bibnamefont
  {Barnett}},\ }\bibfield  {title} {\enquote {\bibinfo {title} {Phase
  properties of the quantized single-mode electromagnetic field},}\ }\href
  {https://dx.doi.org/10.1103/PhysRevA.39.1665} {\bibfield  {journal} {\bibinfo
   {journal} {Phys. Rev. A}\ }\textbf {\bibinfo {volume} {39}},\ \bibinfo
  {pages} {1665} (\bibinfo {year} {1989})}\BibitemShut {NoStop}%
\bibitem [{\citenamefont {Levy-Leblond}(1976)}]{Levy1976}%
  \BibitemOpen
  \bibfield  {author} {\bibinfo {author} {\bibfnamefont {J.-M.}\ \bibnamefont
  {Levy-Leblond}},\ }\bibfield  {title} {\enquote {\bibinfo {title} {Who is
  afraid of nonhermitian operators? a quantum description of angle and
  phase},}\ }\href@noop {} {\bibfield  {journal} {\bibinfo  {journal} {Ann.
  Phys.}\ }\textbf {\bibinfo {volume} {101}},\ \bibinfo {pages} {319} (\bibinfo
  {year} {1976})}\BibitemShut {NoStop}%
\bibitem [{\citenamefont {Pegg}\ and\ \citenamefont
  {Barnett}(1988)}]{PeggBarn88}%
  \BibitemOpen
  \bibfield  {author} {\bibinfo {author} {\bibfnamefont {D.~T.}\ \bibnamefont
  {Pegg}}\ and\ \bibinfo {author} {\bibfnamefont {S.~M.}\ \bibnamefont
  {Barnett}},\ }\bibfield  {title} {\enquote {\bibinfo {title} {Unitary phase
  operator in quantum mechanics},}\ }\href
  {https://doi.org/10.1209/0295-5075/6/6/002} {\bibfield  {journal} {\bibinfo
  {journal} {EPL (Europhysics Letters)}\ }\textbf {\bibinfo {volume} {6}},\
  \bibinfo {pages} {483} (\bibinfo {year} {1988})}\BibitemShut {NoStop}%
\bibitem [{\citenamefont {Wiseman}(1995)}]{Wiseman1995}%
  \BibitemOpen
  \bibfield  {author} {\bibinfo {author} {\bibfnamefont {H.~M.}\ \bibnamefont
  {Wiseman}},\ }\bibfield  {title} {\enquote {\bibinfo {title} {Adaptive phase
  measurements of optical modes: Going beyond the marginal q distribution},}\
  }\href@noop {} {\bibfield  {journal} {\bibinfo  {journal} {Phys. Rev. Lett.}\
  }\textbf {\bibinfo {volume} {75}},\ \bibinfo {pages} {4587} (\bibinfo {year}
  {1995})}\BibitemShut {NoStop}%
\bibitem [{\citenamefont {Jacobs}\ \emph {et~al.}(2008)\citenamefont {Jacobs},
  \citenamefont {Jordan},\ and\ \citenamefont {Irish}}]{Jacobs2008}%
  \BibitemOpen
  \bibfield  {author} {\bibinfo {author} {\bibfnamefont {K.}~\bibnamefont
  {Jacobs}}, \bibinfo {author} {\bibfnamefont {A.~N.}\ \bibnamefont {Jordan}},
  \ and\ \bibinfo {author} {\bibfnamefont {E.~K.}\ \bibnamefont {Irish}},\
  }\bibfield  {title} {\enquote {\bibinfo {title} {Energy measurements and
  preparation of canonical phase states of a nano-mechanical resonator},}\
  }\href@noop {} {\bibfield  {journal} {\bibinfo  {journal} {EPL (Europhysics
  Letters)}\ }\textbf {\bibinfo {volume} {82}},\ \bibinfo {pages} {18003}
  (\bibinfo {year} {2008})}\BibitemShut {NoStop}%
\bibitem [{\citenamefont {Barnett}\ and\ \citenamefont
  {Pegg}(1989)}]{BarnPegg89}%
  \BibitemOpen
  \bibfield  {author} {\bibinfo {author} {\bibfnamefont {S.~M.}\ \bibnamefont
  {Barnett}}\ and\ \bibinfo {author} {\bibfnamefont {D.~T.}\ \bibnamefont
  {Pegg}},\ }\bibfield  {title} {\enquote {\bibinfo {title} {On the hermitian
  optical phase operator},}\ }\href {https://doi.org/10.1080/09500348914550021}
  {\bibfield  {journal} {\bibinfo  {journal} {J. Mod. Opt.}\ }\textbf {\bibinfo
  {volume} {36}},\ \bibinfo {pages} {7} (\bibinfo {year} {1989})}\BibitemShut
  {NoStop}%
\bibitem [{\citenamefont {Fl\"uhmann}\ \emph {et~al.}(2018)\citenamefont
  {Fl\"uhmann}, \citenamefont {Negnevitsky}, \citenamefont {Marinelli},\ and\
  \citenamefont {Home}}]{Fluhmann:2018aa}%
  \BibitemOpen
  \bibfield  {author} {\bibinfo {author} {\bibfnamefont {C.}~\bibnamefont
  {Fl\"uhmann}}, \bibinfo {author} {\bibfnamefont {V.}~\bibnamefont
  {Negnevitsky}}, \bibinfo {author} {\bibfnamefont {M.}~\bibnamefont
  {Marinelli}}, \ and\ \bibinfo {author} {\bibfnamefont {J.~P.}\ \bibnamefont
  {Home}},\ }\bibfield  {title} {\enquote {\bibinfo {title} {Sequential modular
  position and momentum measurements of a trapped ion mechanical oscillator},}\
  }\href {\doibase 10.1103/PhysRevX.8.021001} {\bibfield  {journal} {\bibinfo
  {journal} {Phys. Rev. X}\ }\textbf {\bibinfo {volume} {8}},\ \bibinfo {pages}
  {021001} (\bibinfo {year} {2018})}\BibitemShut {NoStop}%
\bibitem [{\citenamefont {Zhang}\ \emph {et~al.}(2017)\citenamefont {Zhang},
  \citenamefont {Zhao}, \citenamefont {Zheng}, \citenamefont {Yu},
  \citenamefont {Su},\ and\ \citenamefont {Yang}}]{Zhang:2017aa}%
  \BibitemOpen
  \bibfield  {author} {\bibinfo {author} {\bibfnamefont {Y.}~\bibnamefont
  {Zhang}}, \bibinfo {author} {\bibfnamefont {X.}~\bibnamefont {Zhao}},
  \bibinfo {author} {\bibfnamefont {Z.-F.}\ \bibnamefont {Zheng}}, \bibinfo
  {author} {\bibfnamefont {L.}~\bibnamefont {Yu}}, \bibinfo {author}
  {\bibfnamefont {Q.-P.}\ \bibnamefont {Su}}, \ and\ \bibinfo {author}
  {\bibfnamefont {C.-P.}\ \bibnamefont {Yang}},\ }\bibfield  {title} {\enquote
  {\bibinfo {title} {Universal controlled-phase gate with cat-state qubits in
  circuit qed},}\ }\href {\doibase 10.1103/PhysRevA.96.052317} {\bibfield
  {journal} {\bibinfo  {journal} {Phys. Rev. A}\ }\textbf {\bibinfo {volume}
  {96}},\ \bibinfo {pages} {052317} (\bibinfo {year} {2017})}\BibitemShut
  {NoStop}%
\bibitem [{\citenamefont {Kitaev}(2006)}]{Kitaev06protected}%
  \BibitemOpen
  \bibfield  {author} {\bibinfo {author} {\bibfnamefont {A.}~\bibnamefont
  {Kitaev}},\ }\bibfield  {title} {\enquote {\bibinfo {title} {Protected qubit
  based on a superconducting current mirror},}\ }\href@noop {} {\bibfield
  {journal} {\bibinfo  {journal} {arXiv:cond-mat/0609441}\ } (\bibinfo {year}
  {2006})}\BibitemShut {NoStop}%
\bibitem [{\citenamefont {Webster}\ \emph {et~al.}(2015)\citenamefont
  {Webster}, \citenamefont {Bartlett},\ and\ \citenamefont
  {Poulin}}]{Webster15}%
  \BibitemOpen
  \bibfield  {author} {\bibinfo {author} {\bibfnamefont {P.}~\bibnamefont
  {Webster}}, \bibinfo {author} {\bibfnamefont {S.~D.}\ \bibnamefont
  {Bartlett}}, \ and\ \bibinfo {author} {\bibfnamefont {D.}~\bibnamefont
  {Poulin}},\ }\bibfield  {title} {\enquote {\bibinfo {title} {Reducing the
  overhead for quantum computation when noise is biased},}\ }\href {\doibase
  10.1103/PhysRevA.92.062309} {\bibfield  {journal} {\bibinfo  {journal} {Phys.
  Rev. A}\ }\textbf {\bibinfo {volume} {92}},\ \bibinfo {pages} {062309}
  (\bibinfo {year} {2015})}\BibitemShut {NoStop}%
\bibitem [{\citenamefont {Chamberland}\ \emph {et~al.}(2017)\citenamefont
  {Chamberland}, \citenamefont {Iyer},\ and\ \citenamefont
  {Poulin}}]{Chamberland17}%
  \BibitemOpen
  \bibfield  {author} {\bibinfo {author} {\bibfnamefont {C.}~\bibnamefont
  {Chamberland}}, \bibinfo {author} {\bibfnamefont {P.}~\bibnamefont {Iyer}}, \
  and\ \bibinfo {author} {\bibfnamefont {D.}~\bibnamefont {Poulin}},\
  }\bibfield  {title} {\enquote {\bibinfo {title} {Fault-tolerant quantum
  computing in the pauli or clifford frame with slow error diagnostics},}\
  }\href@noop {} {\bibfield  {journal} {\bibinfo  {journal} {arXiv:1704.06662}\
  } (\bibinfo {year} {2017})}\BibitemShut {NoStop}%
\bibitem [{\citenamefont {Terhal}(2015)}]{terhal2015quantum}%
  \BibitemOpen
  \bibfield  {author} {\bibinfo {author} {\bibfnamefont {B.~M.}\ \bibnamefont
  {Terhal}},\ }\bibfield  {title} {\enquote {\bibinfo {title} {Quantum error
  correction for quantum memories},}\ }\href@noop {} {\bibfield  {journal}
  {\bibinfo  {journal} {Reviews of Modern Physics}\ }\textbf {\bibinfo {volume}
  {87}},\ \bibinfo {pages} {307} (\bibinfo {year} {2015})}\BibitemShut
  {NoStop}%
\bibitem [{\citenamefont {Bravyi}\ and\ \citenamefont
  {Kitaev}(2005)}]{Bravyi2005}%
  \BibitemOpen
  \bibfield  {author} {\bibinfo {author} {\bibfnamefont {S.}~\bibnamefont
  {Bravyi}}\ and\ \bibinfo {author} {\bibfnamefont {A.}~\bibnamefont
  {Kitaev}},\ }\bibfield  {title} {\enquote {\bibinfo {title} {Universal
  quantum computation with ideal clifford gates and noisy ancillas},}\
  }\href@noop {} {\bibfield  {journal} {\bibinfo  {journal} {Phys. Rev. A}\
  }\textbf {\bibinfo {volume} {71}},\ \bibinfo {pages} {022316} (\bibinfo
  {year} {2005})}\BibitemShut {NoStop}%
\bibitem [{\citenamefont {Travaglione}\ and\ \citenamefont
  {Milburn}(2002)}]{TravMilb02}%
  \BibitemOpen
  \bibfield  {author} {\bibinfo {author} {\bibfnamefont {B.~C.}\ \bibnamefont
  {Travaglione}}\ and\ \bibinfo {author} {\bibfnamefont {G.~J.}\ \bibnamefont
  {Milburn}},\ }\bibfield  {title} {\enquote {\bibinfo {title} {Preparing
  encoded states in an oscillator},}\ }\href {\doibase
  10.1103/PhysRevA.66.052322} {\bibfield  {journal} {\bibinfo  {journal} {Phys.
  Rev. A}\ }\textbf {\bibinfo {volume} {66}},\ \bibinfo {pages} {052322}
  (\bibinfo {year} {2002})}\BibitemShut {NoStop}%
\bibitem [{\citenamefont {Terhal}\ and\ \citenamefont
  {Weigand}(2016)}]{TerhWeig16}%
  \BibitemOpen
  \bibfield  {author} {\bibinfo {author} {\bibfnamefont {B.~M.}\ \bibnamefont
  {Terhal}}\ and\ \bibinfo {author} {\bibfnamefont {D.}~\bibnamefont
  {Weigand}},\ }\bibfield  {title} {\enquote {\bibinfo {title} {Encoding a
  qubit into a cavity mode in circuit qed using phase estimation},}\ }\href
  {https://dx.doi.org/10.1103/PhysRevA.93.012315} {\bibfield  {journal}
  {\bibinfo  {journal} {Phys. Rev. A}\ }\textbf {\bibinfo {volume} {93}},\
  \bibinfo {pages} {012315} (\bibinfo {year} {2016})}\BibitemShut {NoStop}%
\bibitem [{\citenamefont {Eaton}\ \emph {et~al.}(2019)\citenamefont {Eaton},
  \citenamefont {Nehra},\ and\ \citenamefont {Pfister}}]{Eaton:2019aa}%
  \BibitemOpen
  \bibfield  {author} {\bibinfo {author} {\bibfnamefont {M.}~\bibnamefont
  {Eaton}}, \bibinfo {author} {\bibfnamefont {R.}~\bibnamefont {Nehra}}, \ and\
  \bibinfo {author} {\bibfnamefont {O.}~\bibnamefont {Pfister}},\ }\bibfield
  {title} {\enquote {\bibinfo {title} {Gottesman-kitaev-preskill state
  preparation by photon catalysis},}\ }\href {https://arxiv.org/abs/1903.01925}
  {\bibfield  {journal} {\bibinfo  {journal} {arXiv:1903.01925}\ } (\bibinfo
  {year} {2019})}\BibitemShut {NoStop}%
\bibitem [{\citenamefont {Shi}\ \emph {et~al.}(2019)\citenamefont {Shi},
  \citenamefont {Chamberland},\ and\ \citenamefont {Cross}}]{shi2019fault}%
  \BibitemOpen
  \bibfield  {author} {\bibinfo {author} {\bibfnamefont {Y.}~\bibnamefont
  {Shi}}, \bibinfo {author} {\bibfnamefont {C.}~\bibnamefont {Chamberland}}, \
  and\ \bibinfo {author} {\bibfnamefont {A.~W.}\ \bibnamefont {Cross}},\
  }\bibfield  {title} {\enquote {\bibinfo {title} {Fault-tolerant preparation
  of approximate gkp states},}\ }\href@noop {} {\bibfield  {journal} {\bibinfo
  {journal} {arXiv:1905.00903}\ } (\bibinfo {year} {2019})}\BibitemShut
  {NoStop}%
\bibitem [{\citenamefont {Heeres}\ \emph {et~al.}(2015)\citenamefont {Heeres},
  \citenamefont {Vlastakis}, \citenamefont {Holland}, \citenamefont
  {Krastanov}, \citenamefont {Albert}, \citenamefont {Frunzio}, \citenamefont
  {Jiang},\ and\ \citenamefont {Schoelkopf}}]{Heeres2015}%
  \BibitemOpen
  \bibfield  {author} {\bibinfo {author} {\bibfnamefont {R.~W.}\ \bibnamefont
  {Heeres}}, \bibinfo {author} {\bibfnamefont {B.}~\bibnamefont {Vlastakis}},
  \bibinfo {author} {\bibfnamefont {E.}~\bibnamefont {Holland}}, \bibinfo
  {author} {\bibfnamefont {S.}~\bibnamefont {Krastanov}}, \bibinfo {author}
  {\bibfnamefont {V.~V.}\ \bibnamefont {Albert}}, \bibinfo {author}
  {\bibfnamefont {L.}~\bibnamefont {Frunzio}}, \bibinfo {author} {\bibfnamefont
  {L.}~\bibnamefont {Jiang}}, \ and\ \bibinfo {author} {\bibfnamefont {R.~J.}\
  \bibnamefont {Schoelkopf}},\ }\bibfield  {title} {\enquote {\bibinfo {title}
  {Cavity state manipulation using photon-number selective phase gates},}\
  }\href@noop {} {\bibfield  {journal} {\bibinfo  {journal} {Phys. Rev. Lett.}\
  }\textbf {\bibinfo {volume} {115}},\ \bibinfo {pages} {137002} (\bibinfo
  {year} {2015})}\BibitemShut {NoStop}%
\bibitem [{\citenamefont {Krastanov}\ \emph {et~al.}(2015)\citenamefont
  {Krastanov}, \citenamefont {Albert}, \citenamefont {Shen}, \citenamefont
  {Zou}, \citenamefont {Heeres}, \citenamefont {Vlastakis}, \citenamefont
  {Schoelkopf},\ and\ \citenamefont {Jiang}}]{Krastanov2015}%
  \BibitemOpen
  \bibfield  {author} {\bibinfo {author} {\bibfnamefont {S.}~\bibnamefont
  {Krastanov}}, \bibinfo {author} {\bibfnamefont {V.~V.}\ \bibnamefont
  {Albert}}, \bibinfo {author} {\bibfnamefont {C.}~\bibnamefont {Shen}},
  \bibinfo {author} {\bibfnamefont {C.-L.}\ \bibnamefont {Zou}}, \bibinfo
  {author} {\bibfnamefont {R.~W.}\ \bibnamefont {Heeres}}, \bibinfo {author}
  {\bibfnamefont {B.}~\bibnamefont {Vlastakis}}, \bibinfo {author}
  {\bibfnamefont {R.~J.}\ \bibnamefont {Schoelkopf}}, \ and\ \bibinfo {author}
  {\bibfnamefont {L.}~\bibnamefont {Jiang}},\ }\bibfield  {title} {\enquote
  {\bibinfo {title} {Universal control of an oscillator with dispersive
  coupling to a qubit},}\ }\href {\doibase 10.1103/PhysRevA.92.040303}
  {\bibfield  {journal} {\bibinfo  {journal} {Phys. Rev. A}\ }\textbf {\bibinfo
  {volume} {92}},\ \bibinfo {pages} {040303} (\bibinfo {year}
  {2015})}\BibitemShut {NoStop}%
\bibitem [{\citenamefont {W{\"u}nsche}(1999)}]{Wunsche1999}%
  \BibitemOpen
  \bibfield  {author} {\bibinfo {author} {\bibfnamefont {A.}~\bibnamefont
  {W{\"u}nsche}},\ }\bibfield  {title} {\enquote {\bibinfo {title} {Ordered
  operator expansions and reconstruction from ordered moments},}\ }\href@noop
  {} {\bibfield  {journal} {\bibinfo  {journal} {Journal of Optics B: Quantum
  and Semiclassical Optics}\ }\textbf {\bibinfo {volume} {1}},\ \bibinfo
  {pages} {264} (\bibinfo {year} {1999})}\BibitemShut {NoStop}%
\bibitem [{\citenamefont {B{\'e}ny}\ and\ \citenamefont
  {Oreshkov}(2010)}]{Beny2010}%
  \BibitemOpen
  \bibfield  {author} {\bibinfo {author} {\bibfnamefont {C.}~\bibnamefont
  {B{\'e}ny}}\ and\ \bibinfo {author} {\bibfnamefont {O.}~\bibnamefont
  {Oreshkov}},\ }\bibfield  {title} {\enquote {\bibinfo {title} {General
  conditions for approximate quantum error correction and near-optimal recovery
  channels},}\ }\href@noop {} {\bibfield  {journal} {\bibinfo  {journal} {Phys.
  Rev. Lett.}\ }\textbf {\bibinfo {volume} {104}},\ \bibinfo {pages} {120501}
  (\bibinfo {year} {2010})}\BibitemShut {NoStop}%
\bibitem [{\citenamefont {Bravyi}\ and\ \citenamefont
  {K{\"o}nig}(2013)}]{Bravyi2013}%
  \BibitemOpen
  \bibfield  {author} {\bibinfo {author} {\bibfnamefont {S.}~\bibnamefont
  {Bravyi}}\ and\ \bibinfo {author} {\bibfnamefont {R.}~\bibnamefont
  {K{\"o}nig}},\ }\bibfield  {title} {\enquote {\bibinfo {title}
  {Classification of topologically protected gates for local stabilizer
  codes},}\ }\href@noop {} {\bibfield  {journal} {\bibinfo  {journal} {Phys.
  Rev. Lett.}\ }\textbf {\bibinfo {volume} {110}},\ \bibinfo {pages} {170503}
  (\bibinfo {year} {2013})}\BibitemShut {NoStop}%
\bibitem [{\citenamefont {Pastawski}\ and\ \citenamefont
  {Yoshida}(2015)}]{Pastawski2015}%
  \BibitemOpen
  \bibfield  {author} {\bibinfo {author} {\bibfnamefont {F.}~\bibnamefont
  {Pastawski}}\ and\ \bibinfo {author} {\bibfnamefont {B.}~\bibnamefont
  {Yoshida}},\ }\bibfield  {title} {\enquote {\bibinfo {title} {Fault-tolerant
  logical gates in quantum error-correcting codes},}\ }\href@noop {} {\bibfield
   {journal} {\bibinfo  {journal} {Phys. Rev. A}\ }\textbf {\bibinfo {volume}
  {91}},\ \bibinfo {pages} {012305} (\bibinfo {year} {2015})}\BibitemShut
  {NoStop}%
\bibitem [{\citenamefont {Gottesman}(2009)}]{Gottesman09}%
  \BibitemOpen
  \bibfield  {author} {\bibinfo {author} {\bibfnamefont {D.}~\bibnamefont
  {Gottesman}},\ }\bibfield  {title} {\enquote {\bibinfo {title} {An
  introduction to quantum error correction and fault-tolerant quantum
  computation},}\ }in\ \href@noop {} {\emph {\bibinfo {booktitle} {Quantum
  information science and its contributions to mathematics, Proceedings of
  Symposia in Applied Mathematics}}},\ Vol.~\bibinfo {volume} {68}\ (\bibinfo
  {year} {2009})\ pp.\ \bibinfo {pages} {13--58}\BibitemShut {NoStop}%
\bibitem [{\citenamefont {Sanders}(1992)}]{Sanders92}%
  \BibitemOpen
  \bibfield  {author} {\bibinfo {author} {\bibfnamefont {B.~C.}\ \bibnamefont
  {Sanders}},\ }\bibfield  {title} {\enquote {\bibinfo {title} {Superpositions
  of distinct phase states by a nonlinear evolution},}\ }\href@noop {}
  {\bibfield  {journal} {\bibinfo  {journal} {Phys. Rev. A}\ }\textbf {\bibinfo
  {volume} {45}},\ \bibinfo {pages} {7746} (\bibinfo {year}
  {1992})}\BibitemShut {NoStop}%
\bibitem [{\citenamefont {Knill}(2005{\natexlab{b}})}]{Knill05b}%
  \BibitemOpen
  \bibfield  {author} {\bibinfo {author} {\bibfnamefont {E.}~\bibnamefont
  {Knill}},\ }\bibfield  {title} {\enquote {\bibinfo {title} {Scalable quantum
  computing in the presence of large detected-error rates},}\ }\href@noop {}
  {\bibfield  {journal} {\bibinfo  {journal} {Phys. Rev. A}\ }\textbf {\bibinfo
  {volume} {71}},\ \bibinfo {pages} {042322} (\bibinfo {year}
  {2005}{\natexlab{b}})}\BibitemShut {NoStop}%
\bibitem [{\citenamefont {Dawson}\ \emph {et~al.}(2006)\citenamefont {Dawson},
  \citenamefont {Haselgrove},\ and\ \citenamefont {Nielsen}}]{Dawson06}%
  \BibitemOpen
  \bibfield  {author} {\bibinfo {author} {\bibfnamefont {C.~M.}\ \bibnamefont
  {Dawson}}, \bibinfo {author} {\bibfnamefont {H.~L.}\ \bibnamefont
  {Haselgrove}}, \ and\ \bibinfo {author} {\bibfnamefont {M.~A.}\ \bibnamefont
  {Nielsen}},\ }\bibfield  {title} {\enquote {\bibinfo {title} {Noise
  thresholds for optical cluster-state quantum computation},}\ }\href@noop {}
  {\bibfield  {journal} {\bibinfo  {journal} {Phys. Rev. A}\ }\textbf {\bibinfo
  {volume} {73}},\ \bibinfo {pages} {052306} (\bibinfo {year}
  {2006})}\BibitemShut {NoStop}%
\bibitem [{\citenamefont {Radtke}\ \emph {et~al.}(2017)\citenamefont {Radtke},
  \citenamefont {Oi},\ and\ \citenamefont {Jeffers}}]{Radtke:2017aa}%
  \BibitemOpen
  \bibfield  {author} {\bibinfo {author} {\bibfnamefont {J.~C.~J.}\
  \bibnamefont {Radtke}}, \bibinfo {author} {\bibfnamefont {D.~K.~L.}\
  \bibnamefont {Oi}}, \ and\ \bibinfo {author} {\bibfnamefont {J.}~\bibnamefont
  {Jeffers}},\ }\bibfield  {title} {\enquote {\bibinfo {title} {Linear quantum
  optical bare raising operator},}\ }\href
  {http://stacks.iop.org/0953-4075/50/i=21/a=215501} {\bibfield  {journal}
  {\bibinfo  {journal} {J. Phys. B: At. Mol. Opt. Phys.}\ }\textbf {\bibinfo
  {volume} {50}},\ \bibinfo {pages} {215501} (\bibinfo {year}
  {2017})}\BibitemShut {NoStop}%
\bibitem [{\citenamefont {Nielsen}(2002)}]{Nielsen02}%
  \BibitemOpen
  \bibfield  {author} {\bibinfo {author} {\bibfnamefont {M.~A.}\ \bibnamefont
  {Nielsen}},\ }\bibfield  {title} {\enquote {\bibinfo {title} {A simple
  formula for the average gate fidelity of a quantum dynamical operation},}\
  }\href@noop {} {\bibfield  {journal} {\bibinfo  {journal} {Phys. Lett. A}\
  }\textbf {\bibinfo {volume} {303}},\ \bibinfo {pages} {249} (\bibinfo {year}
  {2002})}\BibitemShut {NoStop}%
\bibitem [{\citenamefont {Albert}\ \emph
  {et~al.}(2018{\natexlab{b}})\citenamefont {Albert}, \citenamefont {Mundhada},
  \citenamefont {Grimm}, \citenamefont {Touzard}, \citenamefont {Devoret},\
  and\ \citenamefont {Jiang}}]{Albert18}%
  \BibitemOpen
  \bibfield  {author} {\bibinfo {author} {\bibfnamefont {V.~V.}\ \bibnamefont
  {Albert}}, \bibinfo {author} {\bibfnamefont {S.~O.}\ \bibnamefont
  {Mundhada}}, \bibinfo {author} {\bibfnamefont {A.}~\bibnamefont {Grimm}},
  \bibinfo {author} {\bibfnamefont {S.}~\bibnamefont {Touzard}}, \bibinfo
  {author} {\bibfnamefont {M.~H.}\ \bibnamefont {Devoret}}, \ and\ \bibinfo
  {author} {\bibfnamefont {L.}~\bibnamefont {Jiang}},\ }\bibfield  {title}
  {\enquote {\bibinfo {title} {Pair-cat codes: autonomous error-correction with
  low-order nonlinearity},}\ }\href@noop {} {\bibfield  {journal} {\bibinfo
  {journal} {arXiv:1801.05897}\ } (\bibinfo {year}
  {2018}{\natexlab{b}})}\BibitemShut {NoStop}%
\bibitem [{\citenamefont {Horodecki}\ \emph {et~al.}(1999)\citenamefont
  {Horodecki}, \citenamefont {Horodecki},\ and\ \citenamefont
  {Horodecki}}]{Horodecki1999}%
  \BibitemOpen
  \bibfield  {author} {\bibinfo {author} {\bibfnamefont {M.}~\bibnamefont
  {Horodecki}}, \bibinfo {author} {\bibfnamefont {P.}~\bibnamefont
  {Horodecki}}, \ and\ \bibinfo {author} {\bibfnamefont {R.}~\bibnamefont
  {Horodecki}},\ }\bibfield  {title} {\enquote {\bibinfo {title} {General
  teleportation channel, singlet fraction, and quasidistillation},}\ }\href
  {\doibase 10.1103/PhysRevA.60.1888} {\bibfield  {journal} {\bibinfo
  {journal} {Phys. Rev. A}\ }\textbf {\bibinfo {volume} {60}},\ \bibinfo
  {pages} {1888} (\bibinfo {year} {1999})}\BibitemShut {NoStop}%
\bibitem [{\citenamefont {Aliferis}\ and\ \citenamefont
  {Cross}(2007)}]{Aliferis2007}%
  \BibitemOpen
  \bibfield  {author} {\bibinfo {author} {\bibfnamefont {P.}~\bibnamefont
  {Aliferis}}\ and\ \bibinfo {author} {\bibfnamefont {A.~W.}\ \bibnamefont
  {Cross}},\ }\bibfield  {title} {\enquote {\bibinfo {title} {Subsystem fault
  tolerance with the bacon-shor code},}\ }\href {\doibase
  10.1103/PhysRevLett.98.220502} {\bibfield  {journal} {\bibinfo  {journal}
  {Phys. Rev. Lett.}\ }\textbf {\bibinfo {volume} {98}},\ \bibinfo {pages}
  {220502} (\bibinfo {year} {2007})}\BibitemShut {NoStop}%
\bibitem [{\citenamefont {Cross}\ \emph {et~al.}(2009)\citenamefont {Cross},
  \citenamefont {Divincenzo},\ and\ \citenamefont
  {Terhal}}]{cross2009comparative}%
  \BibitemOpen
  \bibfield  {author} {\bibinfo {author} {\bibfnamefont {A.~W.}\ \bibnamefont
  {Cross}}, \bibinfo {author} {\bibfnamefont {D.~P.}\ \bibnamefont
  {Divincenzo}}, \ and\ \bibinfo {author} {\bibfnamefont {B.~M.}\ \bibnamefont
  {Terhal}},\ }\bibfield  {title} {\enquote {\bibinfo {title} {A comparative
  code study for quantum fault tolerance},}\ }\href@noop {} {\bibfield
  {journal} {\bibinfo  {journal} {Quantum Information \& Computation}\ }\textbf
  {\bibinfo {volume} {9}},\ \bibinfo {pages} {541} (\bibinfo {year}
  {2009})}\BibitemShut {NoStop}%
\bibitem [{\citenamefont {Napp}\ and\ \citenamefont
  {Preskill}(2013)}]{Napp2013}%
  \BibitemOpen
  \bibfield  {author} {\bibinfo {author} {\bibfnamefont {J.}~\bibnamefont
  {Napp}}\ and\ \bibinfo {author} {\bibfnamefont {J.}~\bibnamefont
  {Preskill}},\ }\bibfield  {title} {\enquote {\bibinfo {title} {Optimal
  bacon-shor codes},}\ }\href
  {http://dl.acm.org/citation.cfm?id=2481614.2481623} {\bibfield  {journal}
  {\bibinfo  {journal} {Quantum Info. Comput.}\ }\textbf {\bibinfo {volume}
  {13}},\ \bibinfo {pages} {490} (\bibinfo {year} {2013})}\BibitemShut
  {NoStop}%
\bibitem [{\citenamefont {Gottesman}(2019)}]{Gottesman2016}%
  \BibitemOpen
  \bibfield  {author} {\bibinfo {author} {\bibfnamefont {D.}~\bibnamefont
  {Gottesman}},\ }\bibfield  {title} {\enquote {\bibinfo {title} {Quantum fault
  tolerance in small experiments},}\ }\href {https://arxiv.org/abs/1610.03507}
  {\bibfield  {journal} {\bibinfo  {journal} {arXiv:1610.03507}\ } (\bibinfo
  {year} {2019})}\BibitemShut {NoStop}%
\bibitem [{\citenamefont {Puri}\ \emph {et~al.}(2019)\citenamefont {Puri},
  \citenamefont {St-Jean}, \citenamefont {Gross}, \citenamefont {Grimm},
  \citenamefont {Frattini}, \citenamefont {Iyer}, \citenamefont {Krishna},
  \citenamefont {Touzard}, \citenamefont {Jiang}, \citenamefont {Blais},
  \citenamefont {Flammia},\ and\ \citenamefont {Girvin}}]{Puri:2019aa}%
  \BibitemOpen
  \bibfield  {author} {\bibinfo {author} {\bibfnamefont {S.}~\bibnamefont
  {Puri}}, \bibinfo {author} {\bibfnamefont {L.}~\bibnamefont {St-Jean}},
  \bibinfo {author} {\bibfnamefont {J.~A.}\ \bibnamefont {Gross}}, \bibinfo
  {author} {\bibfnamefont {A.}~\bibnamefont {Grimm}}, \bibinfo {author}
  {\bibfnamefont {N.~E.}\ \bibnamefont {Frattini}}, \bibinfo {author}
  {\bibfnamefont {P.~S.}\ \bibnamefont {Iyer}}, \bibinfo {author}
  {\bibfnamefont {A.}~\bibnamefont {Krishna}}, \bibinfo {author} {\bibfnamefont
  {S.}~\bibnamefont {Touzard}}, \bibinfo {author} {\bibfnamefont
  {L.}~\bibnamefont {Jiang}}, \bibinfo {author} {\bibfnamefont
  {A.}~\bibnamefont {Blais}}, \bibinfo {author} {\bibfnamefont {S.~T.}\
  \bibnamefont {Flammia}}, \ and\ \bibinfo {author} {\bibfnamefont {S.~M.}\
  \bibnamefont {Girvin}},\ }\bibfield  {title} {\enquote {\bibinfo {title}
  {Bias-preserving gates with stabilized cat qubits},}\ }\href
  {https://arxiv.org/abs/1905.00450} {\bibfield  {journal} {\bibinfo  {journal}
  {arXiv:1905.00450}\ } (\bibinfo {year} {2019})}\BibitemShut {NoStop}%
\bibitem [{\citenamefont {Guillaud}\ and\ \citenamefont
  {Mirrahimi}(2019)}]{Guillaud:2019aa}%
  \BibitemOpen
  \bibfield  {author} {\bibinfo {author} {\bibfnamefont {J.}~\bibnamefont
  {Guillaud}}\ and\ \bibinfo {author} {\bibfnamefont {M.}~\bibnamefont
  {Mirrahimi}},\ }\bibfield  {title} {\enquote {\bibinfo {title} {Repetition
  cat-qubits: fault-tolerant quantum computation with highly reduced
  overhead},}\ }\href {https://arxiv.org/abs/1904.09474} {\bibfield  {journal}
  {\bibinfo  {journal} {arXiv:1904.09474}\ } (\bibinfo {year}
  {2019})}\BibitemShut {NoStop}%
\bibitem [{\citenamefont {Pantaleoni}\ \emph {et~al.}(2019)\citenamefont
  {Pantaleoni}, \citenamefont {Baragiola},\ and\ \citenamefont
  {Menicucci}}]{Pantaleoni:2019aa}%
  \BibitemOpen
  \bibfield  {author} {\bibinfo {author} {\bibfnamefont {G.}~\bibnamefont
  {Pantaleoni}}, \bibinfo {author} {\bibfnamefont {B.~Q.}\ \bibnamefont
  {Baragiola}}, \ and\ \bibinfo {author} {\bibfnamefont {N.~C.}\ \bibnamefont
  {Menicucci}},\ }\bibfield  {title} {\enquote {\bibinfo {title} {Modular
  bosonic subsystem codes},}\ }\href {https://arxiv.org/abs/1907.08210}
  {\bibfield  {journal} {\bibinfo  {journal} {arXiv:1907.08210}\ } (\bibinfo
  {year} {2019})}\BibitemShut {NoStop}%
\bibitem [{\citenamefont {Nigg}\ \emph {et~al.}(2012)\citenamefont {Nigg},
  \citenamefont {Paik}, \citenamefont {Vlastakis}, \citenamefont {Kirchmair},
  \citenamefont {Shankar}, \citenamefont {Frunzio}, \citenamefont {Devoret},
  \citenamefont {Schoelkopf},\ and\ \citenamefont {Girvin}}]{Nigg2012}%
  \BibitemOpen
  \bibfield  {author} {\bibinfo {author} {\bibfnamefont {S.~E.}\ \bibnamefont
  {Nigg}}, \bibinfo {author} {\bibfnamefont {H.}~\bibnamefont {Paik}}, \bibinfo
  {author} {\bibfnamefont {B.}~\bibnamefont {Vlastakis}}, \bibinfo {author}
  {\bibfnamefont {G.}~\bibnamefont {Kirchmair}}, \bibinfo {author}
  {\bibfnamefont {S.}~\bibnamefont {Shankar}}, \bibinfo {author} {\bibfnamefont
  {L.}~\bibnamefont {Frunzio}}, \bibinfo {author} {\bibfnamefont
  {M.}~\bibnamefont {Devoret}}, \bibinfo {author} {\bibfnamefont
  {R.}~\bibnamefont {Schoelkopf}}, \ and\ \bibinfo {author} {\bibfnamefont
  {S.}~\bibnamefont {Girvin}},\ }\bibfield  {title} {\enquote {\bibinfo {title}
  {Black-box superconducting circuit quantization},}\ }\href@noop {} {\bibfield
   {journal} {\bibinfo  {journal} {Phys. Rev. Lett.}\ }\textbf {\bibinfo
  {volume} {108}},\ \bibinfo {pages} {240502} (\bibinfo {year}
  {2012})}\BibitemShut {NoStop}%
\bibitem [{\citenamefont {Berry}\ \emph {et~al.}(2009)\citenamefont {Berry},
  \citenamefont {Higgins}, \citenamefont {Bartlett}, \citenamefont {Mitchell},
  \citenamefont {Pryde},\ and\ \citenamefont {Wiseman}}]{Berry:2009aa}%
  \BibitemOpen
  \bibfield  {author} {\bibinfo {author} {\bibfnamefont {D.~W.}\ \bibnamefont
  {Berry}}, \bibinfo {author} {\bibfnamefont {B.~L.}\ \bibnamefont {Higgins}},
  \bibinfo {author} {\bibfnamefont {S.~D.}\ \bibnamefont {Bartlett}}, \bibinfo
  {author} {\bibfnamefont {M.~W.}\ \bibnamefont {Mitchell}}, \bibinfo {author}
  {\bibfnamefont {G.~J.}\ \bibnamefont {Pryde}}, \ and\ \bibinfo {author}
  {\bibfnamefont {H.~M.}\ \bibnamefont {Wiseman}},\ }\bibfield  {title}
  {\enquote {\bibinfo {title} {How to perform the most accurate possible phase
  measurements},}\ }\href {\doibase 10.1103/PhysRevA.80.052114} {\bibfield
  {journal} {\bibinfo  {journal} {Phys. Rev. A}\ }\textbf {\bibinfo {volume}
  {80}},\ \bibinfo {pages} {052114} (\bibinfo {year} {2009})}\BibitemShut
  {NoStop}%
\bibitem [{\citenamefont {Lau}\ and\ \citenamefont
  {Plenio}(2016)}]{Lau:2016aa}%
  \BibitemOpen
  \bibfield  {author} {\bibinfo {author} {\bibfnamefont {H.-K.}\ \bibnamefont
  {Lau}}\ and\ \bibinfo {author} {\bibfnamefont {M.~B.}\ \bibnamefont
  {Plenio}},\ }\bibfield  {title} {\enquote {\bibinfo {title} {Universal
  quantum computing with arbitrary continuous-variable encoding},}\ }\href
  {\doibase 10.1103/PhysRevLett.117.100501} {\bibfield  {journal} {\bibinfo
  {journal} {Phys. Rev. Lett.}\ }\textbf {\bibinfo {volume} {117}},\ \bibinfo
  {pages} {100501} (\bibinfo {year} {2016})}\BibitemShut {NoStop}%
\bibitem [{\citenamefont {Gao}\ \emph {et~al.}(2019)\citenamefont {Gao},
  \citenamefont {Lester}, \citenamefont {Chou}, \citenamefont {Frunzio},
  \citenamefont {Devoret}, \citenamefont {Jiang}, \citenamefont {Girvin},\ and\
  \citenamefont {Schoelkopf}}]{Gao:2019aa}%
  \BibitemOpen
  \bibfield  {author} {\bibinfo {author} {\bibfnamefont {Y.~Y.}\ \bibnamefont
  {Gao}}, \bibinfo {author} {\bibfnamefont {B.~J.}\ \bibnamefont {Lester}},
  \bibinfo {author} {\bibfnamefont {K.~S.}\ \bibnamefont {Chou}}, \bibinfo
  {author} {\bibfnamefont {L.}~\bibnamefont {Frunzio}}, \bibinfo {author}
  {\bibfnamefont {M.~H.}\ \bibnamefont {Devoret}}, \bibinfo {author}
  {\bibfnamefont {L.}~\bibnamefont {Jiang}}, \bibinfo {author} {\bibfnamefont
  {S.~M.}\ \bibnamefont {Girvin}}, \ and\ \bibinfo {author} {\bibfnamefont
  {R.~J.}\ \bibnamefont {Schoelkopf}},\ }\bibfield  {title} {\enquote {\bibinfo
  {title} {Entanglement of bosonic modes through an engineered exchange
  interaction},}\ }\href {\doibase 10.1038/s41586-019-0970-4} {\bibfield
  {journal} {\bibinfo  {journal} {Nature}\ }\textbf {\bibinfo {volume} {566}},\
  \bibinfo {pages} {509} (\bibinfo {year} {2019})}\BibitemShut {NoStop}%
\bibitem [{\citenamefont {Baragiola}\ \emph {et~al.}(2019)\citenamefont
  {Baragiola}, \citenamefont {Pantaleoni}, \citenamefont {Alexander},
  \citenamefont {Karanjai},\ and\ \citenamefont
  {Menicucci}}]{Baragiola:2019aa}%
  \BibitemOpen
  \bibfield  {author} {\bibinfo {author} {\bibfnamefont {B.~Q.}\ \bibnamefont
  {Baragiola}}, \bibinfo {author} {\bibfnamefont {G.}~\bibnamefont
  {Pantaleoni}}, \bibinfo {author} {\bibfnamefont {R.~N.}\ \bibnamefont
  {Alexander}}, \bibinfo {author} {\bibfnamefont {A.}~\bibnamefont {Karanjai}},
  \ and\ \bibinfo {author} {\bibfnamefont {N.~C.}\ \bibnamefont {Menicucci}},\
  }\bibfield  {title} {\enquote {\bibinfo {title} {All-gaussian universality
  and fault tolerance with the gottesman-kitaev-preskill code},}\ }\href
  {https://arxiv.org/abs/1903.00012} {\bibfield  {journal} {\bibinfo  {journal}
  {arXiv:1903.00012}\ } (\bibinfo {year} {2019})}\BibitemShut {NoStop}%
\bibitem [{\citenamefont {Li}\ \emph {et~al.}(2017)\citenamefont {Li},
  \citenamefont {Zou}, \citenamefont {Albert}, \citenamefont {Muralidharan},
  \citenamefont {Girvin},\ and\ \citenamefont {Jiang}}]{LiZouAlbe17}%
  \BibitemOpen
  \bibfield  {author} {\bibinfo {author} {\bibfnamefont {L.}~\bibnamefont
  {Li}}, \bibinfo {author} {\bibfnamefont {C.-L.}\ \bibnamefont {Zou}},
  \bibinfo {author} {\bibfnamefont {V.~V.}\ \bibnamefont {Albert}}, \bibinfo
  {author} {\bibfnamefont {S.}~\bibnamefont {Muralidharan}}, \bibinfo {author}
  {\bibfnamefont {S.~M.}\ \bibnamefont {Girvin}}, \ and\ \bibinfo {author}
  {\bibfnamefont {L.}~\bibnamefont {Jiang}},\ }\bibfield  {title} {\enquote
  {\bibinfo {title} {Cat codes with optimal decoherence suppression for a lossy
  bosonic channel},}\ }\href
  {https://dx.doi.org/10.1103/PhysRevLett.119.030502} {\bibfield  {journal}
  {\bibinfo  {journal} {Phys. Rev. Lett.}\ }\textbf {\bibinfo {volume} {119}},\
  \bibinfo {pages} {030502} (\bibinfo {year} {2017})}\BibitemShut {NoStop}%
\bibitem [{\citenamefont {Ralph}\ \emph {et~al.}(2003)\citenamefont {Ralph},
  \citenamefont {Gilchrist}, \citenamefont {Milburn}, \citenamefont {Munro},\
  and\ \citenamefont {Glancy}}]{RalpGilcMilb03}%
  \BibitemOpen
  \bibfield  {author} {\bibinfo {author} {\bibfnamefont {T.~C.}\ \bibnamefont
  {Ralph}}, \bibinfo {author} {\bibfnamefont {A.}~\bibnamefont {Gilchrist}},
  \bibinfo {author} {\bibfnamefont {G.~J.}\ \bibnamefont {Milburn}}, \bibinfo
  {author} {\bibfnamefont {W.~J.}\ \bibnamefont {Munro}}, \ and\ \bibinfo
  {author} {\bibfnamefont {S.}~\bibnamefont {Glancy}},\ }\bibfield  {title}
  {\enquote {\bibinfo {title} {Quantum computation with optical coherent
  states},}\ }\href {https://dx.doi.org/10.1103/PhysRevA.68.042319} {\bibfield
  {journal} {\bibinfo  {journal} {Phys. Rev. A}\ }\textbf {\bibinfo {volume}
  {68}},\ \bibinfo {pages} {042319} (\bibinfo {year} {2003})}\BibitemShut
  {NoStop}%
\bibitem [{\citenamefont {Carmichael}\ \emph {et~al.}(1989)\citenamefont
  {Carmichael}, \citenamefont {Singh}, \citenamefont {Vyas},\ and\
  \citenamefont {Rice}}]{Carmichael1989}%
  \BibitemOpen
  \bibfield  {author} {\bibinfo {author} {\bibfnamefont {H.}~\bibnamefont
  {Carmichael}}, \bibinfo {author} {\bibfnamefont {S.}~\bibnamefont {Singh}},
  \bibinfo {author} {\bibfnamefont {R.}~\bibnamefont {Vyas}}, \ and\ \bibinfo
  {author} {\bibfnamefont {P.}~\bibnamefont {Rice}},\ }\bibfield  {title}
  {\enquote {\bibinfo {title} {Photoelectron waiting times and atomic state
  reduction in resonance fluorescence},}\ }\href@noop {} {\bibfield  {journal}
  {\bibinfo  {journal} {Phys. Rev. A}\ }\textbf {\bibinfo {volume} {39}},\
  \bibinfo {pages} {1200} (\bibinfo {year} {1989})}\BibitemShut {NoStop}%
\bibitem [{\citenamefont {Wiseman}\ and\ \citenamefont
  {Milburn}(1993)}]{Wiseman1993}%
  \BibitemOpen
  \bibfield  {author} {\bibinfo {author} {\bibfnamefont {H.~M.}\ \bibnamefont
  {Wiseman}}\ and\ \bibinfo {author} {\bibfnamefont {G.~J.}\ \bibnamefont
  {Milburn}},\ }\bibfield  {title} {\enquote {\bibinfo {title} {Quantum theory
  of field-quadrature measurements},}\ }\href@noop {} {\bibfield  {journal}
  {\bibinfo  {journal} {Phys. Rev. A}\ }\textbf {\bibinfo {volume} {47}},\
  \bibinfo {pages} {642} (\bibinfo {year} {1993})}\BibitemShut {NoStop}%
\end{thebibliography}

%

\end{document}